\documentclass[12pt]{article}

\usepackage{amsthm,amsmath,amsfonts,amssymb,subfigure,mathabx}
\usepackage{multirow,array,float,bm,bbm,enumerate,cite}
\usepackage[numbers]{natbib}
\usepackage[colorlinks,linkcolor=blue,citecolor=blue,urlcolor=blue]{hyperref}
\usepackage{graphicx,rotating}
\usepackage[capitalize]{cleveref}

\allowdisplaybreaks[1]

\newtheorem{definition}{Definition}
\newtheorem{proposition}{Proposition}[section]
\newtheorem{assumption}{Assumption}
\newtheorem{lemma}{Lemma}[section]
\newtheorem{theorem}{Theorem}

\newtheorem{remark}{Remark}
\newtheorem{example}{Example}
\newcommand{\bigO}{\mathcal{O}}
\newcommand{\dee}{\mathrm{d}}

\renewcommand{\Pr}{\mathbb{P}}
\def\EE{\mathbb{E}}

\title{Simultaneous Inference for Time Series Functional Linear Regression}

\author{\small
Yan Cui\\
\small Department of Mathematics, Illinois State University, Normal, IL, U.S.
\and
\small Zhou Zhou\\
\small Department of Statistical Sciences, University of Toronto, Toronto, ON, Canada
}
\date{}

\begin{document}
\maketitle
    
\begin{abstract}
	We consider the problem of joint simultaneous confidence band (JSCB) construction for regression coefficient functions of time series scalar-on-function linear regression when the regression model is estimated by a roughness penalization approach with flexible choices of orthonormal basis functions. A simple and unified multiplier bootstrap methodology is proposed for the JSCB construction which is shown to achieve the correct coverage probability asymptotically. Furthermore, the JSCB is asymptotically robust to inconsistently estimated standard deviations of the model.  The proposed methodology is applied to a time series data set of electricity demands to visually investigate and formally test the overall regression relationship, as well as perform model validation. %A uniform Gaussian approximation and comparison result over all Euclidean convex sets for normalized sums of a class of moderately high-dimensional stationary time series is established. %Finally, the proposed methodology can be applied to simultaneous inference for scalar-on-function linear regression of independent cross-sectional data.
\end{abstract}
\textbf{Keywords}: Convex Gaussian approximation; Functional time series; Joint simultaneous confidence band; Multiplier bootstrap; Roughness penalization.

\section{Introduction}\label{sec1}
%It is increasingly common to encounter time series that are densely observed over multiple oscillation periods or natural consecutive time intervals.
Modern data collection technologies routinely produce high-frequency time series with complex functional structure, where each observation consists of an entire curve recorded over a time interval. Examples arising in contemporary data science applications include electricity demand curves measured across a day, environmental monitoring data such as pollution or temperature profiles, biomedical signals, and financial intraday curves. To address the resulting statistical issues, functional (or curve) time series analysis has undergone unprecedented development over the last two decades.  See \cite{horvath2012inference} and  \cite{bosq2012linear} for excellent book-length treatments of the topic.  We also refer the readers to \cite{hormann2010weakly}, \cite{aue2015forecasting,aue2017functional}, \cite{paparoditis2018sieve}, \cite{panaretos2013fourier} and  \cite{dette2018testing}, among many others, for articles that address various modelling, estimation, forecasting and inference aspects of functional time series analysis from both time and spectral domain perspectives.

The main purpose of this article is to perform simultaneous statistical inference for time series scalar-on-function linear regression. Specifically, consider the following time series functional linear model (FLM):
\begin{equation}\label{model}
Y_i=\beta_0+\sum_{j=1}^p\int_0^1 \beta_j(t)X_{ij}(t)\dee t+\epsilon_i,~~i=1,...,n,
\end{equation}\vspace{-0.1cm}
where $\{\bm{X}_i(t):=(X_{i1}(t),...,X_{ip}(t))^\top\}_{i=1}^n$ is a $p$-variate stationary time series of known functional predictors observed on $[0,1]$, $\{Y_i\}_{i=1}^n$ is a univariate stationary time series of responses, and $\{\epsilon_i\}_{i=1}^n$ is a centered stationary time series of regression errors satisfying $\EE[X_{ij}(t)\epsilon_i]=0$ for all $t\in[0,1]$ and $j=1,2,..., p$. Observe that $\bm{X}_{i}(t)$ and $\epsilon_i$ could be dependent and $\beta_0+\sum_{j=1}^p\int_0^1 \beta_j(t)X_{ij}(t)\dee t$ can be viewed as the best linear forecast of $Y_i$ based on $\bm{X}_{i}(t)$. Despite the rapid development of estimation and forecasting methods for functional time series, rigorous statistical inference remains largely underdeveloped, especially for FLMs like \eqref{model} involving functional predictors under temporal dependence. Therefore, we are interested in constructing asymptotically correct joint simultaneous confidence bands (JSCB) for the regression coefficients  $\bm{\beta}(t):=(\beta_1(t),...,\beta_p(t))^\top$; that is, we aim to find random functions $L_{n,j}(t)$ and $U_{n,j}(t)$, $j=1,2,..., p$, such that
$$\lim_{n\rightarrow\infty}\Pr\Big(L_{n,j}(t)\le \beta_j(t)\le U_{n,j}(t),~\mbox{for all } t\in [0,1]~ \mbox{and } j=1,2,...,p\Big)= 1-\alpha$$
for a pre-specified coverage probability $1-\alpha$. The need for JSCB arises in many situations when one wants to, for instance, rigorously investigate the overall magnitude and pattern of the regression coefficient functions, test various assumptions on the regression relationship and perform diagnostic checking and model validation of \eqref{model} without multiple hypothesis testing problems.
\vspace{-0.5cm}

\subsection{Main contribution}
To better understand the main contributions of this paper, we start by emphasizing that there are two major challenges involved in the JSCB construction of $\bm{\beta}(t)$ for the time series FLM \eqref{model}. First, the estimation of model \eqref{model} is related to an ill-posed inverse problem \citep[Section 2.2]{cardot2011functional} and estimators of $\bm{\beta}(t)$ are typically not tight on $[0,1]$. As a result, it has been a difficult and open problem to investigate the large sample distributional behavior of estimators of $\bm{\beta}(t)$ uniformly across $t$. Second, the rates of convergence for various estimators of  $\bm{\beta}(t)$ are sensitive to the smoothness of $\bm{X}_{i}(t)$ and $\bm{\beta}(t)$, and the penalization parameter used in the regression. Consequently, in practice it is challenging to determine convergence rates of the estimated coefficient functions and hence the appropriate normalizing constants for uniform inference of $\bm{\beta}(t)$.

To address the aforementioned challenges, this article makes two principal contributions. First, we develop a uniform Gaussian approximation theory and comparison results over all Euclidean convex sets for the sum of stationary and weakly dependent time series in moderately high-dimensional settings. To the best of our knowledge, this presents a novel theoretical advancement as such results have not previously been investigated within the contexts of functional time series or functional data. Our findings extend the corresponding results established for independent or $m$-dependent data, as explored in \cite{Bentkus03}, \cite{Fang16}, and among others. Specifically, the uniform convex Gaussian approximation theory is established for a weighted maximum deviation uniformly over all quantiles and a wide class of weight functions (c.f. Theorem 1 in the supplemental material).  Additionally, we derive the explicit comparison bounds on the Kolmogorov distance between the sum of moderately high-dimensional Gaussian vectors, which plays a crucial role in validating the use of multiplier bootstrap procedures. These contributions significantly broaden the range of problems that can be rigorously addressed within functional time series analysis, and offer practical tools for addressing inferential challenges in both functional data and time series domains.

Second, we propose a simple and unified multiplier bootstrap procedure for constructing asymptotically correct JSCBs for $\bm{\beta}(t)$. Leveraging the fact that the convex Gaussian approximation and Gaussian comparison results hold uniformly over all Euclidean convex sets, our approach circumvents the need to explicitly derive the limiting distribution of the corresponding maximum deviation or the convergence rates, which are often analytically intractable or extremely difficult to estimate accurately in time-dependent settings. Considering the roughness penalized estimators of $\bm{\beta}(t)$ for instance, we demonstrate that the multiplier bootstrap accurately approximates their weighted maximum deviations on $[0,1]$ uniformly across all quantiles and a wide class of smooth weight functions under quite general conditions in large samples (c.f. Theorem 1 in \cref{esti}). Furthermore, the constructed JSCBs remain asymptotically correct even when the weight function is estimated inconsistently under some mild conditions, thereby adding an additional layer of robustness to the proposed methodology. The multiplier bootstrap procedure is easy to implement and practically useful, offering a powerful and versatile tool for empirical studies. To the best of our knowledge, this is the first work to provide JSCBs for FLM under general time series dependence, thereby filling an important gap in uncertainty quantification of functional time series analysis.

\subsection{Literature review}
To date, results for the time series scalar-on-function regression \eqref{model} are scarce and mainly focus on the consistency of functional principal component (FPC) based estimators \citep*{hormann2010weakly, hormann2015note}. To our best knowledge, there is no existing literature on the asymptotically correct JSCB construction for $\bm{\beta}(t)$ under time series dependence. On the other hand, there is a wealth of statistics literature dealing with estimation, convergence rates, prediction, and application of FLM \eqref{model} when the data $\{(Y_i, \bm{X}_i(t))\}_{i=1}^n$ are independent and identically distributed (i.i.d.). See, for instance, \cite{Cardot03}, \cite{CKS2009}, \cite{LiHsing07}, \cite{cai2006prediction} and \cite{james2009functional} for a far from exhaustive list of references. In parallel with the FPC based estimators, \cite{yuan2010reproducing},
developed a Reproducing Kernel Hilbert Space based approach for estimating the slope coefficient function in FLMs. Subsequent studies on minimax convergence rates, prediction, and statistical inference include \cite{cai2012minimax}, \cite{SC15}, \cite{dette2021statistical}, among others.
We also refer the readers to  \cite{cardot2011functional},   \cite{morris2015functional}, \cite{wang2016functional} and \cite{reiss2017methods} for excellent recent reviews of the topic and more references. 

Meanwhile, the last two decades also witnessed an increase in statistics literature on the inference of FLM \eqref{model} for i.i.d. data. Since a JSCB is primarily an inferential tool, we shall review this literature in more detail. The main body of the aforementioned literature consists of results related to ${\cal L}^2$-type tests on whether $\bm{\beta}(t)=0$ or a fixed known function; see for instance \cite{hilgert2013minimax}, \cite{lei2014adaptive}, \cite{kong2016classical} and \cite{su2017hypothesis}, among others. Other contributions include confidence interval construction for the conditional mean and hypothesis testing for functional contrasts \citep{SC15} and goodness of fit tests for \eqref{model} versus possibly nonlinear alternatives \citep*{horvath2012test, garcia2014goodness, mclean2015restricted}. In contrast, for independent observations, there appear to be few results in the literature discussing the construction of confidence bands for $\bm{\beta}(t)$. \cite{imaizumi2019simple} proposed a simple methodology to construct a conservative confidence band for the slope function of scalar-on-function linear regression aiming at covering ``most" of the points with a prespecified probability for independent data. Recently, \cite{dette2021statistical} constructed an asymptotically correct simultaneous confidence band for function-on-function linear regression of independent data, and the authors also discussed the scalar-on-function case briefly. However, the implementation of the latter paper requires estimating the convergence rate of the regression, which could be a relatively difficult task in moderate samples. 

On the other hand, the multiplier bootstrap technique has attracted much attention recently. Among others, \cite{CB2005} derived asymptotic consistency of the generalized bootstrap technique for estimating equations. \cite{MK2005} established the validity of the weighted bootstrap technique based on a weighted $M$-estimation framework. Later, \cite{SZ2015} considered a multiplier bootstrap procedure in the construction of likelihood-based confidence sets under possible model misspecification. Non-asymptotic results on the multiplier bootstrap have been used for high dimensional inference on hyper-rectangles and certain classes of simple convex sets that can be well approximated by (possibly higher dimensional) hyper-rectangles after linear transformations; see for instance \cite{CDK13} and \cite{CDK17} for independent data and \cite{zhou2020statistical} for functional time series. 

The paper is organized as follows. In Section \ref{sec2}, we propose the methodology of the JSCB construction based on a roughness penalization approach. Section \ref{sec4} investigates finite sample accuracy of the bootstrap methodology for various basis functions and weighting schemes using Monte Carlo experiments. We analyze a time series dataset on electricity demand curves and daily total demand in Spain in Section \ref{sec5}. The theoretical result on the multiplier bootstrap for roughness penalized estimators of $\bm{\beta}(t)$ is discussed in Section \ref{sec3}.  Additional discussions, simulations, examples, theoretical results and the proofs of all theoretical results are deferred to the supplemental material. Reproducible code for the implementation of the JSCB construction is available at \url{https://github.com/YC-stats/FLM_numerical}.

\section{Methodology}\label{sec2}
Hereafter, for simplicity we shall assume that $Y_i$ and $\bm{X}_i(t)$ are centered and hence $\beta_0=0$.  Let $H=\mathcal{L}^2([0,1])$ be the Hilbert space of all square integrable functions on $[0,1]$ with inner product $\langle x,y \rangle=\int_0^1 x(t)y(t)\dee t$. We also denote by $\mathcal{C}^d([0,1])$ the collection of functions that are $d$-times continuously differentiable with absolutely continuous $d$-th derivative on $[0,1]$. The notation ${\rm Std}(Z)$ stands for the standard deviation of a random variable $Z$. Without further specifications, the constants $C_i,i=1,2,...$ are all independent of $n$.

\subsection{Roughness Penalization Estimation}\label{sec_esti}
In order to facilitate the formulation of roughness penalization estimation, we first introduce some notation. Throughout this paper, we assume that $X_{ij}(t)$ for $j=1,...,p,~i=1,...,n$ is continuous on $[0,1]$ a.s. and hence admits the expansion
$X_{ij}(t)=\sum_{k=1}^\infty \widetilde{x}_{ij,k}\alpha_k(t),$
where $\{\alpha_k(t)\}_{k=1}^\infty$ is a set of pre-selected orthonormal basis functions of $H$. From Theorem 1 of \cite{shang2014survey}, $X_{ij}(t)$ has the standard Karhunen-Lo\`{e}ve type expansion,
\begin{equation}\label{basis3}
X_{ij}(t)=\sum_{k=1}^\infty f_{jk}x_{ij,k}\alpha_k(t),
\end{equation}
where $f_{jk}={\rm Std}(\widetilde{x}_{ij,k})$ is the standard deviation of $\widetilde{x}_{ij,k}$ and $x_{ij,k}=\widetilde{x}_{ij,k}/f_{jk}$ if $f_{jk}\neq 0$. Set $x_{ij,k}=0$ if $f_{jk}=0$. Notice that $f_{jk}$ captures the decay speed of $\widetilde{x}_{ij,k}$ as $k$ increases and the random coefficient $\{x_{ij,k}\}_{k=1}^\infty$ remains at the same magnitude with variance $1$ as $k$ increases when $f_{jk}\neq 0$.
Similarly, one can write $\beta_j(t)=\sum_{k=1}^\infty
\beta_{jk}\alpha_k(t).$ 

The roughness penalization approach to the FLM \eqref{model} is designed to deal with a penalized least squares problem. Following the method proposed by \cite{RS05}, we truncate the number of coefficient functions to finite (but diverging) dimensional spans of a priori set of basis functions 
$\{\alpha_k(t)\}_{k=1}^\infty$ and involves a roughness penalty to obtain an estimator of $\bm{\beta}(t)$. Specifically, define the truncated basis expansion of the coefficient function $\beta_j(t)$ as $\beta_{j,c_{j,n}}(t):=\sum_{k=1}^{c_{j,n}}\beta_{jk}\alpha_k(t)$ for $j=1,...,p$ with the truncation number $c_{j,n}\rightarrow\infty$. We consider the penalty term $\lambda \sum_{j=1}^p \int_0^1 [\beta_j''(t)]^2\dee t$ throughout the paper and denote the total truncation number as $c_n=\sum_{j=1}^p c_{j,n}$. Furthermore, let $\bm{\theta}_c$ be a $c_n$-dimensional block vector where $\bm{\theta}_{j,c_n}=(\theta_{j1},...,\theta_{jc_{j,n}})^\top$ is the $j$-th block with $\theta_{jk}=f_{jk}\beta_{jk}$ for $k=1,...,c_{j,n}$, then the penalized least squares estimator becomes
$$\widetilde{\bm{\theta}}_c=\left[
\frac{\bm{X}_c^\top\bm{X}_c}{n}+\bm{R}(\lambda)\right]^{-1}\frac{\bm{X}_c^\top\bm{Y}}{n},$$ where $\bm{Y}=(Y_1,...,Y_n)^\top$, $\bm{X}_c$ is an $n\times c_n$ block design matrix with its $j$th row $(x_{j1,1},\ldots,x_{j1,c_{1,n}},\\
\ldots, x_{jp,1},\ldots,x_{jp,c_{p,n}})$. The penalty term $\bm{R}(\lambda)$ is a $c_n\times c_n$ block diagonal matrix with $\lambda\bm{R}_j$ as its diagonals, and $\bm{R}_j$ is a $c_{j,n}\times c_{j,n}$ matrix with its $(k,l)$th element $R_j(k,l)=\int_0^1\alpha_k''(t)\alpha_l''(t)
\dee t/(f_{jk}f_{jl})$. We refer readers to Section D.2 of the supplemental material for further details of the roughness penalization approach. Consequently, the roughness penalized estimator is given by $\widetilde{\bm{\beta}}(t)=\bm{C}_f(t)\widetilde{\bm{\theta}}_c,$ where $\bm{C}_f(t)\in \mathbb{R}^{p\times c_n}$ is a block-sparse matrix whose $j$-th row contains $(\alpha_1(t)/f_{j1},...,\alpha_{c_{j,n}}(t)/f_{jc_{j,n}})$ in columns $(j-1)c_{j,n}+1$ to $jc_{j,n}$, and zeros elsewhere for $j=1,...,p$.

\subsection{JSCB Construction}\label{meth}

JSCB construction of $\bm{\beta}(t)$ boils down to evaluating the distributional behavior of the weighted maximum deviation 
\begin{equation*}
\Xi_{n,\bm{g}_n}:=\sqrt{n}\sup_{t\in[0,1]}|\widetilde{\bm{\beta}}(t)-\bm{\beta}(t)|_{\bm{g}_n(t)},
\end{equation*}
where $|\bm{V}(t)|_{\bm{g}(t)}=\max_{1\le j\le p}|V_j(t)/g_j(t)|$ for any function $\bm{V}(t)=(V_1(t),...,V_p(t))^\top$ and weight function $\bm{g}(t)=(g_1(t),..., g_p(t))^\top$. The weight function $\bm{g}_n(t)$ is assumed to belong to a class $\mathcal{G}$ with
\begin{align*}
\mathcal{G}&=\{\bm{f}(t): [0,1] \to \mathbb{R}^p,~\bm{f}=(f_1,...,f_p)^{\top}~\text{is a continuous}~p\text{-dimensional}\\
&\text{~vector~function~satisfying} \inf_{t\in[0,1]}\min_{1\le j\le p}f_j(t)\ge \kappa \text{~for some constant}~\kappa>0\}.
\end{align*}
For a given $\alpha\in(0,1)$, denote by $\xi_{n,\bm{g}_n}(\alpha)$ the $(1-\alpha)$-th quantile of $\Xi_{n,\bm{g}_n}$. Then a JSCB with coverage probability $1-\alpha$ can be constructed as
$\widetilde{\beta}_j(t)\pm \xi_{n,\bm{g}_n}(\alpha) g_{nj}(t)/\sqrt{n}$, $t\in[0,1]$, $j=1,2,..., p$.

Observe that the width of the JSCB is proportional to the weight function $\bm{g}_n(t)$. In practice one could simply choose some fixed weight functions such as $g_{nj}(t)\equiv 1$ which yields equal JSCB width at each $t$ and $j$. Alternatively, when the sample size is sufficiently large and temporal dependence is weak or moderately strong, we recommend choosing 
$g_{nj}(t)={\rm Std}(\widetilde{\beta}_j(t))/\int_0^1 {\rm Std}(\widetilde{\beta}_j(s))\dee s,~ j=1,..., p.$ There are two advantages to this data-driven choice of weights. First, the resulting width of the JSCB reflects the standard deviation of $\widetilde{\bm{\beta}}(t)$ which gives direct visual information on the estimation uncertainty at every $t\in[0,1]$ and $j=1,...,p$. Second, this choice of weight function yields much smaller average width of the JSCB compared to some fixed choices such as  $\bm{g}_n(t)\equiv 1$; see our simulations in Section \ref{sec4} for a finite-sample illustration. On the other hand, ${\rm Std}(\widetilde\beta_j(t))$ has to be estimated in practice. Later in this article, we shall discuss its estimation and also asymptotic robustness of our multiplier bootstrap methodology when the weight function is inconsistently estimated. 

To motivate the multiplier bootstrap, denote $\widetilde{\bm{\Sigma}}_c(\lambda)=\bm{X}_c^\top\bm{X}_c/n+\bm{R}(\lambda)$. For simplicity, we assume that each $\beta_j(t)$ has the same degree of smoothness, thereby we let each $c_{j,n}$ have the same rate of divergence. By elementary calculations and basis expansions, we have
\begin{equation}\label{penal}
\widetilde{\bm{\beta}}(t)-\bm{\beta}(t)=\bm{C}_f(t)\widetilde{\bm{\Sigma}}_c
^{-1}(\lambda)\frac{\bm{X}_c^\top
\widetilde{\bm{\epsilon}}}{n}-
\bm{C}_f(t)\widetilde{\bm{\Sigma}}_c
^{-1}(\lambda)\bm{R}(\lambda)\bm{\theta}_{c_n}+
\bigO(c_{n}^{-d_1}),
\end{equation}
where $\widetilde{\bm{\epsilon}}=(\epsilon_1+\bigO_\Pr(c_n^{-(d_1+d_2+1)}),...,
\epsilon_n+\bigO_\Pr(c_n^{-(d_1+d_2+1)}))^\top$. Hence if $c_n$ is sufficiently large and $\lambda$ is relatively small such that $\widetilde{\bm{\beta}}(t)$ is under-smoothed, i.e., the standard deviation of the estimation (captured by the first term on the right hand side of \eqref{penal}) dominates the bias asymptotically, Eq.\eqref{penal} reveals that the maximum deviation of $\sqrt{n}\widetilde{\bm{\beta}}(t)$ from $\sqrt{n}\bm{\beta}(t)$ on $[0,1]$ is determined by the uniform probabilistic behavior of $\bm{Q}_n^z(t,\lambda):=\bm{C}_f(t)\widetilde{\bm{\Sigma}}_c^{-1}
(\lambda)\bm{Z}_n^c$, where $\bm{Z}_n^c:=\frac{1}{\sqrt{n}}\sum_{i=1}^n\bm{z}_{ci}$ and $\bm{z}_{ci}=\bm{x}_{ci}\epsilon_i$ with $\bm{x}_{ci}$ the $i$-th column of $\bm{X}_c^\top$. 

There are two major difficulties in the investigation of $\bm{Q}_n^z(t,\lambda)$ uniformly in $t$. Firstly, $\{\bm{z}_{ci}\}_{i\in\mathbb{Z}}$ is typically a moderately high-dimensional time series whose dimensionality $c_n$ diverges slowly with $n$, and $\bm{Q}_n^z(t,\lambda)$ is not a tight sequence of stochastic processes on $[0,1]$. Consequently, deriving the explicit limiting distribution of the maximum deviation of $\bm{Q}_n^z(t,\lambda)$ is a difficult task. Second, the convergence rate of $\sup_{t\in[0,1]}|\bm{Q}_n^z(t,\lambda)|_{\bm{g}_n(t)}$ depends on many nuisance parameters such as the smoothness of $\bm{X}_i(t)$ and $\bm{\beta}(t)$, and the diverging rate of the truncation parameters $c_{j,n}$, which are difficult to estimate in practice. To circumvent the aforementioned difficulties, one possibility is to utilize certain bootstrap methods to avoid deriving and estimating the limiting distributions and nuisance parameters explicitly. In this article, we resort to the multiplier/wild/weighted bootstrap to mimic the probabilistic behavior of the process $\bm{Q}_n^z(t,\lambda)$ uniformly over $t$. The inference of  $\sup_{t\in[0,1]}|\bm{Q}_n^z(t,\lambda)|_{\bm{g}_n(t)}$ uniformly over all quantiles and weight functions in $\mathcal{G}$ can be transformed into investigating the probabilistic behavior of $\bm{Z}_n^c$ over a large class of moderately high-dimensional convex sets.  However, these convex sets have complex geometric structures for which results that are based on approximations on hyper-rectangles and their linear transformations are not directly applicable.  As a result, in this article we shall extend the uniform Gaussian approximation and comparison results over all high-dimensional convex sets for sums of independent and $m$-dependent data established in, for instance,  \cite{Bentkus03} and \cite{Fang16} to sums of stationary and short memory time series in order to validate the multiplier bootstrap. These results may be of wider applicability in other moderately high-dimensional time series problems. 

To be more specific, we will consider the multiplier bootstrapped sum given a block size $m$: $\bm{U}_n^{boots}=\frac{1}{\sqrt{n-m+1}}\sum_
    {j=1}^{n-m+1}\left(\frac{1}{\sqrt{m}}\sum_{i=j}^{j+m-1}\bm{z}_{ci}\right)u_j,$ where $\{u_j\}_{j=1}^{n-m+1}$ is a sequence of i.i.d. standard normal random variables which is independent of $\bm{Z}_1^n:=\{\bm{z}_{c1},...,\bm{z}_{cn}\}$.We refer to the method for constructing $\bm{U}_n^{boots}$ as the ``Multiplier Bootstrap''; this approach is built on the ideas of the weighted or wild bootstrap \citep{wu1986jackknife} and moving block bootstrap \citep{Lahiri03}. It convolutes the block sums of the random vectors $\bm{z}_{ci}$, with
i.i.d. standard normal weights $\{u_j\}$. The key observation here is that, thanks to the short range dependence assumption (c.f. \cref{ass_dep} in \cref{physi_rep}), sufficiently long block sums of moderately high dimensional vector time series preserve the main temporal dependence structure and consistently estimates the long-run variance matrix of the series.  Notice that ${\rm Var}(\bm{U}_n^{boots}|\bm{Z}_1^n)=
    \frac{1}{m(n-m+1)}\sum_{j=1}^{n-m+1}(\sum_{i=j}^{j+m-1}\bm{z}_{ci})(\sum_{i=j}^{j+m-1}\bm{z}_{ci}^\top)$ aligns with the classic lagged window estimators for long-run variance explored by \cite{newey1987}. Indeed, the conditional variance of the proposed multiplier bootstrap closely approximates the covariance structure of the random vector $\bm{Z}_n^c$ and its Gaussian analogue, as justified by the convex Gaussian comparison result. Armed with the convex Gaussian approximation theory, we validate the use of the multiplier bootstrap as an effective tool to capture the uniform probabilistic behavior of $\bm{Q}_n^z(t,\lambda)$.
    
    In practice, to implement the multiplier bootstrap, we consider the empirical bootstrap statistic denoted by $\widehat{\bm{Q}}_n^{boots}(t,\lambda)$ and its detailed construction can be found in \cref{algo}. It will be shown in \cref{thm} of \cref{sec3} that under appropriate assumptions,
$$\sup\limits_{\bm{g}_n\in \mathcal{G},x\in\mathbb{R}}\Big|
\Pr\Big(\sup_{t\in[0,1]}\big|\widehat{\bm{Q}}_n^{boots}(t,\lambda)
\big|
_{\bm{g}_n(t)}\le x\bigg|\bm{Z}_1^n\Big)-\Pr\Big(\sup_{t\in[0,1]}\big|
\bm{Q}_n^z(t,\lambda)\big|
_{\bm{g}_n(t)}\le x\Big)\Big| \to 0$$ 
as the sample size and the bootstrapped sample size diverge to infinity. It further demonstrates that the conditional distribution of $\sup_{t\in [0,1]}|\widehat{\bm{Q}}_{n}^{boots}(t,\lambda)|_{\bm{g}_n(t)}$ approximates the law of $\sup_{t\in [0,1]}|\bm{Q}_{n}^{z}(t,\lambda)|_{\bm{g}_n(t)}$ uniformly over all quantiles and weight functions in ${\cal G}$.
\vspace{-0.4cm}

\subsection{Tuning Parameter Selection and The Implementation Algorithm}\label{algo}
In this subsection, we will first discuss the issue of tuning parameter selection for the roughness penalization regression. We have three parameters to choose, that is the auxiliary truncation parameter $c_{j,n}$, the smoothing parameter $\lambda$ and the window size $m$. 
We recommend choosing $c_{j,n}=2d_{j,n}$ where $d_{j,n}$ can be selected via the cumulative percentage of total variance criterion. Specifically, we choose $d_{j,n}$ such that the quantity $\sum_{k=1}^{d_{j,n}}\rho_k/\sum_{k=1}^\infty \rho_k$ exceeds a pre-determined high percentage value (e.g., 85\% used in the simulations), where $\rho_k$ is the $k$th eigenvalue of the empirical covariance operator $\widehat{\rm Cov}(X_{ij}(s),X_{ij}(t))$. The rationale is that with the aid of the roughness penalization, $c_{j,n}$ can be chosen at a relatively large value to reduce the sieve approximation bias without blowing up the variance of the estimation.

In addition, the generalized cross validation (GCV) method can be used to choose $\lambda$, see for examples \cite{Cardot03,RO07}. To be more specific, the GCV criterion for the smoothing parameter is defined as ${\rm GCV}(\lambda)=\frac{1}{n}\sum_{i=1}^n\frac{(Y_i-\widehat{Y}_i)^2}{(1-{\rm Trace}(\bm{H})/n)^2},$
where $\bm{H}=\bm{X}_c\widetilde{\bm{\Sigma}}_c^{-1}
\bm{X}_c^\top/n$ and $\widehat{Y}_i$ is the $i$-th element of the vector $\widehat{\bm{Y}}=\bm{H}\bm{Y}$. Thus one can select $\lambda$ over a range by minimizing the above function. 

For the window size $m$, we suggest using the minimum volatility method, which was proposed by \cite{Politis99}. Denote the estimated conditional covariance matrix 
$\widehat{\bm{\Xi}}^c=\widehat{\bm{\Xi}}^c(m)=
\frac{1}{(n-m+1)m}\sum_{j=1}^{n-m+1} \left(
\sum_{i=j}^{j+m-1}\widehat{\bm{z}}_{ci}\right)
\left(\sum_{i=j}^{j+m-1}\widehat{\bm{z}}_{ci}^\top\right)$ where $\widehat{\bm{z}}_{ci}=\bm{x}_{ci}\widehat\epsilon_i$ with the estimated residual $\widehat{\epsilon}_i$. The rationale behind the minimum volatility method is that the estimator $\widehat{\bm{\Xi}}^c(m)$ becomes stable as a function of $m$ when $m$ is in an appropriate range. Let the grid of candidate window sizes be $\{m_1,...,m_{M_1}\}$. The minimum volatility criterion selects window size $m_{j_0}$ such that it minimizes the function $L(m_j):={\rm SE}\left(\left\{\widehat{\bm{\Xi}}^c(m_{j+k})\right\}_{k=-2}^2\right)$, where ${\rm SE}$ denotes the standard error
$${\rm SE}\left(\left\{\widehat{\bm{\Xi}}^c(m_{j+k})\right\}_{k=-2}^2\right)=
\left[\frac{1}{4}\sum_{k=-2}^2\left|\widehat{\bm{\Xi}}^c(m_{j+k})-
\bar{\bm{\Xi}}^c(m_j)\right|_{F}^2\right]^{1/2},$$
with $\bar{\bm{\Xi}}^c(m_j)=\sum_{k=-2}^2\widehat{\bm{\Xi}}^c(m_{j+k})/5$.

Next, we will describe the detailed steps of the multiplier bootstrap procedure for JSCB construction when the weight function $g_{nj}(t)={\rm Std}(Q_{nj}^z(t,\lambda))\\ /\int_0^1 {\rm Std}(Q_{nj}^z(s,\lambda))\,ds$, $j=1,...,p$. Note that ${\rm Std}(\sqrt{n}{\widetilde{\beta}_j(t)})
/{\rm Std}(Q_{nj}^z(t,\lambda))\\ \rightarrow 1$ in probability under some mild conditions.
\vspace{-0.3cm}

\begin{enumerate}[(a)]
	\item Select the window size $m$, such that $m\to \infty,~m=o(n)$. \label{s1}
	{\vspace{-0.25cm}}
	\item  Choose the number of basis expansion $c_{j,n}$ for each $j=1,...,p$ and choose the smoothing parameter $\lambda$. \label{s2}
	{\vspace{-0.25cm}}
	\item Fit an FLM with estimated coefficients $\widehat{x}_{ij,k}:=\widetilde{x}_{ij,k}/\widehat{f}_{jk}$ where $\widehat{f}_{jk}$ is the sample standard derivation of $\widetilde{x}_{ij,k}$, and obtain the residuals $\widehat{\epsilon}_i=Y_i-
	\sum_{j=1}^p\sum_{k=1}^{c_{j,n}}\widetilde{\theta}_{jk}\widehat{x}_{ij,k}$.  \label{s3}
	{\vspace{-0.3cm}}
	\item  Generate $B$ (say 1000) sets of i.i.d. standard normal random variables $\{u^{(r)}_j\}_{j=1}^{n-m+1}$, $r=1,2,...,B$. For each $r$, 	calculate $\widehat{\bm{Q}}_{n,r}^{boots}(t,\lambda):=\widehat{\bm{C}}_f(t)
	\widehat{\bm{\Sigma}}_c^{-1}(\lambda)\widehat{\bm{U}}_{n,r}^{boots}$, where $\widehat{\bm{U}}_{n,r}^{boots}=\frac{1}{\sqrt{n-m+1}}\sum_
	{j=1}^{n-m+1}\left(\frac{1}{\sqrt{m}}\sum_{i=j}^{j+m-1}
	\widehat{\bm{z}}_{ci}\right)u^{(r)}_j$ with $\widehat{\bm{z}}_{ci}=\widehat{\bm{x}}_{ci}\widehat{\epsilon}_i$, $\widehat{\bm{C}}_f(t)$ and $\widehat{\bm{\Sigma}}_c(\lambda)$ have similar definitions to $\bm{C}_f(t)$ and $\widetilde{\bm{\Sigma}}_c(\lambda)$ with $x_{ij,k}$ and $f_{jk}$ involved replaced by their estimates $\widehat{x}_{ij,k}$ and $\widehat{f}_{jk}$.  \label{s4}
	{\vspace{-0.25cm}}
	\item  Estimate the sample standard deviation of $\{\widehat{Q}_{nj,r}^{boots}(t,\lambda)\}_{r=1}^B$, denoted by $\widehat{\mathrm{Std}}(\widehat{Q}_{nj}^{boots}(t,\lambda))$, where $\widehat{Q}_{nj,r}^{boots}(t,\lambda)$ is the $j$-th entry of $\widehat{\bm{Q}}_{n,r}^{boots}(t,\lambda)$. Estimate $g_{nj}(t)$ by $\widehat g_{nj}(t):=\widehat{{\rm Std}}(\widehat{Q}_{nj}^{boots}(t,\lambda))/ \int_0^1 \widehat{{\rm Std}}(\widehat{Q}_{nj}^{boots}(s,\lambda))\dee s$ and obtain $M_r=\sup_{t\in [0,1]}|\widehat{\bm{Q}}_{n,r}^{boots}(t,\lambda)|_{\widehat{\bm{g}}_n(t)}$ for $r=1,2,...,B$, with $\widehat{\bm{g}}_n(t)=(\widehat g_{n1}(t), \cdots, \widehat g_{np}(t))^\top$. \label{s5}
    {\vspace{-0.8cm}}
	\item  For a given level $\alpha\in (0,1)$, let the $(1-\alpha)$-th sample quantile of the sequence $\{M_{r}\}_{r=1}^B$ be $\hat q_{n,1-\alpha}$. Then the JSCB of $\bm{\beta}(t)$ can be constructed as $\widetilde{\beta}_j(t)\pm \widehat{g}_{nj}(t)\hat q_{n,1-\alpha}/ \sqrt{n}$ for $t\in[0,1]$, $j=1,2,...,p$. \label{s6}
\end{enumerate}

In the rare case where $\widehat g_{nj}(t_0)$ is close to 0 at some $t_0$, one can raise $\widehat g_{nj}(t_0)$ to a certain threshold (say, $\max_{t\in[0,1]}\widehat{{\rm Std}}(\widehat{Q}_{nj}^{boots}(t,\lambda))/ [100\int_0^1 \widehat{{\rm Std}} (\widehat{Q}_{nj}^{boots}(s,\lambda))\,ds]$) while keeping the weight function continuous such that  $\widehat g_{nj}(t)\in {\cal G}$. As we will show in Section \ref{sec4}, the above manipulations do not affect the asymptotic validity of the bootstrap. 

If one is interested in constructing JSCB for a group of parameter functions, say $\beta_{i_1}(t),\cdots, \beta_{i_k}(t)$, then one just needs to focus on the $i_1$th, $i_2$th, $\cdots$, $i_k$th elements of the bootstrap process $\widehat{\bm{Q}}_{n,r}^{boots}(t,\lambda)$, $[\widehat{Q}_{ni_1,r}^{boots}(t,\lambda),\widehat{Q}_{ni_2,r}^{boots}(t,\lambda),\cdots,
\widehat{Q}_{ni_k,r}^{boots}(t,\lambda)]^\top $, to conduct simultaneous inference on those parameter functions. The implementation procedure is very similar to the above, and we shall omit the details.

\section{Simulation Studies}\label{sec4}
Throughout this section, we focus on the case where $p=1$. Three basis functions will be considered, i.e., Fourier bases (Fou.), Legendre polynomial bases (Leg.) and functional principal components (FPC).  Due to page constraints, we refer the readers to Section B of the supplemental material for more numerical experiments, including the JSCB construction under the Brownian motion case, comparison  with other methods for $p=1$, simulation studies for $p>1$ and rejection rates of our JSCB for $p=1$ when it is used as a test.

Recall model (\ref{model}) and restate the basis expansions as
$\beta(t)=\sum_{k=1}^\infty\beta_k\alpha_k(t)$, $X_i(t)=\sum_{k=1}^\infty \widetilde{x}_{ik}\alpha_k(t)$ when $p=1$. Next, denote $\widetilde{\bm{x}}_i=(\tilde{x}_{i1},\tilde{x}_{i2},...)^\top$,  $\bm{\eta}_i=(\eta_{i1},\eta_{i2},...)^\top$ and $\bm{D}$ as an infinite-dimensional tridiagonal matrix with $1$ on the diagonal and $1/5$ on the off-diagonal. We investigate the following model:\\
\noindent
$\bullet$ FMA(1) model. $\widetilde{\bm{x}}_i=\bm{D}(\bm{\eta}_i+\phi_1\bm{\eta}_{i-1})$, and the MA coefficient $\phi_1=0.5$ or $1$. The entries $\{\eta_{ik}\}_{k=1}^{\infty}$ of $\bm{\eta}_i$ are independent $\mathcal{N}(0,k^{-2})$ random variables.

The following basis expansion coefficients of $\beta(t)$ and the error process $\{\epsilon_i\}_{i=1}^n$ in model \eqref{model} are considered:\\ 
\noindent
$\bullet$ $\beta_1=0.8,~\beta_2=0.5,~\beta_3=-0.3$ and $\beta_k=k^{-3}, k\ge 4$. $\{\epsilon_i\}_{i=1}^n$ are dependent on $X_i(t)$. Let $\{s_i\}_{i=1}^n$ be an AR(1) process $s_i=0.2s_{i-1}+e_i$ where $\{e_i\}_{i=1}^n$ are i.i.d. standardized $t$-distributed random variables with $8$ degrees of freedom (denoted by $\sqrt{3/4}t_8$) and set $\epsilon_i=0.5s_i\widehat{x}_{i1}$ where $\widehat{x}_{i1}$ is the first FPC score of $X_i(t)$.  

When using the FPC basis, we adopt the same settings as above except that the component-wise dependence structure of $\widetilde{\bm{x}}_i$ is adjusted so that $\bm{D}$ becomes a diagonal matrix with all diagonal elements equal to $1$. This guarantees the true FPCs of $X_i(t)$ are $\{\alpha_k(t)\}_{k=1}^\infty$. 

\begin{table}
\caption{Simulated coverage probabilities with average JSCB widths in parentheses.}
\label{T1}
\centering
    \small
	\begin{tabular}{|c|c|cc|cc|}
			\hline
			&&\multicolumn{4}{c|}{$n=400$}\\
			\cline{3-6}
			&&\multicolumn{2}{c|}{$1-\alpha=0.95$}&
			\multicolumn{2}{c|}{$1-\alpha=0.90$}\\ 
			\hline
			\centering $\widehat{g}_{nj}$&Basis&$\phi_1=0.5$&$1$&$\phi_1=0.5$&$1$\\
			\hline
			\multirow{3}{*}{1}&Fou.&0.943(1.46)&0.936(1.46)&0.884(1.28)&0.886(1.29)\\
			\cline{2-6}
			&Leg.&0.952(2.95)&0.945(2.86)&0.907(2.55)&0.895(2.48)\\
			\cline{2-6}
			&FPC&0.960(1.65)&0.932(1.69)&0.903(1.45)&0.886(1.49)\\
			\hline
			\multirow{3}{*}{${\rm Std}$}
			&Fou.&0.936(1.18)&0.936(1.22)&0.875(1.07)&0.879(1.10)\\
			\cline{2-6}
			&Leg.&0.947(1.34)&0.940(1.33)&0.884(1.21)&0.885(1.21)\\
			\cline{2-6}
			&FPC&0.938(1.37)&0.924(1.40)&0.870(1.25)&0.858(1.27)\\
			\hline
			&&\multicolumn{4}{c|}{$n=800$}\\
			\cline{3-6}
			&&\multicolumn{2}{c|}{$1-\alpha=0.95$}&
			\multicolumn{2}{c|}{$1-\alpha=0.90$}\\ 
			\hline
			$\widehat{g}_{nj}$&Basis&$\phi_1=0.5$&$1$&$\phi_1=0.5$&$1$\\
			\hline
			\multirow{3}{*}{1}&Fou.&0.957(1.16)&0.958(1.12)&0.903(1.01)&0.914(0.99)\\
			\cline{2-6}
			&Leg.&0.953(2.20)&0.951(2.08)&0.900(1.90)&0.890(1.81)\\
			\cline{2-6}
			&FPC&0.945(1.17)&0.948(1.19)&0.893(1.04)&0.894(1.06)\\
			\hline
			\multirow{3}{*}{${\rm Std}$}
			&Fou.&0.949(0.92)&0.940(0.94)&0.883(0.84)&0.903(0.86)\\
			\cline{2-6}
			&Leg.&0.943(0.98)&0.941(0.97)&0.887(0.89)&0.888(0.88)\\
			\cline{2-6}
			&FPC&0.943(1.00)&0.934(1.01)&0.874(0.90)&0.871(0.92)\\
			\hline
	\end{tabular}
\end{table}
In the simulation studies, the bootstrap procedures discussed in Section \ref{algo} are employed with $B=1000$ to find the critical values $\hat{q}_{n,1-\alpha}$ at levels $\alpha=0.05$ and $0.1$. The simulation results are based on 1000 Monte Carlo experiments. Table 1 reports the simulated coverage probabilities and average JSCB widths with aforementioned three types of basis functions and two types of weight functions; i.e., $g_{nj}(t)\equiv 1$ and $\widehat{g}_{nj}(t)=\widehat{{\rm Std}}(\widehat{Q}_{nj}^{boots}(t,\lambda))\big/ \int_0^1\widehat{{\rm Std}}(\widehat{Q}_{nj}^{boots}(s,\lambda))\dee s$. For a confidence band of $\beta_j(t)$, we compute its average width as $\sum_{k=1}^{N} (2\widehat{g}_{nj}(t_k)\hat{q}_{n,1-\alpha}/\sqrt{n})/N$ according to Step (\ref{s6}) of the multiplier bootstrap procedure in \cref{algo}, where $N$ represents the total number of equally spaced grids into which the unit interval $[0,1]$ is discretized. Here, we choose $N=100$.

From \cref{T1}, we observe reasonably accurate performance of the JSCB for $n = 400$ and most of the
results for $n = 800$ are close to the nominal levels. Armed with the additional simulation results in Section B.1 of the supplemental material, the performances of the JSCB for dependent predictors and errors are similar to those for independent case, which supports our theoretical result that the multiplier bootstrap is robust to dependence between the predictors and errors. Furthermore, we observe from the simulation results that the JSCB is narrower when the weights are selected proportional to the standard deviations of the estimators. We recommend using the data-driven weights when $n\ge 400$ for Fourier and Legendre basis functions under weak dependence, and for all basis functions if $n\ge 800$. In practice, our procedure is computationally efficient as it involves selecting only two tuning parameters. On a standard laptop computer, for 1000 bootstrap replications and 1000 simulation runs, the computation takes roughly 3.5 mins and 5.9 mins for $n=400$ and 800, respectively. 

\section{Empirical Illustrations}\label{sec5}
We consider the daily curves of electricity real demands (MWh) in Spain from January 1st 2015 to December 31st 2017. These data can be obtained from  \href{https://www.esios.ree.es/en}{Red El\'{e}ctrica de Esp\~{a}na system operator}. Since the daily electricity demands on weekdays and weekends differ, in this paper we focus on the weekday curves (from Monday to Friday) with $n=782$ days. The hourly records of electricity demands in year 2011--2012 have been investigated in \cite{aneiros2016short}.

\begin{figure}[hbtp!]
	\vspace{-0.25cm}
	\centering
    \caption{Functional time series plot for log-transformed electricity demands.}
    \label{Fig_plot}
\includegraphics[width=10cm,height=4cm]{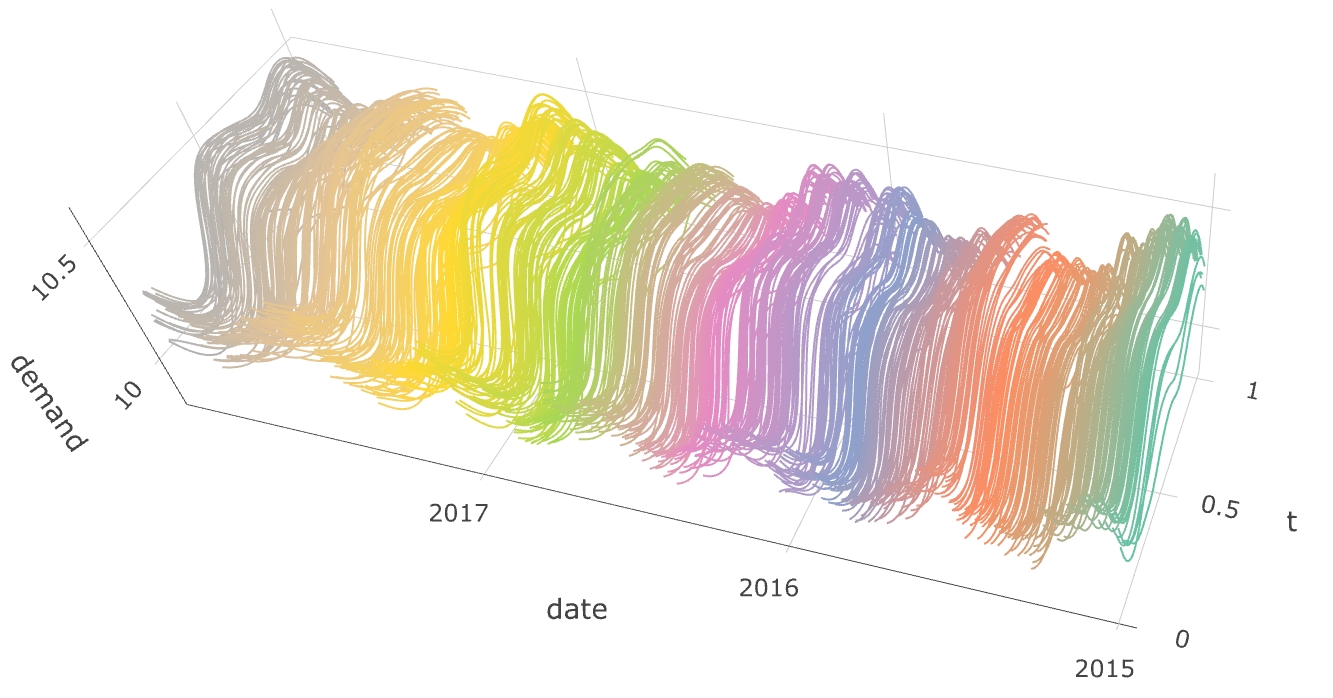}
	\vspace{-0.5cm}
\end{figure}

The original dataset are recorded by 10-minute intervals from 00:00--23:50 on each day, which consists of 144 observations. We consider the daily log-transformed real demand curves by smoothing and rescaling them to a continuous interval $[0,1]$. The plot of the smoothed functional time series is shown in \cref{Fig_plot}. The stationarity test of \cite{horvath14} is also implemented and it turns out that the  test does not reject the stationarity hypothesis at $5\%$ level during the considered period. Next, we aim to investigate the relationship between daily electricity real demand curves and future daily total demands. We explore the FLM: 
$$Y_{i+1}=\beta_0+\sum_{j=1}^3 \int_{0}^1\beta_j(t)X_{i+1-j}(t)\dee t +\epsilon_{i+1},\, i=3,4,...,781,$$ where $X_{i+1-j}(t)$ stands for the daily electricity demand curve on the $(i+1-j)$-th weekday, and $Y_{i+1}$ represents the daily total real demands on the $(i+1)$-th weekday. 

The Legendre polynomial and FPC bases are used in this example. Firstly, we select $d_{j,n}=3$ by cumulative percentage of total variance criterion in \cref{algo}, ensuring that at least 
95\% of the variability in the data is explained. We then set $c_{j,n}=2d_{j,n}=6,~j=1,2,3$. The block size is chosen as $m=16$ by minimum volatility method in \cref{algo} for aforementioned bases, the smoothing parameter $\lambda$ is selected according to the GCV criterion, yielding $8.5\times10^{-12}$ for the Legendre polynomial bases and $4.9\times 10^{-9}$ for the FPC bases. The JSCBs are constructed in 10000 bootstrap replicates.

\begin{figure}[hbtp!]
	\centering
    
	\caption{\label{Fig1}Estimated coefficient functions (black solid curve), $95\%$ JSCBs (blue dotdashed) and $95\%$ point-wise confidence bands (red dashed) with data-driven weights.}
    \vspace{-0.2cm}
	%\parbox{.75\textwidth}{	
		\centering
		\subfigure{
			\begin{minipage}{4cm}
				\centering
				\includegraphics[width=3.5cm,height=3.1cm]{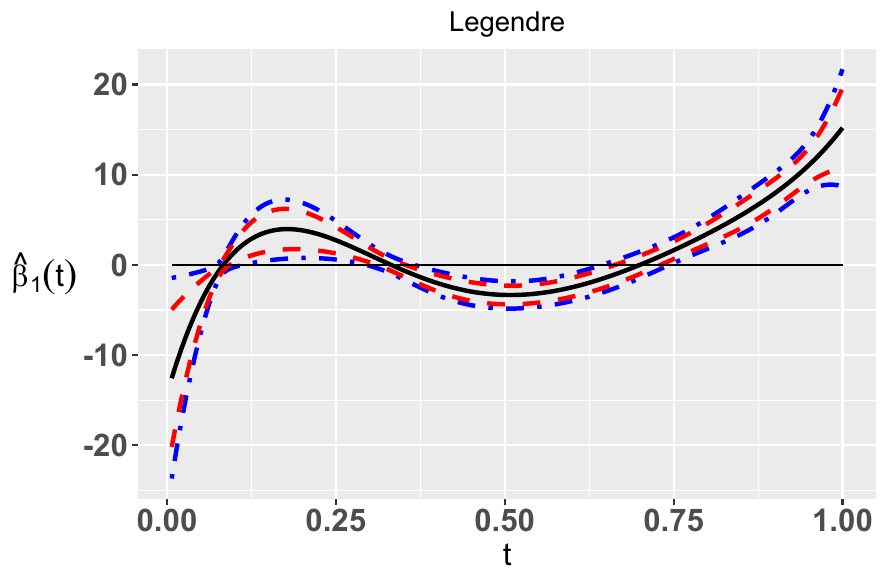}
			\end{minipage}
			\begin{minipage}{4cm}
				\centering
				\includegraphics[width=3.5cm,height=3.1cm]{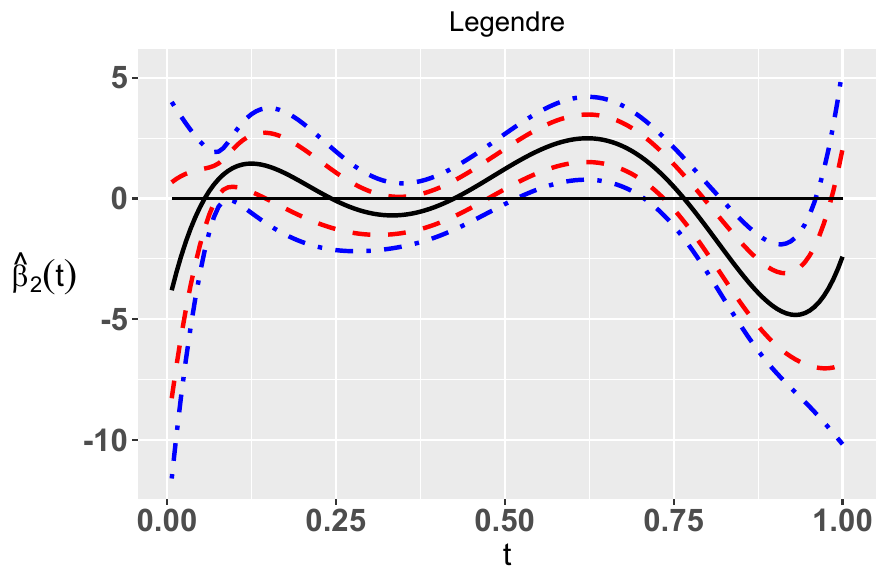}
			\end{minipage}
			\begin{minipage}{4cm}
				\centering
				\includegraphics[width=3.5cm,height=3.1cm]{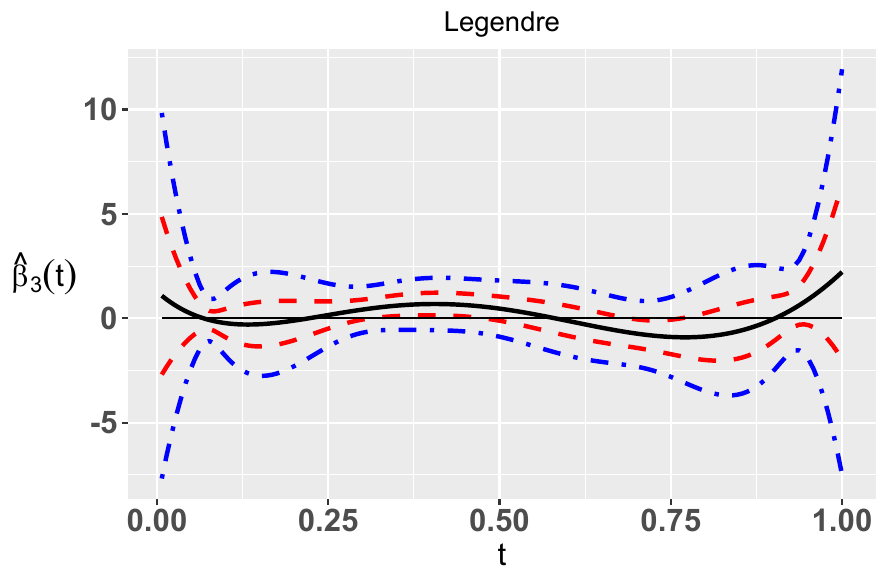}
			\end{minipage}
		}%
        \vspace{-0.1cm}
        
		\subfigure{
			\begin{minipage}{4cm}
				\centering
				\includegraphics[width=3.5cm,height=3.1cm]{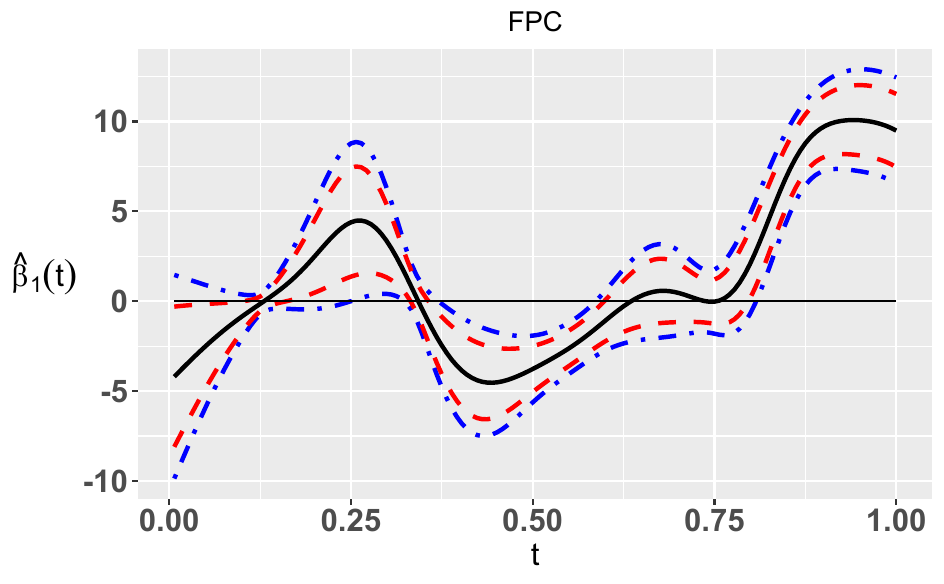}
			\end{minipage}
			\begin{minipage}{4cm}
				\centering
				\includegraphics[width=3.5cm,height=3.1cm]{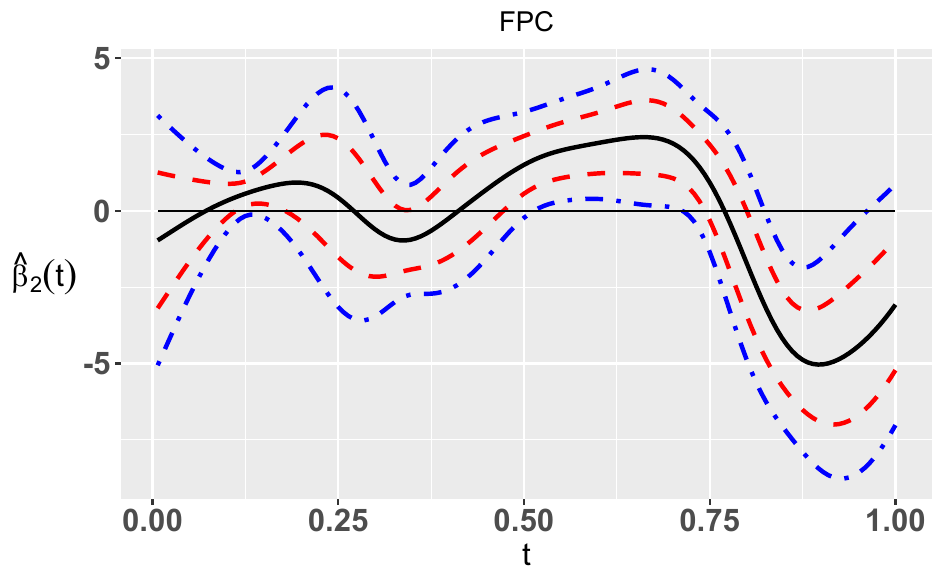}
			\end{minipage}
			\begin{minipage}{4cm}
				\centering
				\includegraphics[width=3.5cm,height=3.1cm]{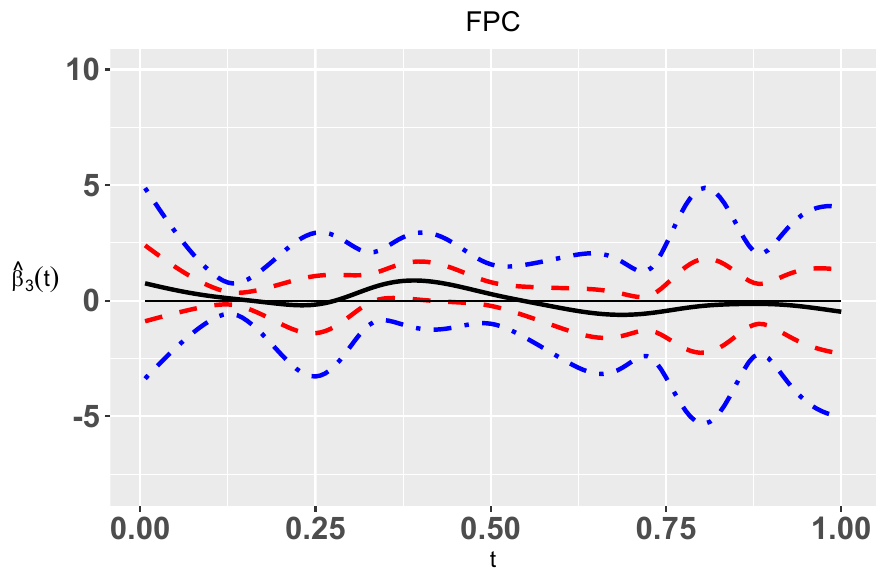}
			\end{minipage}
	}
	\vspace{-0.5cm}
\end{figure}  

We show the plots of JSCBs and point-wise confidence bands for $\beta_1(t),$ $\beta_2(t)$ and $\beta_3(t)$ in \cref{Fig1}. Note that for each time point $t\in[0,1]$, the coefficient $\beta_j(t)$ is asymptotically normal. Therefore, the point-wise confidence bands can be constructed by aggregating all point-wise confidence intervals $\widetilde{\beta}_j(t)\pm \widehat{\rm Std}(\widehat{Q}_{nj}(t,\lambda))q_{n,1-\alpha}/\sqrt{n}$ for each single $t\in[0,1]$, where $q_{n,1-\alpha}$ is the $(1-\alpha)$-th quantile of $\mathcal{N}(0,1)$. From \cref{Fig1}, both $\beta_1(t)$ and $\beta_2(t)$ are significantly non-zero, and $\beta_3(t)$ is insignificant under the JSCB construction for both bases. In particular, for both bases $\beta_1(t)$ is significantly positive at the morning and evening peak times, and significantly negative at off-peak time. By contrast, $\beta_2(t)$ is only significantly positive in the afternoon period and significantly negative in the late evening period, contributing to a slightly weaker impact on the response compared to $\beta_1(t)$. On the other hand, while the JSCB of $\beta_3(t)$ indicates that it is not significantly different from 0, the point-wise confidence bands for $\beta_3(t)$ based on both bases do not fully cover the horizontal axis $\beta_3(t)\equiv 0$, which leads to a spurious significance of $\beta_3(t)$. These findings point to a lag 2 FLM where the electricity demands over the previous two weekdays (especially the peak time from the previous weekday) is highly correlated with the total demands of the next weekday. This also indicates that the FLM successfully captures predictive signals in demand curves when forecasting the next-day total demand. Moreover,  according to the JSCB fluctuations for the coefficients $\beta_1(t)$ and $\beta_2(t)$, the constant coefficient  hypothesis $H_0: \beta_j(t)\equiv C$ for some finite constant $C$ and the linear coefficient hypothesis $H_0: \beta_j(t)=a_jt+b_j$ for some constants $a_j, b_j\in\mathbb{R}, j=1,2$ are both rejected.

The evaluations of JSCBs and computations of average widths are carried out by 144 discretized data points in a curve. The average widths of SCBs for $\beta_1(t)$ and $\beta_2(t)$ are calculated as $4.05,~3.43$ under Legendre polynomials and $4.77,~4.02$ under FPC bases, respectively. For the point-wise confidence bands of $\beta_1(t)$ and $\beta_2(t)$, they turn out to be narrower as $2.80,~2.36$ based on Legendre polynomials and $3.27,~2.66$ for FPC bases. In contrast, we also implement the confidence band constructions proposed by \cite{imaizumi2019simple} and \cite{dette2021statistical}. The former method fails to detect any statistically significant coefficient functions and yields average confidence band widths of 5.90 across 
$\beta_j(t), j=1,2,3$. For the latter approach, following the authors' recommended choice of the truncation level $c=\lceil n^{2/5}\rceil$, all three coefficient functions $\beta_j(t), j=1,2,3$ are identified as statistically significant. However, this comes at the cost of substantially wider confidence bands. To obtain a more appropriate truncation level for this dataset with a large sample size, we employ our cumulative percentage of total variance method to determine $c$. For this case, only $\beta_1(t)$ remains statistically significant, the resulting confidence bands remain considerably wider than those produced by our approach, with average widths of 9.82, 8.38, and 4.25 for $\beta_j(t), j=1,2,3$, respectively.

In line with the AE's suggestion, we extend our FLM to incorporate weather-related predictors. We consider 2015-2017 daily maximum temperatures in Spain obtained from \href{https://www.aemet.es/es/portada}{Agencia
Estatal de Meterolo\'{i}ga} and construct the national daily maximum temperature ($T_i$) as a population-weighted average across provinces using population data from \href{https://www.ine.es/}{Instituto Nacional de Estad\'{i}stica}. To capture the nonlinear impact of temperature on the total demand, we introduce indicator variables  $Z_{i1}=I(T_i\ge r_1)$ and $Z_{i2}=I(T_i\le r_2)$ representing potential air conditioning and heating effects, respectively. By evaluating the linear association between the total demand and temperature variables across candidate thresholds, we select $r_1=31.7^\circ$C and 
$r_2=14.6^\circ$C, corresponding to temperatures at which the use of cooling and heating systems becomes more pronounced. Based on confidence intervals of the coefficients, we find that the temperature predictors are statistically significant at the 5\% level. It suggests that temperature variables provides valuable information for forecasting total electricity demand.

To illustrate the prediction performance of the selected FLM in comparison with classic time series prediction methods, we apply a univariate ARIMAX model with exogenous temperature variables to predict total electricity demand on weekdays. The optimal model specification is ARIMAX(6,0,0) according to the AIC criterion. 

We compare the out-of-sample forecasts for the last seven days' total electricity demands for our FLM with temperature covariates and the above-mentioned ARIMAX(6,0,0) model. Our FLM achieves a mean squared prediction error of $0.0080$, an $58.75\%$ improvement over that of the ARIMAX model which is $0.0127$. These results demonstrate that functional linear regression modeling better captures high-frequency electricity demand patterns than univariate ARIMAX, leading to more accurate predictions in this dataset. 

\begin{figure}[hbtp!]
	\centering
	\caption{\label{prec}The 95\% Prediction intervals (blue) for the log total demand (black) over the last 15 weekdays.}
	\centering
	\includegraphics[width=6cm,height=4cm]{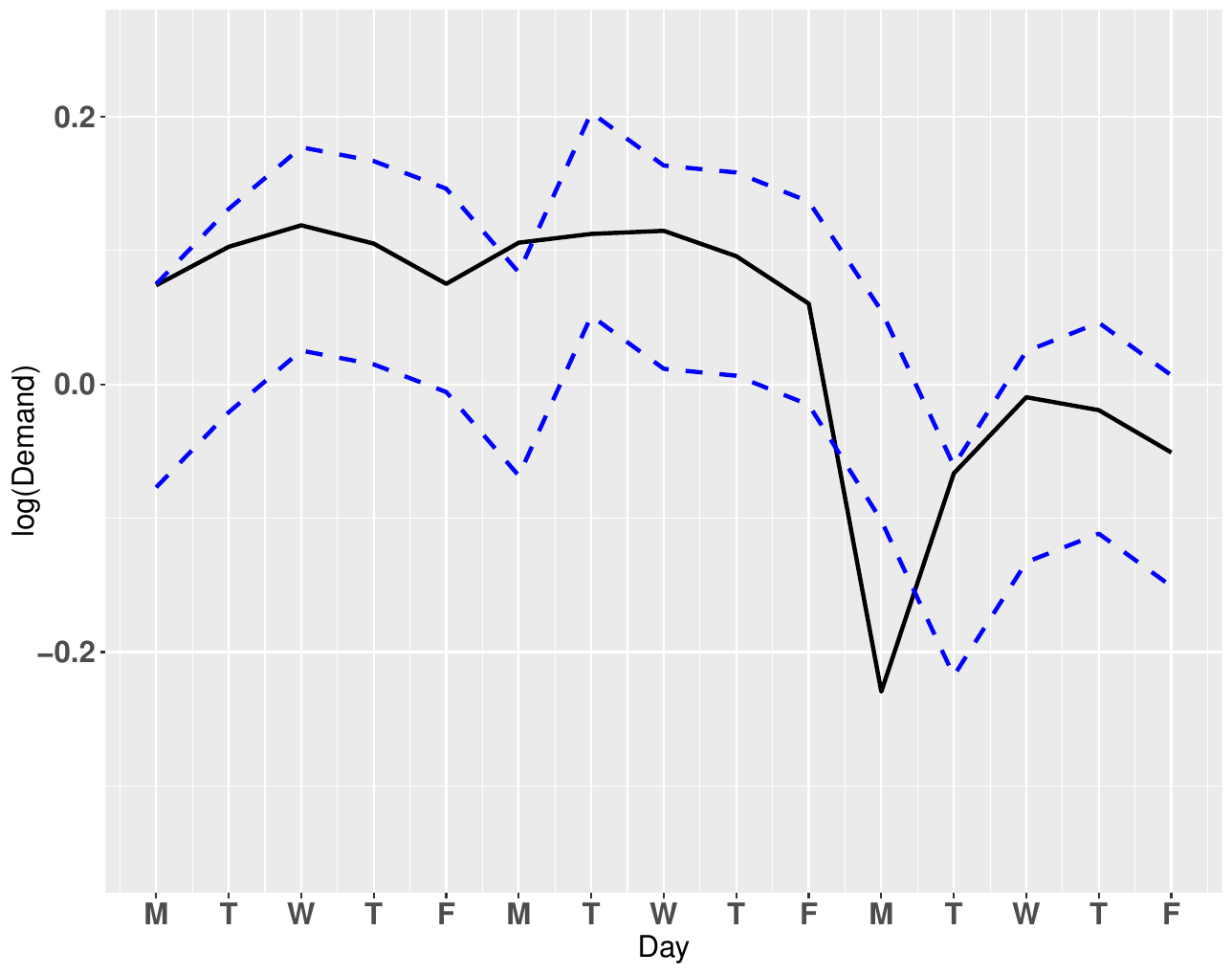}
\end{figure}

Finally, we perform an additional out-of-sample forecast to examine the prediction intervals. Specifically, we trained the model using the first 767 weekdays (with lag 2 structure and two exogenous temperature predictors) and predicted the total demand for the last 15 weekdays (three weeks) of 2017. The resulting 95\% prediction intervals are in \cref{prec}. Among all 15 weekdays, only two observations fall outside the 95\% prediction interval. Interestingly, both correspond to Mondays. This pattern is consistent with the fact that our model is trained exclusively on weekday dynamics and does not explicitly incorporate weekend effects. Since Monday demand may depend partly on weekend consumption patterns, the reduced accuracy for those days is not unexpected.

\section{Theoretical Results}\label{sec3}
In this section, we first model the functional time series $\bm{X}_i(t)$ from a basis expansion and nonlinear system \citep{Wu05} point of view, and then investigate the multiplier bootstrap theory. 

\subsection{Functional Time Series Models}\label{physi_rep}
Based on the basis expansion \eqref{basis3}, we aim to utilize a general time series model for $\{x_{ij,k}\}_{k=1}^\infty$ from a nonlinear system point of view. This will serve as a preliminary for our theoretical investigations. 
\begin{definition}\label{def1}
	Assume that $\{x_{ij,k}\}_{k=1}^\infty$ satisfy $\Vert x_{ij,k}\Vert_q<\infty$, $q> 9$, where $\|Z\|_q:=\EE[|Z|^q]^{1/q}$ for a random variable $Z$. We  say that $\{\bm{X}_i(t)\}_{i\in\mathbb{Z}}$ admits a physical representation if for each fixed $j$ and $k$, the stationary time series $\{x_{ij,k}\}_{i=-\infty}^\infty$ can be written as $%\label{eq:nonlinear}
	x_{ij,k}=G_{jk}(\mathcal{F}_i),$ where $G_{jk}$ is a measurable function and $\mathcal{F}_i=(...,\eta_{i-1},\eta_i)$ is the one-sided shift process with $\eta_i$ being i.i.d. random elements. For $l\ge 0$, define the $l$-th physical dependence measure for the functional time series $\{\bm{X}_i(t)\}$ with respect to the basis  $\{\alpha_k(t)\}_{k=1}^\infty$ and moment $q$ as $
	\delta_x(l,q)=\sup_{1\le j\le p,1\le k<\infty}\Vert G_{jk}(\mathcal{F}_i)-G_{jk}(\mathcal{F}_{i,l})\Vert_q,$
	where $\mathcal{F}_{i,l}=(\mathcal{F}_{i-l-1},\eta_{i-l}^\ast,\eta_{i-l+1},...,
	\eta_i)$ is a coupled process with $\eta_{i-l}^\ast$ an i.i.d. copy of $\eta_{i-l}$.
\end{definition}

Note that in the above definition, $\delta_x(l,q)$ does not depend on $i$. The above formulation of time series $x_{ij,k}$ can be viewed as a physical system where functions $G_{jk}$ are the underlying data generating mechanisms and $\{\eta_i\}$ are the shocks or innovations that drive the system. Meanwhile, $\delta_x(l,q)$ measures the temporal dependence of $\{\bm{X}_i(t)\}$ by quantifying the corresponding changes in the system's output uniformly across all basis expansion coefficients when the shock of the system $l$ steps ahead is changed to an i.i.d. copy. We refer to \cite{Wu05} for more discussions of the physical dependence measures with examples on how to calculate them for a wide range of linear and nonlinear time series models.

Definition \ref{def1} is related to the class of functional time series formulated in \cite{zhou2020statistical}. The difference is that in \eqref{basis3} we separate the standard deviation $f_{jk}$ from $\widetilde{x}_{ij,k}$ and  the functional time series model in \cite{zhou2020statistical} is formulated without this extra step. Standardization of the basis expansion coefficients is needed in the fitting of the FLM \eqref{model} to avoid near singularity of the design matrix.  Furthermore, Definition \ref{def1} is also related to the concept of  $m$-approximable functional time series introduced in \cite{hormann2010weakly}, as both formulations utilize the concepts of Bernoulli shifts and coupling. The difference lies in our adaptation of the basis expansion, similar to that in \cite{zhou2020statistical}, which separates the functional index $t$ and time index $i$ and hence makes it easier technically to investigate the behavior of various estimators of $\bm{\beta}(t)$ uniformly over $t$. Now, we impose an assumption on the speed of decay for the dependence measures $\delta_x(l,q)$.
\begin{assumption}\label{ass_dep}
	There exists some constant $\tau>5$ such that for some universal constant $C_1>0$, the physical dependence measure satisfies
	$\delta_x(l,q) \le C_1(l+1)^{-\tau},~l\ge 0.$
\end{assumption}
Assumption \ref{ass_dep} is a mild short-range dependence assumption which asserts that the temporal dependence of the functional time series $\bm{X}_i(t)$ decays at a sufficiently fast polynomial rate. For independent functional data, the condition $\delta_x(l,q) \le C_1(l+1)^{-\tau}$ is automatically satisfied as there is no temporal dependence ($\delta_x(l,q)=0$). In Section D of the online supplemental material, we will provide two examples on how to calculate $\delta_x(l,q)$ for a class of functional MA$(\infty)$ and functional AR(1) processes, respectively.

\subsection{Validating The Multiplier Bootstrap}\label{esti}
Adopting the nonlinear system point of view \citep{Wu05}, we model the stationary error process $\{\epsilon_i\}_{i=1}^n$ as
	$\epsilon_i=G(\mathcal{F}_i)$ for some measurable function $G$. Observe that both $\bm{X}_i(t)$ and $\epsilon_i$ are generated by the same set of shocks $\{\eta_j\}_{j\in\mathbb{Z}}$ and hence they could be statistically dependent. Further define $\bm{\Sigma}_c(\lambda)= \frac{\EE(\bm{X}_c^\top\bm{X}_c)}{n}+\bm{R}(\lambda)$. To establish the main results, we need the following conditions:

\begin{assumption}\label{ass0}
	For some non-negative integers $d_1$ and $d_2$ with $d_1\ge d_2$, assume that $\beta_j(t)\in \mathcal{C}^{d_1}([0,1])$ and $X_{ij}(t)\in \mathcal{C}^{d_2}([0,1])$ a.s.. Suppose that $|\beta_{jk}|\le C_2k^{-(d_1+1)}$, $\forall j=1,...,p$ and $C_2>0$ is some finite constant. Moreover, assume that the random coefficients $\left\Vert\widetilde{x}_{ij,k}\right\Vert_2\le C_3k^{-(d_2+1)}$ for $i=1,...,n,~j=1,...,p$.
\end{assumption}

\begin{assumption}\label{ass_lower}
		The smallest eigenvalue of $\bm{\Sigma}_c(\lambda)$ is greater or equal to some constant $\sigma>0$, which does not depend on $c_n$ or $n$.
	\end{assumption}
	
	\begin{assumption}\label{ass_eps}%C
	The stationary process $\{\epsilon_i\}_{i=1}^n$ satisfies $\Vert \epsilon_i\Vert_q <\infty,~q>9$, there exist some constants $C_4>0$, and $\tau>5$ such that its physical dependence measure achieves $\delta_\epsilon(l,q)\le C_4(l+1)^{-\tau},~l\ge 0$.
	\end{assumption}
	
	\begin{assumption}\label{ass_moment}
		$\max_{1\le j\le c}\EE|z_{ci,j}|^q\le C_q<\infty$ for some $C_q>0$.
	\end{assumption}
	
	The above conditions are mild and are needed for establishing a Gaussian approximation and comparison theory for roughness-penalized estimators. Let $d\ge 0$  be an integer. It is well-known that for a general $\mathcal{C}^{d}([0,1])$ function, the fastest decay rate for its $k$-th basis expansion coefficient is $O(k^{-d-1})$ for a wide class of basis functions  \citep{chen2007large}. For instance, the Fourier basis (for periodic functions), the weighted Chebyshev polynomials \citep{trefethen2008gauss} and the orthogonal wavelets with degree $m\ge d$ \citep{Meyer92} admit the latter decay rate under some additional mild assumptions on the behavior of the function's $d$-th derivative. Hence Assumption \ref{ass0} essentially requires that the basis expansion coefficients of $\beta_j(t)$ and $X_{ij}(t)$ decay at the fastest rate. On the other hand, we remark that the basis expansion coefficients may decay at slower speeds for some orthonormal bases. An example is Legendre polynomials, where the coefficients decay at an $O(k^{-d-1/2})$ speed \citep{wang2012convergence}. For basis functions whose coefficients decay at slower rates, following the proofs of this paper it is obvious that the multiplier bootstrap method for the JSCB construction is still asymptotically valid under the corresponding restrictions on the tuning parameters. However, in this case the estimates of $\bm\beta(t)$ will converge at a slower speed, and the bootstrap approximation may be less accurate. For the sake of brevity, we shall stick to Assumption \ref{ass0} for our theoretical investigations throughout this paper. \cref{ass_lower} ensures positive definiteness of the design matrix in order to avoid multicollinearity. \cref{ass_eps} is a short range dependent condition on $\{\epsilon_i\}_{i=1}^n$ in accordance with \cref{ass_dep}. Finally, \cref{ass_moment} puts some moment restrictions on the random variable $z_{ci,j}$. We refer the readers to Section E.3 for examples and discussion on these assumptions.

    Let $\widetilde{\bm{R}}_j$ be a $c_{j,n}\times c_{j,n}$ matrix with its $(k,l)$ element $\widetilde{R}_j(k,l)$ $=\int_0^1\alpha_k''(t)\alpha_l''(t)\dee t$. The following additional assumptions are needed for the validation of the multiplier bootstrap.
\begin{assumption}\label{ass_penalty}
	For $j=1,\cdots,p$, we assume $|\widetilde{\bm{R}}_j|\asymp c_{j,n}^{2\gamma}$, where $\gamma$ is a positive constant depending on the basis function, and $|A|$ is the spectral norm (largest singular value) of a matrix $A$.
\end{assumption}
\vspace{-0.5cm}
\begin{assumption}\label{ass_conti}
	For each $k\ge 1$, there exists some constant $\psi\ge 0$ such that $|\alpha_k(t)|_\infty \le C_5k^\psi$ for some positive constant $C_5$, where $|\cdot|_\infty$ denotes the uniform norm of a bounded function, i.e, if $f: \mathcal{X}\to\mathbb{R}$ then $|f|_\infty=\sup_{x\in\mathcal{X}}|f(x)|.$ In addition, for any $t_1,~t_2\in [0,1]$ and $k\ge 1$, there exists a nonnegative constant $\phi$ and some finite constant $C_6$ such that $|\alpha_k(t_1)-\alpha_k(t_2)| \le C_6k^{\phi}|t_1-t_2|.$
\end{assumption}
\vspace{-0.3cm}
\begin{assumption}\label{f_bound}
	For sufficiently large $k$ and $j=1,\cdots,p$, $|f_{jk}|\ge C_7k^{-(d_2+1)}$, where $C_7>0$ is a universal finite constant.
\end{assumption}

Assumption \ref{ass_penalty} is mild and can be easily checked for many frequently-used basis functions, such as the Fourier basis $(\gamma=2)$ and the Legendre polynomial basis $(\gamma=4)$. Meanwhile, Assumption \ref{ass_conti} is satisfied by most frequently-used sieve bases. For instance, the pair $(\psi,\phi)=(0,1)$ for the trigonometric polynomial series, $(\psi,\phi)=(1/2,0)$ for the polynomial spline basis functions and $(\psi,\phi)=(1,5/2)$ for the normalized Legendre polynomial basis. We refer to Section E.3 of the supplemental material for a detailed discussion of the above claims on Assumptions \ref{ass_penalty} and \ref{ass_conti}. Assumption \ref{f_bound} is frequently adopted in the FLM literature, for example \cite{Hall07}, which imposes a lower bound on the decay rate of $f_{jk}$. Similar to our discussion of Assumption \ref{ass0}, for $\mathcal{C}^{d_2}[0,1]$ functions the fastest decay speed of their basis expansion coefficients is of the order $\bigO(k^{-(d_2+1)})$ for a wide class of basis functions. Hence Assumption \ref{f_bound} is mild. 

Define $\bm{\Xi}^c:=\frac{1}{n}\EE\left(\sum_{i=1}^n
\bm{z}_{ci}\right)\left(\sum_{i=1}^n\bm{z}_{ci}^\top\right)$ and the Kolmogorov distance by
\vspace{-0.2cm}
{\small
\begin{equation*}
\widehat{\mathcal{K}}(\widehat{\bm{U}}_n^{boots},
\bm{Z}_n^c)
:=\sup\limits_{\bm{g}_n\in \mathcal{G},x\in\mathbb{R}}\Big|
\Pr\Big(\sup_{t\in[0,1]}\big|\widehat{\bm{Q}}_n^{boots}(t,\lambda)
\big|
_{\bm{g}_n(t)}\le x\bigg|\bm{Z}_1^n\Big)-\Pr\Big(\sup_{t\in[0,1]}\big|
\bm{Q}_n^z(t,\lambda)\big|
_{\bm{g}_n(t)}\le x\Big)\Big|.
\end{equation*}}
Then we have the following theorem on the consistency of the proposed multiplier bootstrap method.

\begin{theorem}\label{thm}
Under Assumptions \ref{ass_dep}--\ref{f_bound}, the smallest eigenvalue of $\bm{\Xi}^c$ is bounded below by some constant $\widehat{b}>0$ and $m=\bigO(n^{1/3})$. Define $\mathcal{B}_n^\epsilon=\{\omega:\Delta_n(\omega):=|\widetilde{
		\bm{\Xi}}^c-\bm{\Xi}^c|_F \le C_8c_nn^{-1/3}h_n\},$ where $\omega$ represents the element in the probability space, $|\cdot|_F$ indicates Frobenius norm, $h_n$ diverges to infinity at an arbitrarily slow rate and $C_8>0$ is a finite constant. Then $\Pr(\mathcal{B}_n^\epsilon)=1-o(1)$. Under the event $\mathcal{B}_n^\epsilon$, we have 
	$$\widehat{\mathcal{K}}(\widehat{\bm{U}}_n^{boots},\bm{Z}_n^c)
	\le C_9\chi_n,$$ 
    where $C_9>0$ is some finite constant and $\chi_n\to 0$ as $n,B\to \infty$. Further suppose the conditions in Section E.1 of the supplemental material hold true, the JSCB achieves
\begin{equation}\label{cover1}
\begin{aligned}
	\lim_{n\to \infty}\lim_{B\to\infty}\Pr\Big(\beta_{j}(t)\in &\left[\widetilde{\beta}_j(t)-\frac{\hat{q}_{n,1-\alpha}
	\widehat{g}_{nj}(t)}{\sqrt{n}},
	\widetilde{\beta}_j(t)+\frac{\hat{q}_{n,1-\alpha}\widehat{g}_{nj}(t)}
	{\sqrt{n}}\right]\\
	&~~\text{for}~\forall t\in [0,1] ~\text{and}~j=1,...,p\Big)=1-\alpha.
	\end{aligned}
    \end{equation}
\end{theorem}

\cref{thm} states that under certain regularity conditions, the JSCB achieves the correct coverage probability asymptotically. In particular, $\chi_n$ captures the rate of the bootstrap approximation to $\bm{Q}_n^z(t,\lambda)$, including the Gaussian approximation error, multiplier bootstrap approximation error, and estimation error; see Section E.1 of the supplemental material for its detailed representation. The rate $m=\bigO(n^{1/3})$ is the optimal one that balances the bias and variance of the bootstrapped covariance matrices. Since Theorem C.1 in Section C of the supplemental material and \cref{thm} above are established uniformly over all weight functions in $\mathcal{G}$, the JSCB achieves asymptotically correct coverage probability without assuming that $\widehat{g}_{nj}(t)$ is a uniformly consistent estimator of $\mbox{Std}(\widetilde\beta_j(t))/\int_0^1\mbox{Std}(\widetilde\beta_j(s))\dee s$ as long as $\widehat{\bm{g}}_{n}(t)\in\mathcal{G}$ almost surely. Hence the multiplier bootstrap is asymptotically robust to inconsistently estimated weight functions. The price one has to pay for inconsistently estimated weight functions is that the average width of the JSCB may be inflated.

\subsection{Data-Driven Basis Functions Based on Functional Principal Components}
A popular data-driven orthonormal basis in functional data analysis is the FPC. Observe that FPCs have to be estimated from the data which inevitably causes estimation errors. When one employs data-driven basis functions such as the FPCs to fit model \eqref{model}, the additional estimation error must be taken into account. Note that $f_{jk}^2=\EE(\widetilde{x}_{ij,k}^2)$ are the eigenvalues of the corresponding covariance operator. Throughout this subsection, for any given $j$, we assume that $f_{j1}>f_{j2}>...>f_{jc_{j,n}}>0$. This assumption implies that the first $c_{j,n}$ eigenvalues are separated, which is commonly used in the theoretical investigation for FPC-based methods. The next proposition establishes the asymptotic validity of the bootstrap for the FPC basis functions under some extra condition. 

\begin{proposition}\label{prop:FPC}
	Under Assumptions \ref{ass_dep}--\ref{f_bound}, the multiplier bootstrap result hold true for the FPC basis functions. Further assume that 
    {\vspace{-0.3cm}}
    \begin{equation}\label{fpc_constraint}
	\lambda^{-\frac{2d_2+3} {4(\gamma+d_2+1)}}>c_n^{4(d_2+1)-d_1}/\sqrt{n}
    {\vspace{-0.8cm}}
	\end{equation} 
	and the conditions in Section E.1 of the supplemental material are satisfied, then \eqref{cover1} holds for the FPC basis.
\end{proposition}

This proposition imposes an extra constraint \eqref{fpc_constraint} on the smoothness of $\bm{\beta}(t)$ and $\bm{X}_{i}(t)$ to make the additional bias term resulting from FPC estimation negligible compared to the standard deviation term. 

\small
\bibliographystyle{plain}
\bibliography{scalar} 
    
\clearpage
\appendix

\begin{center}
    {\Large \textbf{Supplementary Material of "Simultaneous Inference for Time Series Functional Linear Regression"}}
\end{center}

\vspace{1em}
% Redefine section numbering
\renewcommand{\thesection}{\Alph{section}}

\setcounter{section}{0}

\begin{abstract}
		 Section \ref{sec_dis} of this supplemental material discusses several important aspects related to the main article, including the effects of pre-smoothing, practical choices of the basis functions and extensions to regression models with functional response. Section \ref{sec_simu} provides additional simulation results under various numbers of predictors and data generating mechanisms. In Section \ref{gau}, we establish a Gaussian approximation theory for the roughness penalization estimation method, which may be of independent interest. Examples on calculations of the physical dependence measure for a class of functional MA$(\infty)$ models and a class of functional AR(1) models are presented in Section \ref{a}. Additional theoretical results, together with proofs of all main results, are provided in Section \ref{proof}.
	\end{abstract}
	
		\section{Discussion on the Main Article}\label{sec_dis}
We now discuss some issues related to the practical implementation of the regression as well as possible extensions. Firstly, the predictors $\bm{X}_i(t)$ are typically only discretely observed with noises. Hence, pre-smoothing is required to transfer the discretely observed predictors into continuous curves which inevitably produces some smoothing errors.  In this article, for the purposes of brevity and to keep the discussion focused, we assume that the smooth curves of $\bm{X}_i(t)$ are observed. It can be seen from the proofs that the results of the paper hold under the densely observed functional data scenario as long as the smoothing error achieves an order $O_\Pr(\log n/\sqrt{n})$ uniformly. Smoothing for time series or densely observed functional data has been intensively investigated in the literature; see for instance \cite{hall2006properties}, \cite{zhang2007statistical} and \cite{wu2007inference}, among many others. Though the aforementioned references are not exactly intended for functional time series, their results can be extended to the functional time series setting, which we will pursue in a separate future work. On the other hand, we do not expect that our theory and methodology will directly carry over to the sparsely observed functional data setting \citep{yao2005functional}, and the corresponding investigations are beyond the scope of the current paper.    

Secondly, we note that the choice of basis functions is a non-trivial task in practice. In the literature, there are several discussions with respect to the choice of basis functions, or methods for the functional linear regression models in general. See for instance Section 6.1 in \cite{reiss2017methods} and the references therein. Here we shall add some additional notes. For functional time series whose observation curves are clearly periodic, such as the yearly temperature curves, the Fourier basis is a natural choice. Similar choices can be made based on prior knowledge of the shapes of the observation curves in various scenarios. Our limited simulation studies and data analysis in Sections 3 and 4 of the main article suggested many popular classes of basis functions produce similar estimates of the regression curves and comparable inference results, which demonstrates a certain level of robustness towards the basis choices.

Finally, in some real data applications the response time series may be function-valued as well. One prominent example is functional auto-regression \citep{bosq2012linear}. We hope that our multiplier bootstrap methodology as well as the underlying Gaussian approximation and comparison results will shed light on the simultaneous inference problem for FLM with functional responses. We will investigate this direction in a future research endeavor.

\section{Additional Simulation Results}\label{sec_simu}
	In this section we would like to conduct some additional simulation studies that complement those in Section 3 of the main paper. The significance levels are set at $\alpha = 0.05$ and $0.1$. We choose the sample size as $n=400, 800$ and the bootstrap replications $B=1000$ based on 1000 simulation runs throughout this section.

\subsection{Simulation Studies for an FAR(1) Model}

To measure the effects of temporal dependence, we also conduct a simulation study for an FAR(1) model. First, we consider $\beta(t)=\sum_{k=1}^\infty \beta_k\alpha_k(t), X_i(t)=\sum_{k=1}^\infty \widetilde{x}_{ik}\alpha_k(t)$ and denote $\widetilde{\bm{x}}_i=(\widetilde{x}_{i1},\widetilde{x}_{i2},\ldots)^\top$. The autoregressive model can be constructed as\\
\noindent
$\bullet$ $\widetilde{\bm{x}}_i=
\phi_2\bm{D}\widetilde{\bm{x}}_{i-1}+\bm{\eta}_i$ with $\phi_2 \in [0,0.723)$, and we choose the AR coefficient $\phi_2=0,0.2$ or $0.5$ to represent weak to moderately strong dependencies. The entries $\{\eta_{ik}\}_{k=1}^{\infty}$ of $\bm{\eta}_i$ are chosen as independent $\mathcal{N}(0,{\rm e}^{-(k-1)})$ random variables.

Furthermore, the generating mechanism for $\beta(t)$ and error process can be given by\\
\noindent
$\bullet$ $\beta_1=0.8,~\beta_2=0.5,~\beta_3=-0.3$ and $\beta_k=e^{-k}$ for $k\ge 4$. $\{\epsilon_i\}_{i=1}^n$ are independent of $X_i(t)$. Let $\{\epsilon_i\}_{i=1}^n$ follow an AR(1) process $\epsilon_i=0.2\epsilon_{i-1}+e_i$ where $e_i$ is i.i.d. standard normally distributed.
\begin{table}[htbp!]
	\caption{\label{T2}Simulated coverage probabilities with average JSCB widths in parentheses.}
    \small
	\begin{tabular}{|c|c|ccc|ccc|}
		\hline
		&&\multicolumn{6}{c|}{$n=400$}\\
		\cline{3-8}
		&&\multicolumn{3}{c|}{$1-\alpha=0.95$}&
		\multicolumn{3}{c|}{$1-\alpha=0.90$}\\ 
		\cline{1-8}
		\centering $\widehat{g}_{nj}$&Basis& $\phi_2=0$& $0.2$& $0.5$&$\phi_2=0$&$0.2$&$0.5$\\
		\hline
		\multirow{3}{*}{1}&Fou.&0.945(1.77)&0.944(1.79)&0.935(1.72)&
		0.898(1.55)&0.883(1.56)&0.884(1.51)\\
		\cline{2-8}
		&Leg.&0.942(3.10)&0.947(3.16)&0.940(3.05)&
		0.901(2.67)&0.888(2.72)&0.877(2.64)\\
		\cline{2-8}
		&FPC&0.939(1.91)&0.941(1.91)&0.947(1.90)&
		0.893(1.67)&0.889(1.66)&0.892(1.66)\\
		\hline
		\multirow{3}{*}{${\rm Std}$}
		&Fou.&0.938(1.47)&0.932(1.47)&0.923(1.43)&
		0.884(1.32)&0.876(1.32)&0.864(1.29)\\
		\cline{2-8}
		&Leg.&0.934(1.50)&0.923(1.53)&0.922(1.56)&
		0.883(1.35)&0.870(1.37)&0.860(1.40)\\
		\cline{2-8}
		&FPC&0.938(1.58)&0.927(1.57)&0.923(1.56)&
		0.871(1.41)&0.869(1.41)&0.851(1.40)\\
		\hline
		&&\multicolumn{6}{c|}{$n=800$}\\
		\cline{3-8}
		&&\multicolumn{3}{c|}{$1-\alpha=0.95$}&
		\multicolumn{3}{c|}{$1-\alpha=0.90$}\\ 
		\cline{1-8}
		$\widehat{g}_{nj}$&Basis&$\phi_2=0$&$0.2$&$0.5$& $\phi_2=0$& $0.2$& $0.5$\\
		\hline
		\multirow{3}{*}{1}&Fou.&0.960(1.25)&0.944(1.27)&0.940(1.23)&
		0.903(1.10)&0.896(1.11)&0.891(1.08)\\
		\cline{2-8}
		&Leg.&0.959(2.12)&0.946(2.12)&0.940(2.14)&
		0.908(1.83)&0.894(1.83)&0.888(1.85)\\
		\cline{2-8}
		&FPC&0.952(1.28)&0.952(1.30)&0.955(1.26)&
		0.909(1.12)&0.893(1.14)&0.914(1.10)\\
		\hline
		\multirow{3}{*}{${\rm Std}$}
		&Fou.&0.947(1.04)&0.940(1.06)&0.930(1.02)&
		0.890(0.93)&0.878(0.95)&0.871(0.91)\\
		\cline{2-8}
		&Leg.&0.940(1.02)&0.944(1.04)&0.938(1.04)&
		0.895(0.91)&0.886(0.93)&0.886(0.93)\\
		\cline{2-8}
		&FPC&0.942(1.06)&0.939(1.08)&0.934(1.04)&
		0.889(0.95)&0.882(0.97)&0.887(0.93)\\
		\hline
	\end{tabular}
\vspace{-0.3cm}
\end{table}

The simulated coverage probabilities and average widths of the joint simultaneous confidence bands (JSCB)  with two types of weight functions are reported in \cref{T2}. It is obvious to find that when $n=400$ and the data-adaptive weights are used, the coverage probabilities are reasonably close to the nominal levels for most cases under weaker dependence ($\phi_2=0,0.2$). However, under stronger dependence the performances of the three bases weaken slightly when $n=400$. The decrease in estimation accuracy and coverage probability in finite samples under stronger temporal dependence is well-known in time series analysis. This decrease seems to be universal across various inferential tools (such as subsampling, block bootstrap, multiplier bootstrap, and self-normalization) though some methods may be less sensitive to stronger dependence. One explanation is that the variances of the estimators tend to be higher under stronger dependence, which leads to less accurate estimators. This reduced accuracy then results in deteriorated coverage probabilities in small to moderately large samples.

 \subsection{Simulation Studies for Brownian Motions}
 We consider an additional simulation experiment where the functional predictor $X_i(t)$ is a Brownian motion process. We use three types of basis functions, Fourier bases, shifted Jacobi polynomial bases and functional principal components (FPC). First, we list the following approximation representations for the generation of Brownian motions $X_i(t)$ based on two kinds of basis functions:\\
 $\bullet$ Fourier bases:
 $$X_i(t)=Z_0 t+\sum_{k=1}^\infty \frac{\sqrt{2}Z_k}{\pi k}\sin(k\pi t),$$ where $Z_k,k\ge 0$ are independent and identically distributed (i.i.d.) $\mathcal{N}(0,1)$ random variables.\\
 \noindent
 $\bullet$ Shifted Jacobi polynomial bases:
 $$X_i(t)=Z_0 t+\sum_{k=1}^\infty Z_k\phi_k(t),$$
 where $Z_0\sim \mathcal{N}(0,1), Z_k\sim \mathcal{N}(0,k^{-1}(k+1)^{-1})$ for $k\ge 1$ and all $Z_k, k\ge 0$ are independent. $\{\phi_k(t)\}_{k\ge 1}$ are orthogonal basis functions constructed by Jacobi polynomial bases as
 $$\phi_k(t)=\frac{\sqrt{k(k+1)(2k+1)}}{k}P_{k+1}^{(-1,-1)}
 (2t-1),~t\in[0,1]$$ where $P_{k+1}^{(-1,-1)}(\cdot)$ is the $(-1,-1)$-Jacobi polynomial defined by $$P_{k+1}^{(-1,-1)}(t)=\frac{k}{4k+2}[P_{k+1}(t)-P_{k-1}(t)],~t\in[-1,1]$$ with the original Legendre polynomial $P_k(t)$ defined in \cref{eg_derivative2} of \cref{sec_D2}. 

 Now, restate the basis expansion $\beta(t)=\sum_{k=1}^\infty \beta_k\alpha_k(t)$ when $p=1$. We consider the following parameter configurations:\\
 $\bullet$ $\beta_1=0.8, \beta_2=0.5, \beta_3=-0.3$ and $\beta_k={\rm e}^{-k}$ for $k\ge 4$.\\
 \noindent
 $\bullet$ Case (a): $(X_i(t), Y_i)$ are i.i.d. and the error process $\{\epsilon_i\}_{i=1}^n$ are i.i.d. with $\mathcal{N}(0,0.25)$.\\
 \noindent
 $\bullet$ Case (b): $X_i(t)$ are i.i.d. across $i$ and 
 $\{\epsilon_i\}_{i=1}^n$ follow an AR(1) process $\epsilon_i=0.2\epsilon_{i-1}+e_i$, where $e_i$ is i.i.d. with $\mathcal{N}(0,0.25)$. Furthermore, $\{\epsilon_i\}_{i=1}^n$ are independent of $\{X_i(t)\}_{i=1}^n$.

Similar to the simulation studies in Section 3 of the main paper, the evaluations of the JSCB and computations of the average widths are carried out by discretizing the unit interval $[0,1]$ into $100$ equally spaced grids. \cref{bm_table} shows simulated coverage probabilities and average JSCB widths with three types of basis functions based on both constant and data-driven weight functions. From that, one can observe that most of the results, especially for $n=800$, are close to the nominal levels. Specifically, the average widths of JSCB for $\widehat{g}_{nj}(t)=\widehat{\rm Std}(\widehat{Q}_{nj}^{boots}(t,\lambda))\big/ \int_0^1 \widehat{\rm Std}(\widehat{Q}_{nj}^{boots}(s,\lambda))\dee s$ are narrower than those for $g_{nj}(t)\equiv 1$. In summary, we claim that our methodology can be applied to the scenario where the functional data is continuous but non-differentiable everywhere, such as Brownian motion. 
\begin{table}[htbp!]
		\centering
		\caption{\small Simulated coverage probabilities for the Brownian motion case.}
		\small
		\begin{tabular}{|c|c|cc|cc|}
			\hline
			\multicolumn{2}{|c|}{Case (a)}&\multicolumn{2}{c|}{$1-\alpha=0.95$}&
			\multicolumn{2}{c|}{$1-\alpha=0.90$}\\ 
			\hline
			$\widehat{g}_{nj}$&Basis&
                $n=400$&$n=800$&$n=400$&$n=800$\\
			\hline	
			\multirow{3}{*}{1}&Fourier&0.947(0.99)&0.946(0.67)&
			0.898(0.87)&0.889(0.59)\\
			\cline{2-6}
			&Jacobi&0.934(1.66)&0.934(1.27)&
			0.882(1.49)&0.886(1.14)\\
			\cline{2-6}
			&FPC&0.947(0.97)&0.946(0.66)&
			0.895(0.85)&0.889(0.58)\\
			\hline
			\multirow{3}{*}{Std}&Fourier&0.934(0.86)&0.935(0.58)&
			0.870(0.78)&0.879(0.52)\\
			\cline{2-6}
			&Jacobi&0.929(1.60)&0.936(1.22)&
			0.880(1.45)&0.872(1.11)\\
			\cline{2-6}
			&FPC&0.934(0.85)&0.934(0.57)&
			0.872(0.76)&0.879(0.52)\\
			\hline
   \multicolumn{2}{|c|}{Case (b)}&\multicolumn{2}{c|}{$1-\alpha=0.95$}&
			\multicolumn{2}{c|}{$1-\alpha=0.90$}\\ 
   \hline
   $\widehat{g}_{nj}$&
   Basis&$n=400$&$n=800$&$n=400$&$n=800$\\
   \hline
   \multirow{3}{*}{1}&Fourier&0.934(0.99)&0.937(0.69)
			&0.877(0.87)&0.889(0.60)\\
			\cline{2-6}
			&Jacobi&0.936(1.70)&0.942(1.30)&
			0.886(1.52)&0.880(1.16)\\
			\cline{2-6}
			&FPC&0.935(0.97)&0.937(0.68)&
			0.890(0.85)&0.890(0.60)\\
			\hline
   \multirow{3}{*}{Std}&Fourier&0.937(0.86)&0.938(0.59)
			&0.876(0.77)&0.889(0.53)\\
			\cline{2-6}
			&Jacobi&0.930(1.62)&0.942(1.24)&
			0.874(1.47)&0.881(1.13)\\
			\cline{2-6}
			&FPC&0.935(0.84)&0.939(0.59)&
			0.876(0.76)&0.888(0.53)\\
			\hline
		\end{tabular}
		\label{bm_table}
	\end{table}
 
 \subsection{Numerical Experiments Comparing with Other Methods}
In this subsection, we conduct another simulation study, comparing our JSCB construction with the conservative confidence bands proposed by \cite{imaizumi2019simple} and the confidence bands construction investigated by \cite{dette2021statistical}. Since the methodologies of the latter two papers are limited to i.i.d. data, we will perform three distinct sets of comparison experiment with $p=1$, including i.i.d. scenario, weak temporal dependent and moderately strong dependent cases. Here, recall the basis expansions as $\beta(t)=\sum_{k=1}^\infty \beta_k\alpha_k(t), X_i(t)=\sum_{k=1}^\infty \widetilde{x}_{ik}\alpha_k(t)$. Consider \\
\noindent
$\bullet$ $\beta_1=0.8, \beta_2=0.5, \beta_3=-0.3$ and $\beta_k={\rm e}^{-k}$ for $k\ge 4$.\\
 \noindent
 $\bullet$ $\widetilde{x}_{ik}=k^{-1}U_k$ where $\{U_k\}_{k=1}^\infty$ i.i.d. follow the Uniform distribution as ${\rm Unif}(-\sqrt{3},\sqrt{3})$. 
 
 Now, we investigate the following cases:\\
 \noindent
 $\bullet$ Case (c): Consider the above data generation processes for $\beta(t)$ and $X_i(t)$, and the error term $\epsilon_i$ is i.i.d. $\mathcal{N}(0,1)$.\\
 \noindent
 $\bullet$ Case (d): The above data generation processes for $\beta(t)$ and $X_i(t)$ are considered, the error process $\{\epsilon_i\}_{i=1}^n$ follows an AR(1) model $\epsilon_i=0.2\epsilon_{i-1}+e_i$ where $e_i$ is i.i.d. $\mathcal{N}(0,1)$.\\
 \noindent
 $\bullet$ Case (e): The generation process for $\beta(t)$ 
 retains the same of that in Case (d), $X_i(t)$ follows the FAR(1) model in Section 4 of the main paper where the coefficient matrix $\bm{D}$ is diagonal with all diagonal elements being 1 to make sure the true FPCs of $X_i(t)$ are $\{\alpha_k(t)\}_{k=1}^\infty$ and the autoregressive coefficient $\phi=0.5$. The error process follows the same AR(1) model as described in Case (d).

 \begin{table}[t]
		\centering
		\caption{\small Simulated coverage probabilities and average widths for MCP, MCP$^\ast$ and JSCB.}
		\small
		\begin{tabular}{|c|c|cc|cc|}
			\hline
			\multicolumn{2}{|c|}{Case (c)}&\multicolumn{2}{c|}{$\tau_1=0.05$}&
			\multicolumn{2}{c|}{$\tau_1=0.10$}\\ 
			\hline
			Method&Cutoff level/Weight&
                $n=400$&$n=800$&$n=400$&$n=800$\\
			\hline	
			\multirow{2}{*}{MCP}&$\max
   \{\widehat{m}_n,2\}$&0.985(1.94)&0.947(1.51)&
			0.926(1.68)&0.862(1.32)\\
			\cline{2-6}
			&$\widehat{m}_n+1$&0.999(2.72)
   &1(2.04)&0.999(2.40)&1(1.80)\\
			\hline
   \multirow{2}{*}{MCP$^\ast$}&$\max
   \{\widehat{m}_n,2\}$&0.917(1.94)&0.874(1.51)&
			0.795(1.69)&0.820(1.32)\\
			\cline{2-6}
			&$\widehat{m}_n+1$&0.995(2.72)&1(2.04)&
			0.987(2.40)&1(1.80)\\
			\hline
       JSCB&$\widehat{g}_{n1}={\rm Std}$
        &0.940(2.63)&0.941(1.88)
           &0.887(2.38)&0.893(1.70)\\
           \hline
           \multicolumn{2}{|c|}{Case (d)}&\multicolumn{2}{c|}{$\tau_1=0.05$}&
			\multicolumn{2}{c|}{$\tau_1=0.10$}\\ 
			\hline
			Method&Cutoff level/Weight&
                $n=400$&$n=800$&$n=400$&$n=800$\\
			\hline	
			\multirow{2}{*}{MCP}&$\max
   \{\widehat{m}_n,2\}$&0.985(1.97)&0.936(1.53)&
			0.935(1.62)&0.844(1.34)\\
			\cline{2-6}
			&$\widehat{m}_n+1$&1(2.75)&1(2.11)&
			1(2.42)&1(1.86)\\
			\hline
   \multirow{2}{*}{MCP$^\ast$}&$\max
   \{\widehat{m}_n,2\}$&0.928(1.87)&0.858(1.53)&
			0.807(1.62)&0.785(1.34)\\
			\cline{2-6}
			&$\widehat{m}_n+1$&0.995(2.75)&0.999(2.11)&
			0.990(2.42)&0.993(1.86)\\
			\hline
         JSCB&$\widehat{g}_{n1}={\rm Std}$&0.926(2.67)&0.944(1.92)
           &0.871(2.42)&0.887(1.74)\\
           \hline
           \multicolumn{2}{|c|}{Case (e)}&\multicolumn{2}{c|}{$\tau_1=0.05$}&
			\multicolumn{2}{c|}{$\tau_1=0.10$}\\ 
         \hline
           Method&Cutoff level/Weight&
           $n=400$&$n=800$&$n=400$&$n=800$\\
           \hline
         \multirow{2}{*}{MCP}&$\max\{\widehat{m}_n,2\}$&
        0.896(1.79)&0.888(1.34)&
			0.784(1.56)&0.884(1.16)\\
			\cline{2-6}
        &$\widehat{m}_n+1$&1(2.90)&1(2.41)
           &1(2.52)&0.999(2.09)\\
           \hline
           \multirow{2}{*}{MCP$^\ast$}&$\max
   \{\widehat{m}_n,2\}$&0.804(1.79)&0.885(1.34)&
			0.712(1.56)&0.867(1.16)\\
			\cline{2-6}
			&$\widehat{m}_n+1$&0.999(2.90)&0.999(2.41)&
			0.999(2.52)&0.996(2.09)\\
			\hline
           JSCB&$\widehat{g}_{n1}={\rm Std}$&0.930(1.53)&0.936(1.03)
           &0.862(1.37)&0.879(0.92)\\
           \hline
		\end{tabular}
		\label{iid_table}
	\end{table}
 
 In the simulation experiment, we focus on Fourier basis functions $\alpha_k(t)=\{1, \sqrt{2}\cos(\pi t), \\ \sqrt{2}\cos(2\pi t),\cdots\}$ and examine the coverage probabilities of confidence bands for sample sizes $n=400, 800$ with $1000$ replicates. For the conservative confidence bands construction in \cite{imaizumi2019simple}, we choose the significance levels $\tau_1=0.05, 0.1$, the other level $\tau_2=0.1$ and preserve two kinds of choices for the cutoff level (i.e. truncation number used in the PCA-based estimation) $m_n$, denoted by $\max\{\widehat{m}_n,2\}$ and $\widehat{m}_n+1$. To provide further clarity, we evaluate the confidence bands of our method in comparison with methods in \cite{imaizumi2019simple} using 
 \begin{align*}
 {\rm JSCB}&=\Pr\big(\beta(t)\in [\widetilde{\beta}(t)-\widehat{g}_{n1}(t)\hat{q}_{n,1-\tau_1}/\sqrt{n},
 \widetilde{\beta}(t)+\widehat{g}_{n1}(t)\hat{q}_{n,1-\tau_1}/\sqrt{n}], \forall t\in[0,1]\big),\\
 {\rm MCP}&=\Pr\big(\mu\big(\{t\in[0,1],\beta(t)\notin \widehat{\mathcal{C}}(t)\}\big)\le \tau_2\big),\\
 {\rm MCP}^\ast&=\Pr\big(\beta(t)\in \widehat{\mathcal{C}}(t), \forall t\in[0,1]\big),
 \end{align*}
 where $\widehat{g}_{n1}(t)=\widehat{\rm Std}(\widehat{Q}_{n1}(t,\lambda))\big/ \int_0^1\widehat{\rm Std}(\widehat{Q}_{n1}(s,\lambda))\dee s$ is considered in the JSCB construction. The confidence bands in MCP construction are specified as $$\widehat{\mathcal{C}}(t)=\left[\widetilde{\beta}(t)-\frac{\widehat{\sigma}\widehat{c}_{n,1-\tau_1}}{\sqrt{n}}\sqrt{\frac{1}{\tau_2\mu([0,1])}},
 \widetilde{\beta}(t)+\frac{\widehat{\sigma}
 \widehat{c}_{n,1-\tau_1}}{\sqrt{n}}\sqrt{\frac{1}{\tau_2\mu([0,1])}}\right]$$ with $\widehat{c}_{n,1-\tau_1}$ representing the simulated conditional $(1-\tau_1)$-quantile of the corresponding statistic given functional data $\{X_1(t),...,X_n(t)\}$, $\widehat{\sigma}=\sqrt{n^{-1}\sum_{i=1}^n[Y_i-\sum_{k=1}^{\widehat{m}_n} \widehat{x}_{ik}\widehat{\beta}_k]^2}$ where $\widehat{x}_{ik}$ is the estimated FPC score and $\widehat{\beta}_k$ is the estimated coefficient of $\beta_k$. The notation $\mu$ in the equation of MCP denotes the Lebesgue measure.

The simulated coverage probabilities and average widths for JSCB, MCP and MCP$^\ast$ are shown in \cref{iid_table}. For the MCP method proposed in \cite{imaizumi2019simple}, when the cutoff level is chosen as $\max\{\widehat{m}_n,2\}$, the coverage probabilities for Cases (c) and (d) exceed the nominal levels for $n=400$, while most coverage probabilities are below the nominal level for $n=800$. Furthermore, when the cutoff level is specified as $\max\{\widehat{m}_n,2\}$, most of the coverage probabilities under MCP turn out to decrease as the sample size increases, which implies possible instability in simultaneous coverage of this confidence band construction. We note that the observed instability possibly stems from the fact that the MCP method aims at covering the slope function at ``most" of the points $t\in[0,1]$ with a prespecified probability, but is not designed for simultaneous coverage. On the other hand, the simulated coverage probabilities for MCP based on the cutoff level $\widehat{m}_n+1$ appear to be close to 1 in Cases (c)--(e), at the cost of substantially wider confidence bands.

 Note that we also include MCP$^\ast$ construction with a stringent requirement. Instead of allowing a small proportion ($\tau_2$) of the set of points not covered by the confidence bands in MCP construction, MCP$^\ast$ method evaluates the coverage probabilities that all discrete points $\beta(t)$ are simultaneously covered by the confidence bands. From \cref{iid_table}, we observe that for Cases (c)-(d), the MCP$^\ast$ method with the cutoff level $\max\{\widehat{m}_n,2\}$ have narrower average widths than those of our JSCB construction, but at the expense of under coverage probabilities for both $n=400$ and $800$. Furthermore, for Case (e), MCP$^\ast$ produces confidence bands with wider widths and coverage probabilities substantially below the nominal level. While with the cutoff level $\widehat{m}_n+1$, all coverage probabilities of MCP$^\ast$ are close to 1 with wider average widths, yielding a less informative confidence band construction. 

In comparison, our JSCB method shows quite accurate coverage probabilities for i.i.d. data, weak and moderately strong dependent data (Cases (c)-(e)). Particularly for Cases (c)-(d), the JSCB construction under the data-driven weights yields narrower average widths than both MCP and MCP$^\ast$ methods at the cutoff level $\widehat{m}_n+1$. This advantage extends to Case (e), where our JSCB construction achieves narrower widths than competing methods at both cutoff levels. As a result, it demonstrates the better finite sample performance of our JSCB construction especially for dependent functional time series. Again, we remark that MCP and MCP$^\ast$ methods in \cite{imaizumi2019simple} are tailored for i.i.d. functional data, therefore our limited simulation study here merely suggests that their methods cannot be directly extended to the dependent setting. 

As a reviewer suggested, we also include a simulation experiment to compare our method with the approach proposed in \cite{dette2021statistical}. Specifically, their work considers a reproducing kernel Hilbert space (RKHS)-based approach for estimating the slope coefficient function by
$$\widehat{\beta}(t)=\mathop{\arg\min}_{\beta \in \mathcal{H}} \left[\frac{1}{2n}\sum_{i=1}^n \left\{Y_i-\int_0^1 X_i(s)\beta(s)\dee s\right\}^2 + \frac{\lambda}{2}\int_0^1 \left\{\beta^{''}(s)\right\}^2\dee s\right],$$
where $\mathcal{H}=\left\{\beta: [0,1]\to \mathbb{R} \mid \partial^{(\theta)}\beta \text{ is absolutely continuous for }0\le \theta\le m-1; \partial^{(\theta)}\beta \in \mathcal{L}^2([0,1])\right\}$ is the Sobolev space of order $m>1/2$ of functions defined on $[0,1]$. To implement their bootstrap procedure and construct simultaneous confidence bands for $\beta(t)$, we similarly follow Algorithm 4.1 in \cite{dette2021statistical} with appropriate modifications for a scalar response linear regression model. The bootstrap weights $\{M_{i,q}\}_{1\le i\le n, 1\le q\le Q}$ are generated from a two-point distribution, taking $1-1/\sqrt{2}$ with probability $2/3$ and taking $1+\sqrt{2}$ with probability $1/3$ based on $Q=1000$ bootstrap replications. Then the bootstrap estimator for each $1\le q\le Q$ can be obtained by
$$\widehat{\beta}^\ast_{n,q}(t)=\mathop{\arg\min}_{\beta \in \mathcal{H}} \left[\frac{1}{2n}\sum_{i=1}^n M_{i,q} \left\{Y_i-\int_0^1 X_i(s)\beta(s)\dee s\right\}^2 + \frac{\lambda}{2}\int_0^1 \left\{\beta^{''}(s)\right\}^2\dee s\right].$$
Consequently, the uniform $(1-\alpha)$ coverage probability can be defined by
$$\text{UCP}:=\Pr(\beta(t) \in \mathcal{C}^\ast(t), \forall t\in [0,1]),$$
and the simultaneous asymptotic $(1-\alpha)$ confidence band of $\beta(t)$ can be expressed as $$\mathcal{C}^\ast(t)=\left[\widehat{\beta}(t)-\frac{\widehat{\mathcal{Q}}_{1-\alpha}}{\sqrt{n}\lambda^{(2a+1)/(4D)}},\widehat{\beta}(t)+\frac{\widehat{\mathcal{Q}}_{1-\alpha}}{\sqrt{n}\lambda^{(2a+1)/(4D)}}\right],$$
where $\widehat{\mathcal{Q}}_{1-\alpha}$ is the empirical $(1-\alpha)$-quantile of the sample $\{\sqrt{n}
\lambda^{(2a+1)/(4D)}\sup_{t\in[0,1]}|\widehat{\beta}_{n,q}^\ast(t)-\widehat{\beta}(t)|\}_{q=1}^Q$, $a$ and $D$ are some tuning parameters.

The following \cref{RKHS_table} reports the simulated coverage probabilities and average widths. For Case (c), the empirical coverage probabilities provide good approximations to the nominal confidence levels for both $n=400$ and $n=800$, suggesting satisfactory performance of the confidence band construction for independent  functional data. For Case (d), the functional covariates ${X_i(t)}$ are independent, while the responses ${Y_i}$ exhibit weak correlation induced by the temporal dependence in the error process ${\epsilon_i}$. Under this weak dependent setting, the coverage probabilities remain close to the nominal level for both $n=400$ and $n=800$. However, in Case (f) which corresponds to a moderately strong dependence data generating structure, the confidence band construction fails to deliver accurate inference even for larger sample size $n=800$. It is evident that the RKHS-based method produces systematic undercoverage probabilities, indicating its limitation to i.i.d. functional data. On the other hand, the RKHS-based approach yields noticeably wider average widths compared with other methods discussed earlier, which also results in less informative statistical inference.

\begin{table}[htbp!]
		\centering
		\caption{\small Simulated coverage probabilities and average widths in parentheses.}
		\small
		\begin{tabular}{|c|cc|cc|}
			\hline
            & \multicolumn{2}{c|}{$\alpha=0.05$}&
			\multicolumn{2}{c|}{$\alpha=0.10$}\\
            \hline
            & 
            $n=400$&$n=800$&$n=400$&$n=800$\\
			\hline	
			Case (c)&0.937(4.33)&0.950(3.21)&
			0.884(3.78)&0.895(2.80)\\
			\hline
            Case (d)&0.947(4.42)&0.950(3.28)&
			0.890(3.86)&0.898(2.86)\\
			\hline
       Case (e)
        &0.855(5.26)&0.897(3.36)
           &0.786(4.59)&0.826(2.93)\\
           \hline
           \end{tabular}
           \label{RKHS_table}
           \end{table}

	\subsection{Simulation Studies with Order $p> 1$}
	In the simulation studies of the paper, we only consider the case $p=1$. In this subsection we shall consider $p=1,2,3$ with the following parameter setups:\\
	\noindent 
	$\bullet$ For $j=1,2,3$, $\beta_{j1}=0.8,~\beta_{j2}=0.5,~\beta_{j3}=-0.3$ and $\beta_{jk}=k^{-4}$ for $k\ge 4$;\\
	$\bullet$ For every $j=1,2,3$, consider FMA(1) model $\bm{\widetilde{x}}_{ij}=\bm{D}(\bm{\eta}_{ij}+\phi_3\bm{\eta}_{i-1,j})$ where $\bm{D}$ is defined in Section 4 of the paper and the MA coefficient $\phi_3=0.2$. The random vectors $\bm{\widetilde{x}}_{ij}$ are independent across $j$. The innovation entries $\{\eta_{ij,k}\}_{k=1}^{\infty}$ of $\bm{\eta}_{ij}$ are independent $k^{-1.2}\mathcal{N}(0,1)$ random variables.\\
	$\bullet$ The error process $\{\epsilon_i\}_{i=1}^n$ are independent and identically distributed (i.i.d.) with standard normal distribution.
	
	\begin{table}[t]
		\centering
		\caption{\small Simulated coverage probabilities over different sample sizes $n$ and orders $p$, where the weight function $\widehat{g}_{nj}(t)=\widehat{{\rm Std}}(\widehat{Q}_{nj}^z(t,\lambda))/ \int_0^1\widehat{{\rm Std}}(\widehat{Q}_{nj}^z(s,\lambda))\dee s$.}
		\small
		\begin{tabular}{|c|cc|cc|}
			\hline
			&\multicolumn{4}{c|}{$p=1$}\\
			\hline
			&\multicolumn{2}{c|}{$1-\alpha=0.95$}&
			\multicolumn{2}{c|}{$1-\alpha=0.90$}\\ 
			\hline
			Basis&$400$&$800$&$400$&$800$\\
			\hline
			Fou.&0.935(1.78)&0.946(1.29)
			&0.892(1.59)&0.897(1.16)\\
			\hline
			Leg.&0.930(1.72)&0.946(1.24)&
			0.881(1.54)&0.898(1.11)\\
			\hline
			FPC&0.932(1.76)&0.940(1.26)&
			0.897(1.57)&0.878(1.12)\\
			\hline
			&\multicolumn{4}{c|}{$p=2$}\\
			\hline
			&\multicolumn{2}{c|}{$1-\alpha=0.95$}&
			\multicolumn{2}{c|}{$1-\alpha=0.90$}\\ 
			\hline
			Basis&$400$&$800$&$400$&$800$\\
			\hline	
			Fou.&0.928(1.95)&0.942(1.41)&
			0.873(1.77)&0.874(1.36)\\
			\hline
			Leg.&0.928(1.88)&0.941(1.36)&
			0.853(1.71)&0.888(1.24)\\
			\hline
			FPC&0.931(1.92)&0.935(1.37)&
			0.872(1.76)&0.882(1.25)\\
			\hline
			&\multicolumn{4}{c|}{$p=3$}\\
			\hline
			&\multicolumn{2}{c|}{$1-\alpha=0.95$}&
			\multicolumn{2}{c|}{$1-\alpha=0.90$}\\ 
			\hline
			Basis&$400$&$800$&$400$&$800$\\
			\hline
			Fou.&0.920(2.03)&0.949(1.48)&
			0.859(1.87)&0.886(1.36)\\
			\hline
			Leg.&0.923(1.97)&0.936(1.42)&
			0.855(1.81)&0.873(1.30)\\
			\hline
			FPC&0.925(2.01)&0.935(1.43)&
			0.859(1.85)&0.875(1.32)\\
			\hline
		\end{tabular}
		\label{T_p}
	\end{table}
	
	We list the simulated results for various orders $p$ in Table \ref{T_p}. From it, we find that for all three basis functions, the coverage probabilities for $p=1, 2, 3$ are similar although the coverage probabilities for $p=2$ and $3$ tend to be slightly smaller than those of $p=1$. In particular, all coverage probabilities for $p=1, 2$ and $3$ are reasonably close to the nominal levels when $n=800$.
	
	\subsection{Statistical Power of the JSCB Test}
	Next, we shall perform simulation studies with $p=1$ to investigate the accuracy and power of the JSCB when it is used as a test. Specifically, we shall perform the significance test $\beta(t)\equiv 0$ versus $\beta(t)\not\equiv 0$. Consider the following setting under scenario:\\
	$\bullet$ $\beta_k=\delta(-1)^k k^{-4},~k\ge 1$ and $\delta$ ranges from 0 to $0.5$ with sample size $n=800$.\\
	$\bullet$ $\widetilde{\bm{x}}_i=\bm{\Phi}\widetilde{\bm{x}}_{i-1}+\bm{\eta}_i,$
	where $\bm{\Phi}$ is an infinite-dimensional tridiagonal matrix with $1/5$ on the diagonal and $1/15$ on the off-diagonal, $\eta_{ik}\sim 2k^{-1.2}\mathcal{N}(0,1)$ for $k\ge 1$.\\
	$\bullet$ Let the error process $\{\epsilon_i\}_{i=1}^n$ follow an AR(1) process $\epsilon_i=0.2\epsilon_{i-1}+e_i$ where $e_i$ is i.i.d. standard normally distributed. Moreover $\{\epsilon_i\}$ are independent of $X_i(t)$. 
	
\begin{figure}[htbp!]
		\vspace{-0.25cm}
		\centering
		%\subfigure[Statistical power at $\alpha=0.05$]{
		\begin{minipage}{6cm}
			\centering
			\includegraphics[width=5cm,height=5cm]{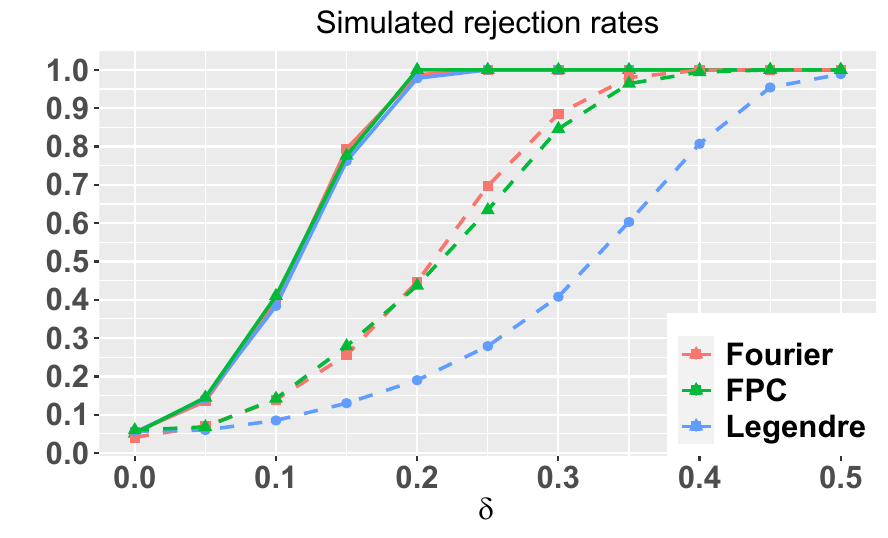}
		\end{minipage}
		\begin{minipage}{6cm}
			\centering
			\includegraphics[width=5cm,height=5cm]{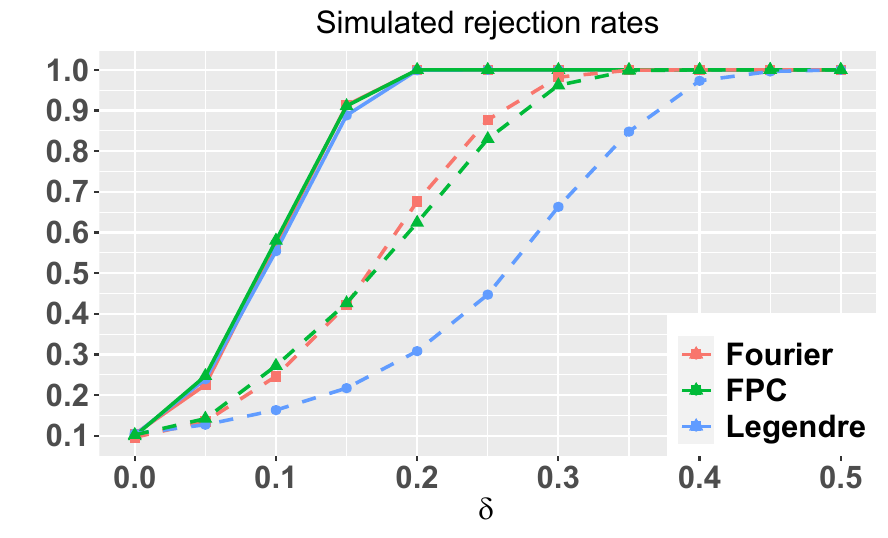}
		\end{minipage}
		\centering
		\caption{Simulated rejection probabilities at nominal levels $\alpha=0.05$ (Left) and $\alpha=0.1$ (right) with fixed constant weight function $1$ (dashed) and data-driven weight function (solid).}
		\label{F1}
		\vspace{-0.3cm}
	\end{figure}

	\cref{F1} shows the simulated rejection probabilities for the test with three types of basis functions at nominal levels $\alpha=0.05, 0.1$. From it, we observe that the power performances of the three basis functions are quite similar with data-driven weight functions in the sense that as $\delta$ increases, the simulated power increases fast. On the other hand, the power curves of the constant weight function increase slower than those of the adaptive weight function. This is consistent with our simulation results that the JSCB is narrower on average when the weights are proportional to the standard deviation of the estimators.

    %\subsection{Additional Discussion on Real Data Example}

    \section{Gaussian Approximation Theory}\label{gau}
	Throughout the supplemental material, we will consistently use the following notation. For a random variable $Z$, denote $\Vert Z\Vert_q := (\EE|Z|^q)^{1/q}$ as its $L^q$ norm. For a square integrable random function $X(t)\in \mathcal{L}^2[0,1]$, we use $|X(t)|_{\mathcal{L}^2}:=(\int_0^1 X^2(t)\dee t)^{1/2}$ to stand for its $\mathcal{L}^2$ norm. Furthermore, we denote $|\cdot|$ as the spectral norm (largest singular value) for a matrix or the Euclidean norm for a random vector. The notations $|\cdot|_F$ and
		 $\Vert\cdot\Vert_{\Psi}$ indicate the Frobenius norm and Orlicz norm respectively, the notation $|\cdot|_{\max}$ signifies the largest element of a matrix. We define $|f(x)|_\infty:=\sup_{x\in\mathcal{X}}|f(x)|$ to state the supremum norm of $f(x)$ and the symbol $C$ denotes a generic finite constant whose value may vary from place to place. 
	
	In this section we shall establish a Gaussian approximation theory for the weighted maximum deviations of $\widetilde{\bm{Q}}_n^z(t,\lambda):=\bm{C}_f(t)\bm{\Sigma}_c^{-1}(\lambda)\bm{Z}_n^c$ uniformly over all quantiles and a wide class of weight functions. As we mentioned in the main article, this result is based on uniform Gaussian approximation results over all Euclidean convex sets for sums of stationary and weakly dependent time series of moderately high dimensions which we will establish in Section \ref{proof} of this supplemental material when we prove the results of this section. The result extends the corresponding findings for independent and $m$-dependent data established in \cite{Bentkus03}, \cite{Fang15} and \cite{Fang16} among others, which may be of separate interest.
	
	Define $\bm{U}_n^c=\frac{1}{\sqrt{n}}\sum_{i=1}^n\bm{u}_{ci}$ where $\{\bm{u}_{ci}\}_{i=1}^n$ is a sequence of $c_n$-dimensional Gaussian random vectors which is independent of $\{\bm{z}_{ci}\}_{i=1}^n$ and preserves the covariance structure of $\{\bm{z}_{ci}\}_{i=1}^n$. Further denote $\widetilde{\bm{Q}}_n^u(t,\lambda)=\bm{C}_f(t)
	\bm{\Sigma}_c^{-1}(\lambda)\bm{U}_n^c$ and define the distance of interest as
	\begin{align*}
	&\mathcal{K}(\bm{Z}_n^c,\bm{U}_n^c)\\
	=&\sup_{\bm{g}_n\in\mathcal{G}}
	\sup_{x\in\mathbb{R}}\left|
	\Pr\left(\sup_{t\in[0,1]}
    |\widetilde{\bm{Q}}_n^z(t,\lambda)|_{\bm{g}_n(t)}\le x\right)-\Pr\left(\sup_{t\in[0,1]}|\widetilde{\bm{Q}}_n^u(t,\lambda)|_{\bm{g}_n(t)}\le x\right)\right|.
	\end{align*}
	Now, we state the Gaussian approximation result for the roughness penalization estimator.
	
	\begin{theorem}\label{thm2}
		Under Assumptions 1--8 of the main article and suppose the smallest eigenvalue of $\bm{\Xi}^c$ is bounded below by some constant $b>0$, there exists a constant $C>0$ such that
		\begin{equation}\label{pGaus1}
		\mathcal{K}(\bm{Z}_n^c,\bm{U}_n^c)\le C\left(c_n^{\frac{7}{4}}n^{-\frac{1}{2}+\frac{9}{2q}+\frac{2}{\tau-1}}+
		c_n^{\frac{3}{8}} \lambda^{\frac{d_2+\psi+2}{2(2\gamma + d_2-\psi)}}\right).
		\end{equation}
	\end{theorem}
	
	\begin{proof}
		See \cref{sec_proof} for the proof.
	\end{proof}

	The above theorem shows that when $\lambda$, $q$ and $\tau$ are sufficiently large and $c_n$ is sufficiently small, the distribution of $\sup_{t\in [0,1]}|\widetilde{\bm{Q}}_n^z(t,\lambda)|_{\bm{g}_n(t)}$ can be well approximated by that of $\sup_{t\in [0,1]}|\widetilde{\bm{Q}}_n^u(t,\lambda)|_{\bm{g}_n(t)}$ uniformly over all quantiles and weight functions in $\mathcal{G}$. 

    The constraints on $c$ and $\lambda$ are also mild. For example, if $\bm{X}_i(t),~\bm{\beta}(t) \in \mathcal{C}^1$ and $\gamma=4$ based on the normalized Legendre polynomial bases, \cref{thm} in \cref{full_ass} shows that $\widetilde{\bm{\beta}}(t)$ is an under-smoothed estimator as long as $c_n\gg (n/\log(n))^{\frac{1}{5}}$ and $\lambda\ll \left(\log(n)/n\right)^{\frac{12}{5}}$. Hence $\widetilde{\bm{\beta}}(t)$ is under-smoothed and at the same time \eqref{pGaus1} goes to 0 for a relatively wide range of $c$ and $\lambda$.
	
	\section{Calculating Physical Dependence Measures for Two Classes of Functional Time Series Models}\label{a}
	Here, we show two examples on how to calculate $\delta_x(l,q)$ for a class of functional MA$(\infty)$ processes and functional AR(1) processes, respectively.
	\subsection{FMA$(\infty)$ Model}
	\begin{example}[Functional MA$(\infty)$ model] \label{exam1}
		Let $\eta_i(s)$ be i.i.d. centered and continuous Gaussian random functions with $\sup_{s\in[0,1]}\EE\Vert\eta_i(s)\Vert_2<\infty$.  For each integer $m\ge 0$, let $B_m(t,s)=a_mB_m^\ast(t,s)$ where  $\{a_m\}$ is a positive deterministic sequence with $\sum_{m=0}^\infty a_m<\infty$ and $B_m^\ast(\cdot,\cdot)$ is a $\mathcal{C}([0,1]^2)$ deterministic function such that $|B^*_m(t,s)|\le C$ for all $t,s$ and $m$ and some finite constant $C$. Consider the functional MA$(\infty)$ model,
		\begin{equation}\label{exp1}
		X_i(t)=\sum_{m=0}^\infty \int_0^1 B_m(t,s)\eta_{i-m}(s)\dee s.
		\end{equation}
		Let  $C(t,s):=\EE(\eta_i(t)\eta_i(s))$ be the covariance function of $\eta_i(t)$. Let $v_1(t), v_2(t),\cdots$ and the corresponding $u_1\ge u_2\ge \cdots$ be the eigenfunctions and eigenvalues of $C(t,s)$. By the basis expansion method, we can write $X_i(t)=\sum_{j=1}^\infty\widetilde{x}_{i,j}\alpha_j(t),~B_m^\ast(t,s)=
		\sum_{j=1}^\infty\sum_{k=1}^\infty b_{j,k}^m\alpha_j(t)v_k(s)$ and $\eta_i(s)=\sum_{j=1}^\infty\eta_{i,j}v_j(s)$.  Next, by substituting the above into \eqref{exp1}, we obtain
		$$\widetilde{x}_{i,j}=\sum_{m=0}^\infty a_m\left(\sum_{k=1}^\infty
		b_{j,k}^m\eta_{i-m,k}\right)=:f_jx_{i,j}.$$ Here $f^2_j=\sum_{m=0}^\infty a_m^2\theta_{jm}$ is the variance of the random coefficient $\widetilde{x}_{i,j}$, where $\theta_{jm}:=\sum_{k=1}^\infty
		(b_{j,k}^m)^2 u_k$. Let $\eta_i=(\eta_{i,j})_{j\ge 1}$. Then we see that $x_{ij,k}$ can be written in the form of physical dependence measure in Definition 1 of the paper. The following lemma bounds the physical dependence measures for \eqref{exp1}.
		\begin{lemma}\label{l1}
			The dependence measures $\delta_x(l,q)=O(a_l)$ for any given $q\ge 2$ if
			\begin{eqnarray}\label{suf_con}
			\sum_{k=0}^\infty a_k^2\theta_{jk}\ge C\theta_{jm}
			\end{eqnarray}
			for sufficiently large $j$ and $m$ and some positive constant $C$ that does not depend on $j$ or $m$.
		\end{lemma}
	\end{example}

    Assumption \eqref{suf_con} is a mild condition in general. For instance, it is easy to see that \eqref{suf_con} holds if $\{B^*_m(t,s)\}_{m=0}^\infty$ is finitely generated; that is, for each non-negative integer $m$, $B^*_m(t,s)$ can only choose from $r$ candidate functions $\{\tilde B_j(t,s),j=1,2,\cdots, r\}$ for some $r<\infty$.  Note that functional MA($r$) models belong to the finitely generated category when $r$ is finite. For another example,  \eqref{suf_con} holds if $B^*_m(t,s)$ admits the decomposition $B^*_m(t,s)=\gamma_m(t)\kappa_m(s)$ for some uniformly bounded and continuous functions $\gamma_m(\cdot)$ and $\kappa_m(\cdot)$. We refer to Lemma \ref{lem:suf_con} in the following for the proof. We make a further note that another sufficient condition for \eqref{suf_con} is that, for some non-negative integer $k_0$, $\theta_{jk_0} \ge C_0 \theta_{jk}$ for sufficiently large $j$ and $k$ and some positive constant $C_0$ that does not depend on $j$ or $k$. For many frequently used basis functions such as the Fourier, wavelet and orthogonal polynomial bases, the decay speed of $\theta_{jk}$ with respect to $j$ is determined by the smoothness of $B^*_k(t,s)$ in $t$ and $\theta_{jk_0} \ge C_0\theta_{jk}$ is satisfied when there exists a $k_0$ such that $B^*_{k_0}(t,s)$ is at most as smooth as $B^*_{k}(t,s)$ in $t$ for all sufficiently large $k$ under some extra mild basis-specific assumptions.
	\bigskip
	
	\noindent
	\textbf{Proof of \cref{l1}.}
	Note that $\theta_{jm}=\EE[\int_{0}^1\int_0^1 B^*_m(t,s)\alpha_j(t)\eta_0(s)\dee t \dee s]^2\le C$ for some finite constant $C$ that does not depend on $j$ or $m$. Hence inequality \eqref{suf_con} holds for all $j$ and sufficiently large $m$ if it holds for sufficiently large $j$ and $m$.
	Observe that $\widetilde{x}_{i,j}$ has the following MA$(\infty)$ representation
	\begin{eqnarray*}
		\widetilde{x}_{i,j}=\sum_{m=0}^\infty a_m \eta_{i,j}^{(m)},
	\end{eqnarray*}	
	where $\eta_{i,j}^{(m)}=\sum_{k=1}^\infty b_{j,k}^m\eta_{i-m,k}$. Therefore a direct application of the definition of the physical dependence measure yields that
	\begin{eqnarray*}
		\delta_x(l,q)\le a_l\sup_{j}\Big[\|\sum_{k=1}^\infty b_{j,k}^{l}(\eta_{i-l,k}-\eta^*_{i-l,k})\|_q/f_j\Big]
	\end{eqnarray*}
	provided $f_j\neq 0$, where $\eta^*_{i-l,k}$ is an i.i.d. copy of $\eta_{i-l,k}$. If $f_j=0$, then it is clear that $\widetilde{x}_{i,j}=0$ almost surely and the temporal dependence at the $j$-th basis expansion level is uniformly 0. Now observe that  $\eta_{i-l,k}$ are independent Gaussian random variables across $k$. Hence $\sum_{k=1}^\infty b_{j,k}^{l}(\eta_{i-l,k}-\eta^*_{i-l,k})$ is normally distributed with mean 0 and variance $2\theta_{jl}$. Furthermore, the $L^q$ norm of a centered Gaussian random variable is proportional to its standard deviation. Therefore $\|\sum_{k=1}^\infty b_{j,k}^{l}(\eta_{i-l,k}-\eta^*_{i-l,k})\|_q=C_q\theta^{1/2}_{jl}$ for some constant $C_q$. Hence $\delta_x(l,q)=O(a_l)$ by inequality \eqref{suf_con} since $f^2_j=\sum_{m=0}^\infty a^2_m\theta_{jm}$. \qed
	
	\begin{lemma}\label{lem:suf_con}
		Inequality \eqref{suf_con} holds if for each $m$, $B^*_m(t,s)$ admits the decomposition $B^*_m(t,s)=\gamma_m(t)\kappa_m(s)$ for some uniformly bounded and continuous functions $\gamma_m(\cdot)$ and $\kappa_m(\cdot)$.
	\end{lemma}
	\begin{proof}
		Write $\gamma_m(t)=\sum_{i=1}^\infty \gamma^m_i\alpha_i(t)$ and  $\kappa_m(s)=\sum_{i=1}^\infty \kappa^m_iv_i(s)$. Then $b_{j,k}^m=\gamma^m_j\kappa^m_k$. Therefore inequality \eqref{suf_con} holds if
		\begin{eqnarray}\label{surf_con_eq1}
		\sum_{k=0}^\infty a_k^2\tilde\theta_{k}\ge C_1\tilde\theta_{m}
		\end{eqnarray}
		for sufficiently large $m$, where $\tilde\theta_{k}=\sum_{r=1}^\infty (\kappa^k_r)^2u_r$. Observe that $\tilde\theta_k=\EE[\int_0^1\kappa_k(s)\eta_0(s)\,ds]^2\le C$ for some constant $C$ that does not depend on $k$ since $\kappa_k(s)$ is uniformly bounded by assumption.  Therefore the right hand side of \eqref{surf_con_eq1} is bounded above by $C_1C$. Therefore \eqref{surf_con_eq1} holds unless all $\theta_k$ are 0. But if all $\theta_k$ are 0, \eqref{surf_con_eq1} is trivial.
	\end{proof}
	
	\subsection{FAR(1) Model}
	In this paper, we focus on the discussion when the physical dependence measure is of polynomial decay, that is, Assumption 2 of the main article holds true. However, all our results can be extended to the case when it is of exponential decay
	\begin{equation}\label{expon}
	\delta_x(l,q)\le C\rho^l, 0<\rho<1.
	\end{equation}
	Next, we will demonstrate an example of FAR(1) model to verify this exponential decay \eqref{expon} of the dependence measures. 
	\begin{example}[Functional AR(1) model] \label{exam2}
		Let $\epsilon_i(t)$ be i.i.d. centered and continuous Gaussian random functions with $\sup_{t\in[0,1]}\EE\Vert\epsilon_i(t)\Vert_2<\infty$. Consider the following model \begin{equation}\label{model}
		X_i(t)=\int_0^1 B(t,s)X_{i-1}(s)\dee s+\epsilon_i(t),
		\end{equation}
		where $B(t,s): [0,1]^2\to \mathbb{R}$ is a continuous, symmetric function satisfying $\int_0^1\int_0^1B^2(t,s)\dee t\dee s<\infty$ and $\int_0^1\int_0^1B(t,s)x(t)x(s)\dee t\dee s\ge 0$ with any random function $x(t)\in \mathcal{L}^2([0,1])$. Thus $B(t,s)$ is called a symmetric and positive-definite kernel on $[0,1]^2$. Define $C(t,s):=\EE(\epsilon_i(t)\epsilon_i(s))$ as the covariance function of $\epsilon_i(t)$. Let $v_1(t), v_2(t),\cdots$ and the corresponding $u_1\ge u_2\ge \cdots$ be the eigenfunctions and eigenvalues of $C(t,s)$. By the basis expansion method, we can write
		$X_i(t)=\sum_{k=1}^\infty \widetilde{x}_{i,k}v_k(t),~
		B(t,s)=\sum_{k=1}^\infty b_kv_k(t)v_k(s)$ and 
		$\epsilon_i(t)=\sum_{k=1}^\infty \epsilon_{i,k}v_k(t)$. Now, we can rewrite \eqref{model} as 
		\begin{equation}\label{basis_expan}
		\widetilde{x}_{i,k}=b_k\widetilde{x}_{i-1,k}+\epsilon_{i,k}=
		\sum_{m=0}^\infty b_k^m\epsilon_{i-m,k}.
		\end{equation}
		Here denote $f^2_k={\rm Var}(\widetilde{x}_{i,k})=\left(\sum_{m=0}^\infty b_k^{2m}\right)u_k$. If we let $\rho:=\sup_k|b_k|\in (0,1)$, then $f_k^2\ge u_k$. Further observe that $\epsilon_{i,k}$ are independent Gaussian random variables across $k$, hence $(\epsilon_{i-l,k}-\epsilon_{i-l,k}^\ast)$ is normally distributed with mean 0 and variance  $2u_k$, then we have 
		\begin{align*}
		\Vert x_{i,k}-x_{i,k}^\ast\Vert_q&=\Vert b_k^l(\epsilon_{i-l,k}-\epsilon_{i-l,k}^\ast)/f_k\Vert_q\\
		&\le C_q\sqrt{2u_k}b_k^l/\sqrt{u}_k\le C\rho^l.
		\end{align*}
	\end{example}

\section{Additional Theoretical Results and Proofs} \label{proof}

    \subsection{Full Assumptions for Theorem 1 of the Main Article}\label{full_ass}
    
    Here, we provide the complete set of assumptions and detailed approximation order in Theorem 1 of the main article.
    \begin{theorem}{(Detailed version of Theorem 1 in Section 5 of the main paper)}
    \label{thm}
Under Assumptions 1--8 in the main article, the smallest eigenvalue of $\bm{\Xi}^c$ is bounded below by some constant $\widehat{b}>0$ and $m=\bigO(n^{1/3})$. Define $\mathcal{B}_n^\epsilon=\{\omega:\Delta_n(\omega):=|\widetilde{
		\bm{\Xi}}^c-\bm{\Xi}^c|_F \le Cc_nn^{-1/3}h_n\},$ where $\omega$ represents the element in the probability space, $|\cdot|_F$ indicates Frobenius norm, $h_n$ diverges to infinity at an arbitrarily slow rate and $C>0$ is a finite constant. Then $\Pr(\mathcal{B}_n^\epsilon)=1-o(1)$. Under the event $\mathcal{B}_n^\epsilon$, we have 
	\begin{equation}\label{approx_order}
	    \widehat{\mathcal{K}}(\widehat{\bm{U}}_n^{boots},\bm{Z}_n^c)
	\le C\left(c_n^{\frac{7}{4}}n^{-\frac{1}{2}+\frac{9}{2q}+\frac{2}{\tau-1}}+c_n^{\frac{3}{8}}
		\lambda^{\frac{d_2+\psi+2}{2(2\gamma+d_2-\psi)}}+
		c_n^{\frac{5}{8}}n^{-\frac{1}{6}}h_n^{1/2}+
		c_n^{\frac{9}{8}}n^{-\frac{1}{4}+\frac{1}{q}}\right),
    \end{equation}
    where $C>0$ is some finite constant.  Further assume 
	\vspace{-0.2cm}
	\begin{enumerate}[(i)]
		\item $c_n\gg \left(n/\log(n)\right)^{\frac{1}{2(d_1+d_2-\psi)+3}}$ and $\lambda\ll \left(\log(n)/n\right)^{\frac{2(\gamma+d_2+1)} {2(d_1+d_2-\psi)+3}}$,\label{c4} 
		\item $\widehat{\bm{g}}_n(t)\in \mathcal{G}$ almost surely, \label{c5}
		\item $c_n^{\frac{7}{4}}n^{-\frac{1}{2}+\frac{9}{2q}+\frac{2}{\tau-1}}+c_n^{\frac{3}{8}}
		\lambda^{\frac{d_2+\psi+2}{2(2\gamma+d_2-\psi)}}+
		c_n^{\frac{5}{8}}n^{-\frac{1}{6}}h_n^{1/2}+
		c_n^{\frac{9}{8}}n^{-\frac{1}{4}+\frac{1}{q}}$
		$\to 0$ as $n\to \infty$, \label{c6}
	\end{enumerate} 
    the JSCB achieves
\begin{equation*}
\begin{aligned}
	\lim_{n\to \infty}\lim_{B\to\infty}\Pr\Big(\beta_{j}(t)\in &\left[\widetilde{\beta}_j(t)-\frac{\hat{q}_{n,1-\alpha}
	\widehat{g}_{nj}(t)}{\sqrt{n}},
	\widetilde{\beta}_j(t)+\frac{\hat{q}_{n,1-\alpha}\widehat{g}_{nj}(t)}
	{\sqrt{n}}\right]\\
	&~~\text{for}~\forall t\in [0,1] ~\text{and}~j=1,...,p\Big)=1-\alpha.
	\end{aligned}
    \end{equation*}
\end{theorem}

Note that \eqref{approx_order} establishes the rate of the bootstrap approximation to the weighted maximum deviation of $\bm{Z}_n^c$. More specifically, the first two terms on the right hand side of \eqref{approx_order} captures the magnitude of Gaussian approximation error, the third term represents the multiplier bootstrap approximation error and the last term is related to the estimation error. Condition (\ref{c4}) in \cref{thm} imposes a lower bound on $c$ and upper bound on $\lambda$ in order to obtain an under-smoothed estimator (hence the estimation bias is asymptotically negligible). The constraints on $c$ and $\lambda$ in Condition (\ref{c4}) are mild. For example, if $\bm{X}_i(t),~\bm{\beta}(t) \in \mathcal{C}^1$, $\gamma=4$ based on the normalized Legendre polynomials and $q,\tau\to \infty$, the parameter should be chosen as $(n/\log(n))^{\frac{1}{5}} \ll c_n\ll n^{\frac{2}{9}}$ and $\lambda\ll \left(\log(n)/n\right)^{\frac{12}{5}}$ such that the approximation error goes to 0. Condition (\ref{c5}) is a mild assumption on the weight function and for Condition (\ref{c6}), we assume that the approximation error vanishes as the sample size diverges to infinity.

\subsection{Discussion on the Roughness Penalization Estimation}\label{add_sec_D}
The roughness penalization approach to the functional linear model in the main article solves the penalized least squares problem
 \begin{equation}\label{ss}
\widetilde{\bm{\beta}}(t)=
\mathop{\arg\min}_{\bm{\beta}(t)}\left\{\frac{1}{n}\sum_{i=1}^n
\left[Y_i-\sum_{j=1}^p\int_0^1
\beta_j(t)X_{ij}(t)\dee t\right]^2+\lambda
\sum_{j=1}^p\int_0^1[\beta''_j(t)]^2\dee t\right\},
\end{equation}
where $\beta_j''(t)$ is the second derivative of the coefficient function $\beta_j(t)$, see \cite{SC15} and references therein. In functional data analysis, researchers usually work on the optimization problem in a finite-dimensional subspace \citep{LiHsing07}, which makes the procedure easily implementable. In this paper, we truncate the number of coefficient functions to finite (but diverging) dimensional spans of $\{\alpha_k(t)\}_{k=1}^\infty$ and recall $\beta_{j,c_{j,n}}(t)=\sum_{k=1}^{c_{j,n}}
\beta_{jk}\alpha_k(t)$ with each $c_{j,n}\to \infty$, then the estimation can be summarized in the following proposition. 
\begin{proposition}\label{prop_estimation}
    The estimation of $\bm{\beta}(t)$ can be achieved by minimizing the following penalized least-squares criterion function with respect to $\beta_{jk}$
    \begin{equation}\label{pena_eq}
    \frac{1}{n}\sum_{i=1}^n\left[Y_i-\sum_{j=1}^p\sum_{k=1}^{c_{j,n}}\beta_{jk}
\widetilde{x}_{ij,k}\right]^2+\lambda\sum_{j=1}^p
\sum_{k,l=1}^{c_{j,n}}\beta_{jk}
\beta_{jl}\widetilde{R}_j(k,l),
\end{equation}
where $\lambda$ ($\lambda\rightarrow 0$) is a smoothing parameter and $\widetilde{R}_j(k,l)=\int_0^1 \alpha_k''(t)\alpha_l''(t)\dee t$. 
\end{proposition}
Note that the smoothing parameter $\lambda$ measures the rate of exchange between fit to the data and smoothness of the estimator, as measured by the residual sum of squares in the first term of \eqref{pena_eq}, and variability of the functions $\beta_{j,c_{j,n}}(t)$ in the second term of \eqref{pena_eq}. Here we assume roughness penalty functions are associated with the same smoothing parameter $\lambda$ for every $j$ as is common in practice, see \cite{FL01}.

On the other hand, one can convert Eq. \eqref{pena_eq} to its equivalent matrix form $\frac{1}{n}[\bm{Y}-\bm{X}_c\bm{\theta}_c]^\top
[\bm{Y}-\bm{X}_c\bm{\theta}_c]
+\bm{\theta}_c^\top\bm{R}(\lambda)\bm{\theta}_c$ where $\bm{Y}, \bm{X}_c, \bm{\theta}_c$ and $\bm{R}(\lambda)$ are defined in Section 2.1 of the main article. Then the penalized least squares estimator can be obtained as $$\widetilde{\bm{\theta}}_c=\left[
\frac{\bm{X}_c^\top\bm{X}_c}{n}+\bm{R}(\lambda)\right]^{-1}\frac{\bm{X}_c^\top\bm{Y}}{n}.$$

\subsection{Discussion on Assumptions 3--7 of the Main Article}\label{sec_D2}
In this subsection, we will provide some examples to verify Assumptions in Section 5.1 of the main article. We consider a simple functional linear model
 \begin{equation}\label{simp_flm}
 Y_i=\int_0^1 \beta(t)X_i(t)\dee t+\epsilon_i,
 \end{equation}
 where $(X_i(t),Y_i)$ are stationary and $\EE[X_i(t)\epsilon_i]=0$ for all $t\in[0,1]$. Note that model \eqref{simp_flm} is a special case of the general functional linear regression model (1) in the main paper with $p=1$ and $\beta_0=0$. First, we claim Assumption 3 in the paper is mild and introduce two examples to justify it.

 \begin{example}\label{ass_31}
 Under model \eqref{simp_flm} and consider the standard Karhunen-Lo\`{e}ve type expansion $X_i(t)=\sum_{k=1}^\infty f_kx_{i,k}\phi_k(t)$, where $\{\phi_k(t)\}_{k=1}^\infty$ are the functional principal components and $\{x_{i,k}\}_{k=1}^\infty$ are the scaled functional principal component scores with $\EE[x_{i,k}^2]\equiv 1$ for all $i$ and $k$. By utilizing the roughness penalization method discussed in Proposition 1 of the paper to estimate unknown coefficients with the truncation number $c_n^\ast~(c_n^\ast\to \infty)$, we have
\begin{equation}\label{eq_sigma}
 \bm{\Sigma}_c=\frac{1}{n}\sum_{i=1}^n
 \EE[\bm{x}_i\bm{x}_i^\top]+\bm{R}(\lambda),
 \end{equation}
 where $\bm{x}_i=(x_{i,1},...,x_{i,c_n^\ast})^\top$. It is obvious to find that $x_{i,j}$ and $x_{i,k}$ are uncorrelated if $j\neq k$, therefore $\EE[\bm{x}_i\bm{x}_i^\top]$ is diagonal and $\lambda_{\min}(\sum_{i=1}^n\EE[\bm{x}_i\bm{x}_i^\top]/n)=1$. Consequently since $\bm{R}(\lambda)$ is semi-positive definite by its definition, we have $\lambda_{\min}(\bm{\Sigma}_c)\ge 1$, which satisfies Assumption 3. 
 \end{example}
 
 \begin{example}\label{ass_32}
 Similar to \cref{ass_31}, we consider the model \eqref{simp_flm} and the standard basis expansion $X_i(t)=\sum_{k=1}^\infty f_kx_{i,k}\alpha_k(t)$, where $\{\alpha_k(t)\}_{k=1}^\infty$ are predetermined basis functions. By constructing penalized least squares estimation, we can similarly calculate $\bm{\Sigma}_c=\frac{1}{n}\sum_{i=1}^n\EE[\bm{x}_i\bm{x}_i^\top]+\bm{R}(\lambda),$
 where $\bm{x}_i=(x_{i,1},...,x_{i,c_n^\ast})^\top$is the vector of the scaled random coefficients in the basis expansion with the truncation number $c_n^\ast$. Since $\EE[x_{i,k}^2]\equiv 1$ for all $i$ and $k$, we assume that there exists a permutation of the components in $\bm{x}_i$, denoted by $\{x_{i,j_1},...,x_{i,j_{c_n^\ast}}\}$, such that ${\rm Corr}(x_{i,j_s}x_{i,j_{s+k}})=
 \EE[x_{i,j_s}x_{i,j_{s+k}}]\le (|k|+1)^{-\tau}$ for $\tau>2$. Consequently by Weyl's inequality and Gershgorin circle theorem, the smallest eigenvalue of the first term in \eqref{eq_sigma} turns out to be
 \begin{align*}
 \lambda_{\min}\left(\frac{1}{n}\sum_{i=1}^n \EE[\bm{x}_i\bm{x}_i^\top]\right)&\ge \frac{1}{n}
 \sum_{i=1}^n\lambda_{\min}(
 {\rm Corr}(\bm{x}_i,\bm{x}_i))\\
 &\ge 1-\sum_{k\neq 0}\left|
 {\rm Corr}(x_{i,j}x_{i,j+k})\right|
 \ge 1-\frac{2^{-\tau+1}}{\tau-1},~~\tau>2.
 \end{align*} 
 On the other hand, $\bm{R}(\lambda)$ has nonnegative eigenvalues by its definition. Combining these two terms in \eqref{eq_sigma} with the first term strictly positive definite, we can conclude that it is mild to assume the smallest eigenvalue of $\bm{\Sigma}_c$ to be bounded below from zero by a universal constant.
 \end{example}

For the short-range dependence condition in Assumptions 4 of the main paper, we refer to \cite{Wu05} for several examples, including linear processes, nonlinear time series models, etc. To keep the supplemental material concise, we will not repeat the details herein. 

Next, we give a remark to illustrate the moment conditions in Assumptions 4--5. 

\begin{remark}\label{moment}
    We acknowledge that the relatively strict moment constraints such as $\Vert x_{ij,k}\Vert_q<\infty$ in Definition 1 and $\max_j\EE|z_{ci,j}|^q\le C_q$ in Assumption 5 for $q>9$ are used for the purpose of technical proofs. On the other hand, we believe that this assumption can be relaxed in many practical applications. To validate this, we perform a simulation study for a classic functional time series model where the error process has moderate heavy-tailed behavior. Specifically, we generate an FMA(1) model with an error process drawn from a $t$-distribution with $8$ degrees of freedom. In this context, $\Vert \epsilon_i\Vert_q$ is infinite when $q\ge 8$; nevertheless, our numerical results demonstrate reasonably good performance. 
\end{remark}
In the following, we present several examples of Assumption 6 in Section 5 of the main paper.
	\begin{example}[Integral of the second derivative]\label{eg_derivative2}
		~
		\begin{enumerate}
			\item Fourier bases. Consider $\alpha_k(t)=\{1,\sqrt{2}\cos(\pi t),\sqrt{2}\cos(2\pi t),\cdots\}$ for $t\in [0,1]$. Its second derivative is $$\alpha_k''(t)=
			\begin{cases}
			0,&k=0,\\
			-\sqrt{2}k^2\pi^2\cos(k\pi t),&k\ge 1.
			\end{cases}$$ 
			Then, we have $$\int_0^1 [\alpha_k''(t)]^2\dee t= 
			\begin{cases}
			0,&k=0,\\
			k^4\pi^4,&k\ge 1.
			\end{cases}$$
			As a result, one can obtain $|\widetilde{\bm{R}}_j|\asymp c_n^4$.
			\item Normalized Legendre polynomials. The Legendre polynomial of degree $n$ can be obtained using Rodrigue's formula
		$$P_n(t)=\frac{1}{2^n n!}\frac{\dee^n}{\dee t^n}(t^2-1)^n,~-1\le t \le 1.$$ For $t\in [0,1]$, the normalized Legendre polynomials turn out to be
		\begin{align*}
		\alpha_k(t)=\begin{cases}
		1,&k=0,\\
		\sqrt{2k+1}P_k(2t-1),&k>0.
		\end{cases}
		\end{align*}	
		To derive the spectral norm of $\widetilde{\bm{R}}_j$, first recall the Bonnet's recursion formula for Legendre polynomials, $$(n+1)P_{n+1}(x)=(2n+1)xP_n(x)-nP_{n-1}(x).$$ By elementary calculations, we have
		$$\frac{\dee}{\dee x}P_{n+1}(x)=(2n+1)P_n(x)+[2(n-2)+1]P_{n-2}(x)+[2(n-4)+1]P_{n-4}(x)+\cdots.$$
		Consequently, due to the fact that $$\sum_{i=1}^n i^2=\frac{n(n+1)(2n+1)}{6},~\sum_{i=1}^n i^5=\frac{n^6}{6}+\frac{n^5}{2}+\frac{5}{12n^4}-\frac{1}{12n^2},$$ we can deduce that
		$$P_n''(t)=\sum_{j=1}^{\lfloor{(n+1)/2}\rfloor}\sum_{k=1}^
		{\lfloor{(n-2j-1)/2}\rfloor} [2(n-2j)+3][2(n-2j-2k)+5]P_{n-2j-2k+2}(t).$$
		Next, we can calculate
		\begin{align*}
		&\int_{-1}^1[P_n''(t)]^2\dee t\\
		=&\int_{-1}^1\left\{ \sum_{j=1}^{\lfloor{(n+1)/2}\rfloor}\left(\sum_{k=1}^{\lfloor{(n-2j-1)/2}\rfloor}
		[2(n-2j)+3][2(n-2j-2k)+5]P_{n-2j-2k+2}(t)\right)^2\right.\\
		+& \sum_{j_1\neq j_2}^{\lfloor{(n+1)/2}\rfloor}\left(\sum_{k_1=1}^{\lfloor{(n-2j_1-1)/2}\rfloor}
		[2(n-2j_1)+3][2(n-2j_1-2k_1)+5]P_{n-2j_1-2k_1+2}(t)\right)\\
		\times& \left.\left(\sum_{k_2=1}^{\lfloor{(n-2j_2-1)/2}\rfloor}
		[2(n-2j_2)+3][2(n-2j_2-2k_2)+5]P_{n-2j_2-2k_2+2}(t)\right)\right\} \dee t
		\asymp n^7.
		\end{align*}
		Therefore, it yields that $\int_{0}^{1} [\alpha_k''(t)]^2 \dee t
		\asymp (\sqrt{k})^2k^7= k^8$ and $|\widetilde{\bm{R}}_j|
		\asymp c_n^8$ for $j=1,...,p$. 
	\end{enumerate}
\end{example}
	
	In the following, we will show some examples on Assumption 7 in Section 5.1. Recall $\bm{\alpha}(t)=(\alpha_1(t),\alpha_2(t),\cdots,\alpha_k(t))^\top$ and define $\zeta_k:=\sup_{t\in [0,1]}|\bm{\alpha}(t)|$ whose upper bound has been discussed in many works, see \cite{Newey97}, \cite{belloni2015som} and references therein. For example, $\zeta_k\le \sqrt{k}$ for tensor-products of univariate polynomial spline, trigonometric polynomial or wavelet bases and $\zeta_k\le k$ for tensor-products of power series or orthogonal polynomial bases. With the above result and the relationship $|\alpha_k(t)|_\infty\le \zeta_k$, then the statement of the first part of Assumption 7 will be verified easily. As for the second part, we list some commonly used basis functions as follows:
	\begin{example}[Supremum of the first derivative]\label{eg_derivative1}
		~
	\begin{enumerate}
		\item Fourier bases. Consider the Fourier bases with the same representation in \cref{eg_derivative2} and their first derivatives are $$\alpha_k'(t)=
		\begin{cases}
		0,&k=1,\\
		-\sqrt{2}k\pi\sin(k\pi t),&k\ge 1.
		\end{cases}$$
		Then, we have $\sup_{t\in [0,1]}|\alpha_k'(t)|=\bigO(k)$.
		\item Univariate spline series of order 3.  With a finite number of equally spaced knots $l_1,\cdots,l_{k-4}$ in $[0,1]$, $\alpha_k(t)=\{1,t,t^2,t^3,(t-l_1)^3_+,\cdots,(t-l_{k-4})^3_+\}$. Then the first derivative of spline basis function can be given as $$\alpha_k'(t)=\{0,1,2t,3t^2,3[(t-l_1)^3_+]^{2/3},\cdots\}.$$ Therefore, we conclude $\sup_{t\in [0,1]}|\alpha_k'(t)|=\bigO(1)$.
		\item Normalized Legendre polynomials. For $t\in [0,1]$, recall $$\alpha_k(t)=\{1,\sqrt{3}(2t-1),\sqrt{5/4}[3(2t-1)^2-1],\cdots\}.$$
		By the discussion in the proof of \cref{eg_derivative2}, we have $$\alpha_k'(t)\le 2\sqrt{2k+1}\sup_{x\in[-1,1]}|P_k(2t-1)|\sum_{m=1}^k(2m-1)=\bigO(k^{5/2}),$$ where $P_k(\cdot)$ is the Legendre polynomial basis function on $[-1,1]$.
	\end{enumerate}
	\end{example}
Consequently, the Assumption 7 of the paper is mild and can be satisfied by most basis functions.

	\subsection{Proof of the Results in Section \ref{gau}}\label{sec_proof}
	
	\textbf{Proof of Theorem \ref{thm2}}.
	On the outset, recall the matrix $\bm{\Sigma}_c^{-1}(\lambda)=
	\left(\EE\frac{\bm{X}_c^\top\bm{X}_c}{n}+\bm{R}(\lambda)\right)^{-1}\in\mathbb{R}^{c_n\times c_n}$. To streamline the proof, we assume that $\bm{R}(\lambda)$ is $c_n\times c_n$ diagonal block matrix with its $j$th block being $\bm{R}_j(\lambda)\in\mathbb{R}^{c_{j,n}\times c_{j,n}}$ for $j=1,...,p$. Denote the eigenvalues of $\bm{R}_j(\lambda)$ and $\bm{\Sigma}_c^{-1}(\lambda)$ as $v_1\ge v_2\ge \cdots\ge v_{c_{j,n}}$ and $\rho_1\ge \rho_2\ge\cdots\ge\rho_{c,n}$. Since $\EE\frac{\bm{X}_c^\top\bm{X}_c}{n}$ is positive semi-definite, we have $\rho_i\le Cv_i^{-1}$ for $i=1,...,c_{j,n}$. Moreover, the diagonal entries $R_{j,ii}(\lambda)$ increases as $i$ increases, then the eigenvalues of $\bm{\Sigma}_{c}^{-1}(\lambda)$ will approximate to zero at some truncation number. More specifically, define the truncated matrix as $\bar{\bm{R}}_j^{-1}(\lambda)$, which is similar to $\bm{R}_j^{-1}(\lambda)$ for $j=1,...,p$ with its diagonal components at locations larger than $(k_0,k_0)$ being zeros and $k_0=\lfloor\lambda^{-\frac{1}{2\gamma+d_2-\psi}}\rfloor$. Consequently, we can construct the truncated version of $\bm{\Sigma}_c^{-1}(\lambda)$ as $$\bar{\bm{\Sigma}}_c^{-1}(\lambda)=\bar{\bm{R}}^{-1}(\lambda)\left[\bm{I}_{c_n} +
		\EE\frac{\bm{X}_c^\top\bm{X}_c}{n}
		\bar{\bm{R}}^{-1}(\lambda)\right]^{-1}.$$

    Note that 
    \begin{align*}
	&\mathcal{K}(\bm{Z}_n^c,\bm{U}_n^c)\\
	\le &\sup_{\bm{g}_n\in\mathcal{G}}
	\sup_{x\in\mathbb{R}}\left|
	\Pr\left(\sup_{t\in[0,1]}
    |\bm{C}_f(t)\bm{\Sigma}_c^{-1}(\lambda)\bm{Z}_n^c|_{\bm{g}_n(t)}\le x\right)-\Pr\left(\sup_{t\in[0,1]}
    |\bm{C}_f(t)\bar{\bm{\Sigma}}_c^{-1}(\lambda)\bm{Z}_n^c|_{\bm{g}_n(t)}\le x\right)\right|\\
    & + \sup_{\bm{g}_n\in\mathcal{G}}
	\sup_{x\in\mathbb{R}}\left|
	\Pr\left(\sup_{t\in[0,1]}
    |\bm{C}_f(t)\bar{\bm{\Sigma}}_c^{-1}(\lambda)\bm{Z}_n^c|_{\bm{g}_n(t)}\le x\right)-\Pr\left(\sup_{t\in[0,1]}
    |\bm{C}_f(t)\bar{\bm{\Sigma}}_c^{-1}(\lambda)\bm{U}_n^c|_{\bm{g}_n(t)}\le x\right)\right|\\
    & + \sup_{\bm{g}_n\in\mathcal{G}}
	\sup_{x\in\mathbb{R}}\left|
	\Pr\left(\sup_{t\in[0,1]}
    |\bm{C}_f(t)\bar{\bm{\Sigma}}_c^{-1}(\lambda)\bm{U}_n^c|_{\bm{g}_n(t)}\le x\right)-\Pr\left(\sup_{t\in[0,1]}
    |\bm{C}_f(t)\bm{\Sigma}_c^{-1}(\lambda)\bm{U}_n^c|_{\bm{g}_n(t)}\le x\right)\right|\\
    =:& I + II + III.
    \end{align*}
	
	First consider the term II, let $\bar{\bm{Q}}_n^z(t,\lambda):=
    \bm{C}_f(t)
	\bar{\bm{\Sigma}}_c^{-1}(\lambda)\bm{Z}_n^c$ and $\bar{\bm{Q}}_n^u(t,\lambda)
    :=\bm{C}_f(t)
	\bar{\bm{\Sigma}}_c^{-1}(\lambda)\bm{U}_n^c$. Denote $$A_x^{g_n}=\left\{\bm{S}\in \mathbb{R}^{c_n}:
 \sup_{t\in[0,1]}\max_{1\le j\le p} |\bm{E}_j^\top\bm{C}_f(t)\bar{\bm{\Sigma}}^{-1}_c(\lambda)\bm{S}/g_{nj}(t)
	|\le x\right\}$$ where $\bm{E}_j\in\mathbb{R}^p$ contains 1 at the $j$th location and 0 at others, further let $\mathcal{A}_n=\{A_x^{g_n}: x\in\mathbb{R}, g_n(t)\in \mathcal{G}\}$. Since it is easy to check that  $A_x^{g_n}$ is a convex set and $\mathcal{A}_n$ is a collection of convex sets, we can write II as the Kolmogorov distance between $\bm{Z}_n^c$ and $\bm{U}_n^c$ on $\mathcal{A}_n$,
	$$\bar{\mathcal{K}}(\bm{Z}_n^c,\bm{U}_n^c)=\sup_{\bm{g}_n\in\mathcal{G}}
	\sup_{x\in\mathbb{R}}\left|
	\Pr(\bm{Z}_n^c\in A_x^{g_n})-\Pr(\bm{U}_n^c\in A_x^{g_n})\right| =
	\sup_{A\in\mathcal{A}_n}\left|\Pr(\bm{Z}_n^c\in A)-\Pr(\bm{U}_n^c\in A)\right|,$$
	where $A$ is also a convex set. Our aim is to apply the idea of \cite{Fang16} to obtain the above Gaussian approximation result. First, we follow the smoothing technique of \cite{Bentkus03}. For $A\in\mathcal{A}_n$, let $h_A(x)=\bm{1}_A(x)$ where $\bm{1}_A(x)$ stands for the indicator function of event $A$, and define the smoothed function
	$$h_{A,\epsilon_1}(\bm{\omega})=\psi\left(\frac{{\rm dist} (\bm{\omega},A)}{\epsilon_1}\right),$$ where ${\rm dist}(\bm{\omega},A)=\inf_{\bm{\nu}\in A}|\bm{\omega}-\bm{\nu}|$ and
	
	\begin{equation*}
	\psi(x)=\begin{cases}
	1,~&x<0,\\
	1-2x^2,~&0\le x<\frac{1}{2},\\
	2(1-x)^2,~&\frac{1}{2}\le x<1,\\
	0,~&x\ge 1.
	\end{cases}
	\end{equation*}
	From Lemma 2.3 (iv) of \cite{Bentkus03}, we have $|\nabla h_{A,\epsilon_1}|\le 2\epsilon_1^{-1}$ for all $\bm{\omega}\in \mathbb{R}^{c_n}$. Then, we recall the following main results in the literature.
	\begin{lemma}[Lemma 4.2 of \cite{Fang15}]\label{lemma3}
		For any $d$-dimensional random vector $\bm{W}$, $$\mathcal{K}(\bm{W},\bm{Z})\le 4d^{\frac{1}{4}}\epsilon_1+\sup_{A\in\mathcal{A}_n}\left|\EE\left[
		h_{A,\epsilon_1}(\bm{W})-h_{A,\epsilon_1}(\bm{Z})\right]\right|,$$
		where $\bm{Z}$ is a $d$-dimensional standard Gaussian vector.
	\end{lemma}
	
	\begin{lemma}[Remark 2.2 of \cite{Fang16}]\label{lemma4}
		Let $\bm{W}=\sum_{i=1}^n\bm{X}_i$ be a sum of $d$-dimensional random vectors such that $\EE(\bm{X}_i)=0$ and ${\rm Cov}(\bm{W})=\bm{\Sigma}_w$. Suppose $\bm{W}$ can be decomposed as follows:\\
		1. $\forall i \in [n]$, $\exists i\in N_i \subset [n]$, such that $\bm{W}-\bm{X}_{N_i}$ is independent of $\bm{X}_i$, where $[n] = \{1,\cdots,n\}$.\\
		2. $\forall i \in [n]$, $j\in N_i$, $\exists N_i \subset N_{ij} \subset [n]$, such that $\bm{W}-\bm{X}_{N_{ij}}$ is independent of
		$\{\bm{X}_i, \bm{X}_j\}$.\\
		3. $\forall i \in [n]$, $j \in N_i,~k \in N_{ij},~\exists N_{ij} \subset N_{ijk} \subset [n]$ such that $\bm{W}-\bm{X}_{N_{ijk}}$ is independent of
		$\{\bm{X}_i,\bm{X}_j,\bm{X}_k\}$.
		
		Suppose further that for each $i\in [n],~j\in N_i,~k\in N_{ij},~
		|\bm{X}_i|\le \beta,~|N_i|\le n_1,~|N_{ij}|\le n_2,~|N_{ijk}|\le n_3$. Then there exists a universal constant $C$ such that
		$$\mathcal{K}(\bm{W},\bm{\Sigma}_w^{1/2}\bm{Z})\le Cd^{1/4} n| \bm{\Sigma}_w^{-1/2}|^3\beta^3n_1
		(n_2+\frac{n_3}{d}),$$ where $\bm{Z}$ is a $d$-dimensional standard Gaussian random vector.
	\end{lemma}
	
	Recall $\bm{z}_{ci}=\bm{x}_{ci}\epsilon_i$, since $\{x_{ij,k}\}$ and $\{\epsilon_i\}$ are both stationary processes, we can rewrite $\bm{z}_{ci}$ into a physical representation of stationary multivariate time series, i.e.,
	$$\bm{z}_{ci}=\bm{H}(\mathcal{F}_i),$$ where $\bm{H}=(H_1,...,H_{c_n})^\top$ is a measurable vector function. Here, we define the dependence measure on the element $\{z_{ci,j}\}_{j=1}^{c_n}$ of the process $\{\bm{z}_{ci}\}_{i=1}^n$ as $$\delta_z(l,q/2):=\max_{1\le k\le c}\Vert H_k(\mathcal{F}_i)-H_k(\mathcal{F}_{i,l})\Vert_{q/2},$$ 
	where $\mathcal{F}_{i,l}$ is defined in Definition 1 of the main article. Under Assumptions 1 and 4 in the main paper, we can derive that
	\begin{align*}
	\delta_z(l,q/2)\le&\max_{1\le j\le p}\max_{1\le k\le c_j}\Vert x_{ij,k}-x_{ij,k}^\ast\Vert_{q}\Vert \epsilon_i\Vert_{q}+\max_{1\le j\le p}\max_{1\le k\le c}\Vert x_{ij,k}^\ast\Vert_{q}\Vert \epsilon_i-\epsilon_i^\ast\Vert_{q}\\
	\le& C(l+1)^{-\tau},~~\tau>5,
	\end{align*}
 where $x_{ij,k}^\ast$ and $\epsilon_i^\ast$ are i.i.d. copies of $x_{ij,k}$ and $\epsilon_i$, respectively. To deduce the error bound of Gaussian approximation, we need to make use of truncation approximation, $m$-dependent approximation techniques and finally apply the aforementioned two lemmas. Now define the truncated version of $\bm{z}_{ci}$ as
	\begin{equation*}
	\bar{\bm{z}}_{ci}=\begin{cases}
	\bm{z}_{ci},~&|\bm{z}_{ci}|\le c_n^{\frac{1}{2}}n^{\frac{3}{2q}},\\
	\bm{0}_{c_n},~&\text{otherwise}.
	\end{cases}
	\end{equation*}
	Given a large constant $M=M(n)$, let the $M$-dependent approximation of $\bar{\bm{z}}_{ci}$ be
$$\bar{\bm{z}}_{ci}^M=\EE(\bar{\bm{z}}_{ci}|\eta_{i-M},\cdots,\eta_i),
	~i=1,\cdots,n.$$
	Consequently, we will define the partial sum of the truncated series as $\bar{\bm{Z}}_n^c=\sum_{i=1}^n\bar{\bm{z}}_{ci}/\sqrt{n}$ and the $M$-dependent version as $\bar{\bm{Z}}_n^M=\sum_{i=1}^n\bar{\bm{z}}_{ci}^M/\sqrt{n}$. Further let $\bar{\bm{Z}}_n^\ast=\bar{\bm{Z}}_n^c-\EE\bar{\bm{Z}}_n^c$, $\tilde{\bm{Z}}_n^M=\bar{\bm{Z}}_n^M-\EE\bar{\bm{Z}}_n^c$ and let $\tilde{\bm{U}}_n^M$ be a Gaussian random vector preserving the covariance structure of $\tilde{\bm{Z}}_n^M$. With \cref{lemma3} and the fact $|\nabla h_{A,\epsilon_1}|\le 2\epsilon_1^{-1}$, we have
	\begin{align}
	\bar{\mathcal{K}}(\bm{Z}_n^c,\bm{U}_n^c)\le&~4c_n^{\frac{1}{4}}\epsilon_1
	+\sup_{A\in\mathcal{A}_n}
 \left|\EE\left[h_{A,\epsilon_1}(\bm{Z}_n^c)- h_{A,\epsilon_1}(\bm{U}_n^c)\right]\right| \notag\\
	\le&~4c_n^{\frac{1}{4}}\epsilon_1+\sup_{A\in\mathcal{A}_n}
 \left|\EE\left[
	h_{A,\epsilon_1}(\bm{Z}_n^c)-h_{A,\epsilon_1}(\bar{\bm{Z}}_n^\ast)
	\right]\right|+\sup_{A\in\mathcal{A}_n}\left|
	\EE\left[h_{A,\epsilon_1}(\bar{\bm{Z}}_n^\ast)- h_{A,\epsilon_1}(\tilde{\bm{Z}}_n^M)\right]\right|\notag \\
	&{}+\sup_{A\in\mathcal{A}_n}\left|
	\EE\left[h_{A,\epsilon_1}(\tilde{\bm{Z}}_n^M)- h_{A,\epsilon_1}(\tilde{\bm{U}}_n^M)\right]\right|
	+\sup_{A\in\mathcal{A}_n}\left|
	\EE\left[h_{A,\epsilon_1}(\tilde{\bm{U}}_n^M)- h_{A,\epsilon_1}(\bm{U}_n^c)\right]\right| \notag\\
	\le&~4c_n^{\frac{1}{4}}\epsilon_1+\frac{C}{\epsilon_1}\EE|\bm{Z}_n^c-
	\bar{\bm{Z}}_n^\ast|+\frac{C}{\epsilon_1}\EE|\bar{\bm{Z}}_n^\ast-
	\tilde{\bm{Z}}_n^M|\notag \\
	&{}+\sup_{A\in\mathcal{A}_n}\left|
	\EE\left[h_{A,\epsilon_1}(\tilde{\bm{Z}}_n^M)- h_{A,\epsilon_1}(\tilde{\bm{U}}_n^M)\right]\right|+
	\frac{C}{\epsilon_1}\EE|\tilde{\bm{U}}_n^M-\bm{U}_n^c|\notag\\
	:=&~4c_n^{\frac{1}{4}}\epsilon_1+
	\frac{C\epsilon_2}{\epsilon_1}+\frac{C\epsilon_3}
	{\epsilon_1}+\frac{C\epsilon_4}
	{\epsilon_1}+
 \sup_{A\in\mathcal{A}_n}\left|\EE\left[h_{A,\epsilon_1}
	(\tilde{\bm{Z}}_n^M)-h_{A,\epsilon_1}(\tilde{\bm{U}}_n^M)\right]\right|, \label{combine}
	\end{align}
	where $\epsilon_2:=\EE|\bm{Z}_n^c-\bar{\bm{Z}}_n^\ast|$, $\epsilon_3:=\EE|\bar{\bm{Z}}_n^\ast-
	\tilde{\bm{Z}}_n^M|$ and $\epsilon_4:=\EE|\tilde{\bm{U}}_n^M-\bm{U}_n^c|$.\\
	(1) Truncation approximation.
	
	We shall first control the truncation error $\epsilon_2$. Note that $\EE\bm{z}_i=0$, then we have
 \begin{align*}
	|\bm{Z}_n^c-\bar{\bm{Z}}_n^\ast|&=
	\frac{1}{\sqrt{n}}\left|\sum_{i=1}^n(\bm{z}_{ci}-\EE\bm{z}_{ci}
	-\bar{\bm{z}}_{ci}+\EE\bar{\bm{z}}_{ci})\right|\\
	&\le \frac{1}{\sqrt{n}}\left|\sum_{i=1}^n(\bm{z}_{ci}-
	\bar{\bm{z}}_{ci})\right|+\frac{1}{\sqrt{n}}
	\left|\sum_{i=1}^n\EE(\bm{z}_{ci}-\bar{\bm{z}}_{ci})\right|\\
	:&={\rm I}+{\rm II}.
	\end{align*}
	For ${\rm I}$, notice that for any $i=1,...,n$,
	\begin{align*}
&\Pr\left(|\bm{z}_{ci}|>c_n^{\frac{1}{2}}n^{\frac{3}{2q}}\right)=
	\EE\left[\bm{1}\{|\bm{z}_{ci}|>c_n^{\frac{1}{2}}n^{\frac{3}{2q}}\}\right]\\
	\le &\EE\left[\left(\frac{|\bm{z}_{ci}|}
	{c_n^{\frac{1}{2}}n^{\frac{3}{2q}}}\right)^q\right]=
	c_n^{-\frac{q}{2}}n^{-\frac{3}{2}}\EE|\bm{z}_{ci}|^q \\
	=&c_n^{-\frac{q}{2}}
	n^{-\frac{3}{2}}\EE\left|\sum_{j=1}^{c_n} z_{ci,j}^2\right|^{q/2}
	\le c_n^{-\frac{q}{2}}
	n^{-\frac{3}{2}}c_n^{q/2-1}\sum_{j=1}^{c_n}\EE|z_{ci,j}|^q
	\le C_qn^{-\frac{3}{2}},
	\end{align*}
	where the second to last inequality follows from the inequality $\EE|X_1+\cdots+X_c|^{q/2}\le c_n^{q/2-1}\sum_{j=1}^{c_n}\EE|X_j|^{q/2}$ for random variables $\{X_j\}_{j=1}^{c_n}$ and Assumption 5 of the main paper. Hence, we have $\Pr\left(|\bm{Z}_n^c-\bar{\bm{Z}}_n^c|=0\right)=1-o(n^{-1/2})$. This implies there exists an order, say $n^{-2}$ such that $\Pr\left(\left|\frac{1}{\sqrt{n}}\sum_{i=1}^n
	(\bm{z}_{ci}-\bar{\bm{z}}_{ci})\right|>n^{-2}\right)\to 0$. As a result, ${\rm I}=o_\Pr(n^{-2})$.
	
	For ${\rm II}$, since for any $i=1,...,n$,
	\begin{align*}
	\EE(\bm{z}_{ci}-\bar{\bm{z}}_{ci})
	\le& \EE\left[|\bm{z}_{ci}|
	\bm{1}\{|\bm{z}_{ci}|>c_n^{\frac{1}{2}}
	n^{\frac{3}{2q}}\}\right]\\
	\le& \EE\left[|\bm{z}_{ci}|\left(\frac{|\bm{z}_{ci}|}
	{c_n^{\frac{1}{2}}n^{\frac{3}{2q}}}\right)^{q-1}\right]\\
	=&c_n^{-\frac{q-1}{2}}n^{-\frac{3}{2}+\frac{3}{2q}}\EE|\bm{z}_{ci}|^q\\ 
	\le&C_q c_n^{\frac{1}{2}}n^{-\frac{3}{2}+\frac{3}{2q}},
	\end{align*}
	where the second inequality uses the fact that for the nonnegative random variable $y$ and some number $a>0$, the inequality $y\bm{1}\{y\ge a\}\le y\left(\frac{y}{a}\right)^p$ for any $p>0$ holds true. Consequently, ${\rm II}=\frac{1}{\sqrt{n}}\left|\sum_{i=1}^n\EE(\bm{z}_{ci}-
	\bar{\bm{z}}_{ci})\right|=\bigO(c_n^{\frac{1}{2}}n^{-1+\frac{3}{2q}})$. Now, by choosing $\beta=c_n^{\frac{1}{2}}n^{-\frac{1}{2}+\frac{3}{2q}}$, then $\epsilon_2=\bigO(c_n^{\frac{1}{2}}n^{-1+\frac{3}{2q}})$.
	\bigskip
	
	\noindent
	(2) $M$-dependence approximation.
	
	Next we will deduce the approximation rate between our original process and its $M$-dependent sequence, i.e., control $\epsilon_3$ in \eqref{combine}. Recall the physical dependence measure $\delta_z(l,q/2)$ of ${z}_{ci,j}$ and denote $\Theta_{M,q/2}=\sum_{l=M}^\infty\delta_z(l,q/2)$. Let
	$$\bar{\bm{Z}}_n^c-\bar{\bm{Z}}_n^M=\frac{1}{\sqrt{n}}\sum_{i=1}^n
	(\bar{\bm{z}}_{ci}-\bar{\bm{z}}_{ci}^M)=:\frac{1}{\sqrt{n}}\sum_{i=1}^n
	\bar{\bm{z}}_i^\Delta.$$ It is readily seen that $\{\bar{\bm{z}}_i^\Delta,i=1,...,n\}$ is a sequence of martingale differences, then we have
	\begin{align*}
	&\left\Vert\bar{\bm{Z}}_n^\ast-\tilde{\bm{Z}}_n^M\right\Vert_{q/2}^2=
	\left\Vert\bar{\bm{Z}}_n^c-\bar{\bm{Z}}_n^M\right\Vert_{q/2}^2\\ =&
	\left\Vert \frac{1}{\sqrt{n}}\sum_{i=1}^n \bar{\bm{z}}_i^\Delta \right\Vert_{q/2}^2
	=\left\{\EE\left[\sum_{j=1}^{c_n}\left(\frac{1}{\sqrt{n}}\sum_{i=1}^n
	\bar{z}_{i,j}^\Delta\right)^2
 \right]^{q/4}\right\}^{4/q}\\
	\le&\left\{c_n^{q/4-1}\sum_{j=1}^{c_n}\EE\left|\frac{1}{\sqrt{n}}\sum_{i=1}^n
	\bar{z}_{i,j}^\Delta\right|^{q/2}\right\}^{4/q}\\
	\le& c_n\left\Vert \max_{1\le j\le c_n}\Big | \frac{1}{\sqrt{n}}\sum_{i=1}^n\bar{z}_{i,j}^\Delta\Big |\right\Vert_{q/2}^2 \le Cc_n\Theta_{M,q/2}^2,
	\end{align*}
	where $\bar{z}_{i,j}^\Delta$ is the entrywise of vector $\bar{\bm{z}}_i^\Delta$, the first inequality is due to the fact that $\EE|X_1+\cdots+X_{c_n}|^{q/4}\le c_n^{q/4-1}\sum_{j=1}^{c_n}\EE|X_j|^{q/4}$ with $X_j$ being random variables. The second inequality above is followed by Lemma A.1 of \cite{LiuLin09} using Burkholder's inequality. As a result, $\left\Vert\bar{\bm{Z}}_n^\ast-\tilde{\bm{Z}}_n^M\right\Vert_{q/2}
	\le Cc_n^{1/2}M^{-\tau+1}$.
	
	Therefore, when we choose $M$ appropriately to satisfy $$\frac{M^{-\tau+1}}{n^{-1+\frac{3}{2q}}}\to 0,$$ for example $M=\bigO(n^{\frac{1}{\tau-1}})$ for $\tau>5$, then we have $|\bar{\bm{Z}}_n^\ast-\tilde{\bm{Z}}_n^M|=o_\Pr(c_n^{1/2}n^{-1})$. Consequently, $\epsilon_3=o(\epsilon_2)$.
	\medskip
	
	\noindent
	(3) Using \cref{lemma4} to obtain the final result.
	
	At last, we will employ \cref{lemma4} so as to deal with the last term of \cref{combine}. Let $\tilde{\bm{Z}}_n^M=\sum_{i=1}^n\bar{\bm{z}}_{ci}^\dagger/\sqrt{n}$, then by the truncation and $m$ dependence approximation techniques, we have $|\bar{\bm{z}}_{ci}^\dagger/\sqrt{n}|\le \beta$ and
	$\EE\bar{\bm{z}}_{ci}^\dagger/\sqrt{n}=0$. Recall $\bm{\Xi}^c=
	\EE\left(\sum_{i=1}^n\bm{z}_{ci}\right)
	\left(\sum_{i=1}^n\bm{z}_{ci}^\top\right)/n$, denote $\bm{\Xi}_M^c:={\rm Cov}(\tilde{\bm{Z}}_n^M)=\EE\left[\sum_{i=1}^n(\bar{\bm{z}}_{ci}^M-
	\EE\bar{\bm{z}}_{ci})\right]\left[\sum_{i=1}^n(\bar{\bm{z}}_{ci}^M
	-\EE\bar{\bm{z}}_{ci})^\top\right]/n$. Next we will find that the difference of covariance matrix between $\tilde{\bm{Z}}_n^M$ and $\bm{Z}_n^c$ based on Frobenius norm turns to be
	\begin{align*}
	&\left|\bm{\Xi}^c-\bm{\Xi}_M^c\right|_F\\
	\le &\frac{1}{n}\left\{\left|\EE\left[\sum_{i=1}^n\left(\bm{z}_{ci}-
	\bar{\bm{z}}_{ci}^M\right)\right]\left(\sum_{i=1}^n\bm{z}_{ci}^\top\right)
	\right|_F+\left|\EE\left(\sum_{i=1}^n\bar{\bm{z}}_{ci}^M\right)
	\left[\sum_{i=1}^n\left(\bm{z}_{ci}-\bar{\bm{z}}_{ci}^M\right)^\top\right]
	\right|_F\right.\\
	&+\left.\left|\sum_{i=1}^n\EE
	(\bm{z}_{ci}-\bar{\bm{z}}_{ci})\sum_{i=1}^n\EE
	(\bm{z}_{ci}-\bar{\bm{z}}_{ci})^\top\right|_F\right\}\\
	\le &C(c_nn^{-2}+c_n^{3/2}M^{-\tau+1}+c_n^2n^{-2+3/q})=\bigO(c_n^{3/2}n^{-1}),
	\end{align*}
	where the last inequality follows by the error bound for II and \cref{proof1} and Cauchy Schwarz inequality. Further denote $\tilde{U}_{nj}^M$ and $U_{nj}^c$ are components of vector $\tilde{\bm{U}}_n^M$ and $\bm{U}_n^c$, respectively. Then, 
	\begin{align*}
	\epsilon_4&=\EE\sqrt{\sum_{j=1}^c(\tilde{U}_{nj}^M-U_{nj}^c)^2}\le \sqrt{\EE\sum_{j=1}^c(\tilde{U}_{nj}^M-U_{nj}^c)^2}\\
	&=\sqrt{{\rm Tr}\left[(\bm{\Xi}_M^c)^{1/2}-
		(\bm{\Xi}^c)^{1/2}\right]^2}=\left|(\bm{\Xi}_M^c)^{1/2}-(\bm{\Xi}^c)^{1/2}
	\right|_F\\
	&=\bigO(c_n^{3/2}n^{-1}).
	\end{align*} 
	Since we assume that the smallest eigenvalue of $\bm{\Xi}^c$ is bounded below by some constant $b>0$, we can derive that
	\begin{align*}
	\lambda_{\min}(\bm{\Xi}_M^c)&\ge
	\lambda_{\min}(\bm{\Xi}_M^c-\bm{\Xi}^c)+
	\lambda_{\min}(\bm{\Xi}^c)\\
	&= -\lambda_{\max}(\bm{\Xi}^c-\bm{\Xi}_M^c)+
	\lambda_{\min}(\bm{\Xi}^c)\\
	&\ge b-Cc^{3/2}n^{-1}>0.
	\end{align*}
	Hence, the small eigenvalue of $\bm{\Xi}_M^c$ is also bounded below by some positive constant, then $\left|(\bm{\Xi}_M^c)^{-1}\right|\le C$.
	Further note that $n_1=M,~n_2=2M,~n_3=3M$ and together with the Eq. (4.24) in \cite{Fang16}, we have
	\begin{equation}\label{key}
	\bar{\mathcal{K}}(\tilde{\bm{Z}}_n^M,\tilde{\bm{U}}_n^M)\le 4c_n^{\frac{1}{4}}\epsilon_1
	+2Cn\beta^3M^2\frac{1}{\epsilon_1}[c_n^{\frac{1}{4}}(\epsilon_1+3M\beta)+
	\bar{\mathcal{K}}(\tilde{\bm{Z}}_n^M,\tilde{\bm{U}}_n^M)].
	\end{equation}
	By substituting $\beta=c_n^{\frac{1}{2}}n^{-\frac{1}{2}+\frac{3}{q}}$ and optimizing $\epsilon_1$, we can choose $\epsilon_1=\bigO(c_n^{\frac{3}{2}}
	n^{-\frac{1}{2}+\frac{9}{2q}}M^2)$. Then the second term in \cref{combine} turns out to be $\bigO(c_n^{-1}n^{-\frac{1}{2}-\frac{3}{q}}M^{-2})$ and the fourth term is $\bigO(n^{-\frac{3}{2}-\frac{3}{2q}}M^{-2})$. By substituting $M=\bigO(n^{\frac{1}{\tau-1}})$, we have $$II=\bar{\mathcal{K}}(\bm{Z}_n^c,\bm{U}_n^c)=\bigO(
	c_n^{\frac{7}{4}}n^{-\frac{1}{2}+\frac{9}{2q}+\frac{2}{\tau-1}}).$$  
	
	Next for the term III, we denote 
    $\bm{W}_n^c:=
    \bm{\Sigma}_c^{-1}\bm{U}_n^c \in\mathbb{R}^{c_n}$, $\bar{\bm{W}}_n^c:=\bar{\bm{\Sigma}}_c^{-1}\bm{U}_n^c\in\mathbb{R}^{c_n}$ and $\bm{\Delta}(\lambda)=\bar{\bm{\Sigma}}_c^{-1}(\lambda)-
	\bm{\Sigma}_c^{-1}(\lambda)$. In a similar fashion, consider 
    $$B_x^{g_n}=\left\{\bm{S}\in \mathbb{R}^{c_n}:
 \sup_{t\in[0,1]}\max_{1\le j\le p} |\bm{E}_j^\top\bm{C}_f(t)\bm{S}/g_{nj}(t)
	|\le x\right\}$$ and let $\mathcal{B}_n=\{B_x^{g_n}: x\in\mathbb{R}, g_n(t)\in \mathcal{G}\}$. According to \cref{lemma3} and the fact $|\nabla h_{B,\epsilon}|\le 2\epsilon_1^{-1}$, we have
    \begin{align}
        III=&\mathcal{K}(\bar{\bm{W}}_n^c,
        \bm{W}_n^c)\le 4c_n^{1/4}\epsilon_1 + \sup_{B\in \mathcal{B}_n}|\EE[h_{B,\epsilon_1}(\bar{\bm{W}}_n^c)-h_{B,\epsilon_1}(\bm{W}_n^c)]| \notag \\
        \le & 4c_n^{1/4}\epsilon_1 + \frac{C}{\epsilon_1} \EE\left|[\bar{\bm{\Sigma}}_c^{-1}(\lambda)-\bm{\Sigma}_c^{-1}(\lambda)]\bm{U}_n^c\right| \notag \\
        \le & 4c_n^{1/4}\epsilon_1 + \frac{C}{\epsilon_1} \sqrt{\EE\left[\text{Tr}\left(
        \bm{\Delta}^2(\lambda)
        \bm{U}_n^c(\bm{U}_n^c)^\top
        \right)\right]} \notag\\
        \le & 4c_n^{1/4}\epsilon_1 + 
        \frac{C}{\epsilon_1} |\bm{\Delta}(\lambda)|_F |\bm{\Xi}^c|^{1/2}, \label{gau_third}
        %\le & 4c_n^{1/4}\epsilon_1 + \frac{C}{\epsilon_1} \lambda^{\frac{2d_2+2\psi-3}{2(2\gamma+d_2-\psi)}}.
    \end{align}
    where the third inequality follows by the Jensen's inequality and the cyclic invariance of the trace. The last inequality can be obtained by the fact ${\rm Tr}(\bm{AB})\le {\rm Tr}(\bm{A})|\bm{B}|$ for positive semi-definite matrix $\bm{A}$ and arbitrary matrix $\bm{B}$. Now it suffices to derive the upper bound of $|\bm{\Delta}(\lambda)|_F$. Notice that
    {\small
   \begin{align*}
   \bm{\Delta}(\lambda) =& \bar{\bm{R}}^{-1}(\lambda)\left[\bm{I}_{c_n} +
		\EE\frac{\bm{X}_c^\top\bm{X}_c}{n}
		\bar{\bm{R}}^{-1} (\lambda)\right]^{-1} -\bar{\bm{R}}^{-1}(\lambda)\left[\bm{I}_{c_n} +
		\EE\frac{\bm{X}_c^\top\bm{X}_c}{n}
		\bm{R}^{-1}(\lambda)\right]^{-1}\\
        =& \left(\bar{\bm{R}}^{-1}(\lambda)-\bm{R}^{-1}(\lambda)\right)
        \left[\bm{I}_{c_n} +
		\EE\frac{\bm{X}_c^\top\bm{X}_c}{n}
		\bar{\bm{R}}^{-1} (\lambda)\right]^{-1} \\
        & \quad - \bm{R}^{-1}(\lambda)\left\{
        \left[\bm{I}_{c_n} +
		\EE\frac{\bm{X}_c^\top\bm{X}_c}{n}
		\bm{R}^{-1}(\lambda)\right]^{-1} - \left[\bm{I}_{c_n} +
		\EE\frac{\bm{X}_c^\top\bm{X}_c}{n}
		\bar{\bm{R}}^{-1}(\lambda)\right]^{-1}\right\}\\
        =& \left(\bar{\bm{R}}^{-1}(\lambda)-\bm{R}^{-1}(\lambda)\right)
        \left[\bm{I}_{c_n} +
		\EE\frac{\bm{X}_c^\top\bm{X}_c}{n}
		\bar{\bm{R}}^{-1} (\lambda)\right]^{-1} \\
        & \quad - \bm{\Sigma}_c^{-1}(\lambda)\EE\frac{\bm{X}_c^\top
        \bm{X}_c}{n}
        \left(\bar{\bm{R}}^{-1}(\lambda)-\bm{R}^{-1}(\lambda)\right)
        \left[\bm{I}_{c_n} +
		\EE\frac{\bm{X}_c^\top\bm{X}_c}{n}
		\bar{\bm{R}}^{-1}(\lambda)\right]^{-1}
   \end{align*}
   }
   Consequently, we can obtain
   \begin{align*}
   |\bm{\Delta}(\lambda)|_F \le & 
   \left|\bar{\bm{R}}^{-1}(\lambda)-\bm{R}^{-1}(\lambda)\right|_F 
   \left|\left[\bm{I}_{c_n} +
		\EE\frac{\bm{X}_c^\top
        \bm{X}_c}{n}
		\bar{\bm{R}}^{-1} (\lambda)\right]^{-1}
        \right| \\
        &\quad + \left|\bar{\bm{R}}^{-1}(\lambda)-\bm{R}^{-1}(\lambda)\right|_F \left|\bm{\Sigma}_c^{-1}(\lambda)\right| \left|\EE\frac{\bm{X}_c^\top
        \bm{X}_c}{n}\right| \left|\left[\bm{I}_{c_n} +
		\EE\frac{\bm{X}_c^\top
        \bm{X}_c}{n}
		\bar{\bm{R}}^{-1} (\lambda)\right]^{-1}
        \right|\\
        \le & C\lambda^{\frac{2d_2+2\psi+3}{2(2\gamma+d_2-\psi)}}.
   \end{align*}
   Then by balancing the two terms in \eqref{gau_third}, we have 
   $$III\le Cc_n^{\frac{1}{8}} \lambda^{\frac{d_2+\psi+3/2}{2(2\gamma + d_2-\psi)}}.$$

Finally for the error term I, we adopt the similar notation in \cite{Fang16}. Specifically, define
$$B^\epsilon = \{x\in \mathbb{R}^{c_n}: \text{dist}(x,B)\le \epsilon\},\quad 
B^{-\epsilon} = \{x\in B: \text{dist}(x,\mathbb{R}^{c_n}\backslash B)> \epsilon\}.
$$
Further denote $\bm{V}_n^c=\bm{\Sigma}_c^{-1}(\lambda)\bm{Z}_n^c$ and $\bar{\bm{V}}_n^c=\bar{\bm{\Sigma}}_c^{-1}(\lambda)\bm{Z}_n^c$. Elementary calculations show that
\begin{align*}
        I \le & \Pr(|\bm{V}_n^c-\bar{\bm{V}}_n^c|> \epsilon)
        +\sup_{B\in \mathcal{B}_n}
        \left\{\Pr(\bar{\bm{V}}_n^c \in B^\epsilon\backslash B), 
        \Pr(\bar{\bm{V}}_n^c \in B\backslash B^{-\epsilon})\right\}\\
        :=& V_1 + V_2.
    \end{align*}
    By the Markov's inequality, we have
    \begin{align}\label{error_v1}
    V_1 &\le \frac{1}{\epsilon} \EE\left|[\bm{\Sigma}_c^{-1}(\lambda)-\bar{\bm{\Sigma}}_c^{-1}(\lambda)]\bm{Z}_n^c\right| \notag\\
    &\le \frac{1}{\epsilon} \left|
    \bm{\Sigma}^{-1}_c(\lambda)-\bar{\bm{\Sigma}}_c^{-1}(\lambda)\right|\EE\left| \frac{1}{\sqrt{n}}\sum_{i=1}^n \bm{z}_{ci}\right| \notag\\
    &\le \frac{C}{\epsilon} c_n^{\frac{1}{2}}\lambda^{\frac{d_2+\psi+2}{2\gamma+d_2-\psi}}
    \end{align}
    On the other hand, notice that
    $$\Pr(\bar{\bm{V}}_n^c \in B^\epsilon\backslash B)=\Pr(\bar{\bm{V}}_n^c \in B^\epsilon)- 
    \Pr(\bar{\bm{V}}_n^c \in B).$$ 
    Then we have
    \begin{align*}
        & \Pr(\bar{\bm{V}}_n^c \in B^{\epsilon}\backslash B)-
        \Pr(\bar{\bm{W}}_n^c \in B^{\epsilon}\backslash B)\\
        \le & \left|\Pr(\bar{\bm{V}}_n^c \in B^{\epsilon})-
        \Pr(\bar{\bm{W}}_n^c \in B^{\epsilon})\right|+\left|
        \Pr(\bar{\bm{V}}_n^c \in B)-\Pr(\bar{\bm{W}}_n^c \in B)\right|\\
        \le & 2\mathcal{K}(\bar{\bm{V}}_n^c, \bar{\bm{W}}_n^c)=2\bar{\mathcal{K}}(\bm{Z}_n^c, \bm{U}_n^c).
    \end{align*}
    Similarly,
    $$\Pr(\bar{\bm{V}}_n^c \in B\backslash B^{-\epsilon})-
        \Pr(\bar{\bm{W}}_n^c \in B\backslash B^{-\epsilon})
        \le  2\mathcal{K}(\bar{\bm{V}}_n^c, \bar{\bm{W}}_n^c)=2\bar{\mathcal{K}}(\bm{Z}_n^c, \bm{U}_n^c).$$ Then by Lemma 4.2 of \cite{Fang16}, it yields that
        $$V_2 \le \sup_{B \in \mathcal{B}_n}\{\Pr(\bar{\bm{W}}_n^c \in B^{\epsilon}\backslash B),
        \Pr(\bar{\bm{W}}_n^c \in B\backslash B^{-\epsilon})\} + 2\bar{\mathcal{K}}(\bm{Z}_n^c, \bm{U}_n^c)\le 4c_n^{\frac{1}{4}}\epsilon + 2\bar{\mathcal{K}}(\bm{Z}_n^c,\bm{U}_n^c).$$
        As a result, we choose $\epsilon=\bigO(
        c_n^{\frac{1}{8}}\lambda^{\frac{d_2+\psi+2}{2(2\gamma+d_2-\psi)}})$ to balance terms $V_1$ and $V_2$, and finally obtain $I\le C\left(c_n^{\frac{3}{8}}
        \lambda^{\frac{d_2+\psi+2}{2(2\gamma+d_2-\psi)}}+c_n^{\frac{7}{4}}n^{-\frac{1}{2}+\frac{9}{2q}+
        \frac{2}{\tau-1}}\right)$.
        
        Putting the above error bounds for I, II and III together, we conclude that 
        $$\mathcal{K}(\bm{Z}_n^c, \bm{U}_n^c)\le C\left(
        c_n^{\frac{7}{4}}n^{-\frac{1}{2}+\frac{9}{2q}+
        \frac{2}{\tau-1}} + c_n^{\frac{3}{8}} \lambda^{\frac{d_2+\psi+2}{2(2\gamma + d_2-\psi)}}\right).$$

	\subsection{Proof of Main Results in Section 5.2 of the Article}\label{sec2}
    
	To prove Theorem 1 in Section 5.2 of the main article, we need two intermediate results on the theoretical bootstrap approximation and consistency of estimators, respectively.
	\bigskip
	
	\noindent
	$\bullet$ {\bf Theoretical bootstrap approximation}.
	\medskip
	
	\noindent
	First define the conditional variance of $\bm{U}_n^{boots}$ and the corresponding Kolmogorov distance as
	\begin{align*}
		&\widetilde{\bm{\Xi}}^c=\EE[\bm{U}_n^{boots}(\bm{U}_n^{boots})^\top\big| \bm{Z}_1^n]=\frac{1}{(n-m+1)m}\sum_{j=1}^{n-m+1} \left(\sum_{i=j}^{j+m-1}\bm{z}_{ci}\right)\left(\sum_{i=j}^{j+m-1}
		\bm{z}_{ci}^\top\right),\\
		&\mathcal{K}'(\bm{U}_n^{boots},\bm{Z}_n^c)
		=\sup \limits_{\bm{g}_n\in \mathcal{G},x\in\mathbb{R}}\Big|
		\Pr\Big(\sup_{t\in[0,1]}\big|\bm{Q}_n^{boots}(t,\lambda)\big|_{\bm{g}_n(t)}\le x\bigg|\bm{Z}_1^n\Big)-\Pr\Big(\sup_{t\in[0,1]}\big|
		\widetilde{\bm{Q}}_n^z
        (t,\lambda)\big|_{\bm{g}_n(t)}\le x\Big)\Big|,
	\end{align*}
    where $\bm{Q}_n^{boots}(t,\lambda) = \bm{C}_f(t)\widetilde{\bm{\Sigma}}_c^{-1}(\lambda)\bm{U}_n^{boots}$ and $\widetilde{\bm{Q}}_n^z(t,\lambda)=\bm{C}_f(t)\bm{\Sigma}_c^{-1}(\lambda)\bm{Z}_n^c$. Then we will obtain the following proposition, which establishes the rate of the bootstrap approximation to the weighted maximum deviation of $\bm{Z}_n^c$.
	
	\begin{proposition}\label{prop_boots}
	Assume that the smallest eigenvalue of $\bm{\Xi}^c$ is bounded below by some constant $b>0$ and $m=\bigO(n^{1/3})$. For some finite constant $C>0$, define $\mathcal{B}_n^c=\left\{\omega:\Delta_n^c(\omega):=
	\left|\widetilde{\bm{\Xi}}^c-\bm{\Xi}^c\right|_F\le Cc_nn^{-1/3}h_n\right\},$ where $\omega$ represents the element in the probability space, $h_n$ diverges to infinity at an arbitrarily slow rate, then $\Pr(\mathcal{B}_n^c)=1-o(1)$. Under Assumptions 1--8 of the main article, on the event $\mathcal{B}_n^c$, we have 
		\begin{align}\label{eq_compare}
			&\mathcal{K}'(\bm{U}_n^{boots},\bm{Z}_n^c)\notag \\
			\le &C\left(c_n^{7/4}n^{-\frac{1}{2}+\frac{9}{2q}+\frac{2}{\tau-1}}+
			c_n^{\frac{3}{8}} \lambda^{\frac{d_2+\psi+2}{2(2\gamma + d_2-\psi)}}
			+c_n^{5/8}n^{-1/6}h_n^{1/2}\right),
		\end{align}
		where $C>0$ is some finite constant. Further suppose
		\begin{enumerate}[(i)]%[label=(\roman*)]
			\item $c_n\gg \left(n/\log(n)\right)^{\frac{1}{2(d_1+d_2-\psi)+3}}$ and $\lambda\ll \left(\log(n)/n\right)^{\frac{2(\gamma+d_2+1)} {2(d_1+d_2-\psi)+3}}$,\label{c4} 
			\item $\widehat{\bm{g}}_n(t)\in \mathcal{G}$ almost surely, \label{c5}
			\item $c_n^{7/4}n^{-\frac{1}{2}+\frac{9}{2q}+\frac{2}
				{\tau-1}}+c_n^{\frac{3}{8}} \lambda^{\frac{d_2+\psi+2}{2(2\gamma + d_2-\psi)}}+c_n^{5/8}n^{-1/6}h_n^{1/2}\to 0$ as $n\to \infty$, \label{c6}
		\end{enumerate}
		then the JSCB based on the roughness penalization approach achieves	
		\begin{equation}\label{cover2}
			\begin{split}
				\lim_{n\to \infty}\lim_{B\to\infty}\Pr\Big(\beta_{j}(t)\in \left[\widetilde{\beta}_j(t)-\hat{q}_{n,1-\alpha}
				\widehat{g}_{nj}(t)/\sqrt{n},
				\widetilde{\beta}_j(t)+\hat{q}_{n,1-\alpha}\widehat{g}_{nj}(t)
				/\sqrt{n}\right]&\\
				~\text{for}~\forall t\in [0,1] ~\text{and}~j=1,...,p\Big) =1-&\alpha.
			\end{split}
		\end{equation}
	\end{proposition}
From the above \cref{prop_boots}, we conclude that with Condition \eqref{c6}, Gaussian approximation rate and the bootstrap approximation rate in \eqref{eq_compare} both converge to 0. In particular, if $\bm{X}_i(t)\in \mathcal{C}^1,~\bm{\beta}(t) \in \mathcal{C}^2$, $\gamma=2$ based on Fourier bases and $q$ and $\tau$ go to infinity, the above theorem shows that $\widetilde{\bm{\beta}}(t)$ is an under-smoothed estimator as long as $c_n\gg (n/\log(n))^{\frac{1}{9}}$ as well as $\lambda\ll \left(\log(n)/n\right)^{\frac{8}{9}}$. Further with the constraint $c_n\ll n^{\frac{1}{7}}$, the right hand side of \eqref{eq_compare} goes to 0.
\medskip

\begin{proof}
	Recall $\bm{\Xi}^c=\EE[\bm{U}_n^c(\bm{U}_n^c)^\top]$.
	By \cref{prop3} in \cref{sec_c5}, we have $\left\Vert\widetilde{\bm{\Xi}}^c_{jk}-\bm{\Xi}^c_{jk}\right\Vert_2
	=\bigO\left(\frac{1}{m}+\sqrt{\frac{m}{n}}\right)$. By choosing $m=\bigO(n^{1/3})$, $|\widetilde{\bm{\Xi}}^c-\bm{\Xi}^c|_F=\bigO(c_nn^{-1/3})=o(c_nn^{-1/3}h_n)$, then $\Pr(\mathcal{B}_n^c)=1-o(1)$. On the other hand, note that
	\begin{align*}
	&\mathcal{K}'(\bm{U}_n^{boots},\bm{Z}_n^c) \\
	= &\sup_{x\in \mathbb{R}}\sup_{\bm{g}_n\in\mathcal{G}}
	\bigg|\Pr\big(\sup_{t\in[0,1]}\big|\bm{C}_f(t)
	\widetilde{\bm{\Sigma}}_c^{-1}(\lambda)
	\bm{U}_n^{boots}\big|_{\bm{g}_n(t)}\le x\big| \bm{Z}_1^n\big)-
\Pr\big(\sup_{t\in[0,1]}\big|\bm{C}_f(t)\bm{\Sigma}_c^{-1}(\lambda)
	\bm{Z}_n^c\big|_{\bm{g}_n(t)}\le x\big)\bigg|  \\
	\le &\sup_{x\in \mathbb{R}}\sup_{\bm{g}_n\in\mathcal{G}}
	\bigg|\Pr\big(\sup_{t\in[0,1]}\big|\bm{C}_f(t)
	\widetilde{\bm{\Sigma}}_c^{-1}(\lambda)\bm{U}_n^{boots}\big|_{\bm{g}_n(t)}\le x
	\big| \bm{Z}_1^n \big)
    - \Pr\big(\sup_{t\in[0,1]}\big|\bm{C}_f(t)
	\bm{\Sigma}_c^{-1}(\lambda)\bm{U}_n^{boots}\big|_{\bm{g}_n(t)}\le x
	\big| \bm{Z}_1^n \big)\bigg|\\
    & {} + \sup_{x\in \mathbb{R}}\sup_{\bm{g}_n\in\mathcal{G}}
	\bigg|\Pr\big(\sup_{t\in[0,1]}\big|\bm{C}_f(t)
	\bm{\Sigma}_c^{-1}(\lambda)\bm{U}_n^{boots}\big|_{\bm{g}_n(t)}\le x
	\big| \bm{Z}_1^n \big)
    - \Pr\big(\sup_{t\in[0,1]}\big|\bm{C}_f(t)
	\bm{\Sigma}_c^{-1}(\lambda)\bm{U}_n^c\big|_{\bm{g}_n(t)}\le x
	\big)\bigg|\\
    & {} + \sup_{x\in \mathbb{R}}\sup_{\bm{g}_n\in\mathcal{G}}
	\bigg|\Pr\big(\sup_{t\in[0,1]}\big|\bm{C}_f(t)
	\bm{\Sigma}_c^{-1}(\lambda)\bm{U}_n^c\big|_{\bm{g}_n(t)}\le x \big)
    - \Pr\big(\sup_{t\in[0,1]}\big|\bm{C}_f(t)
	\bm{\Sigma}_c^{-1}(\lambda)\bm{Z}_n^c\big|_{\bm{g}_n(t)}\le x
	   \big)\bigg|\\
    =:&P_1 + P_2 + \mathcal{K}(\bm{Z}_n^c, \bm{U}_n^c)
    \end{align*}
    
By the Gaussian approximation result established \cref{thm2}, we have
$$\mathcal{K}(\bm{Z}_n^c, \bm{U}_n^c) \le C\left(
        c_n^{\frac{7}{4}}n^{-\frac{1}{2}+\frac{9}{2q}+
        \frac{2}{\tau-1}} + c_n^{\frac{3}{8}} \lambda^{\frac{d_2+\psi+2}{2(2\gamma + d_2-\psi)}}\right).$$

Similar to the discussion in the proof of \cref{thm2}, denote $\widetilde{\bm{V}}_n^m(\lambda)=
\widetilde{\bm{\Sigma}}_c^{-1}(\lambda)\bm{U}_n^{boots}$ and $\bm{V}_n^m(\lambda)=
\bm{\Sigma}_c^{-1}(\lambda)\bm{U}_n^{boots}$. Then following by Lemmas \ref{lemma3} and \ref{lemma1}, we have
\begin{align*}
	P_1\le & 4c_n^{1/4}\epsilon_1+\sup_{B\in\mathcal{B}_n}
	\left|\EE\left\{[h_{B,\epsilon_1}(\widetilde{\bm{V}}_n^{m}(\lambda))-h_{B,\epsilon_1}(\bm{V}_n^m(\lambda))]\mid \bm{Z}_1^n\right\} \right|\\
	\le & 4c_n^{1/4}\epsilon_1+\frac{2}{\epsilon_1}
	\EE\left\{ \left| \big[\widetilde{\bm{\Sigma}}_c^{-1}(\lambda)-\bm{\Sigma}_c^{-1}(\lambda)\big]\bm{U}_n^{boots}\right|  \mid \bm{Z}_1^n \right\} \\
	\le & 4c_n^{1/4}\epsilon_1 + \frac{C}{\epsilon_1} \left|
    \widetilde{\bm{\Sigma}}_c^{-1}(\lambda)-\bm{\Sigma}_c^{-1}(\lambda)\right| \sqrt{\frac{1}{n-m+1}\sum_{j=1}^{n-m+1}\left|\frac{1}{\sqrt{m}}\sum_{i=j}^{j+m-1} \bm{z}_{ci}\right|_F^2}\\
        \le & 4c_n^{1/4}\epsilon_1 + 
        \frac{C}{\epsilon_1} \frac{c_n^{3/2}\log(n)}{\sqrt{n}}
	\end{align*}
    We choose $\epsilon_1=\bigO(c^{5/8}n^{-1/4}\log^{1/2}(n))$ to balance the above two error bounds and conclude $P_1\le Cc^{7/8}n^{-1/4}\log^{1/2}(n)$.

	Next, we will use the result in Lemma \ref{lemma3} again to derive the error bound for $P_2$. For any $\epsilon_1>0$, we have
	\begin{align*}
	P_1&\le 4c_n^{1/4}\epsilon_1+\sup_{A\in\mathcal{A}_n}
	|\EE[h_{A,\epsilon_1}(\bm{U}_n^{boots})-h_{A,\epsilon_1}(\bm{U}_n^c)]|\\
	&\le 4c_n^{1/4}\epsilon_1+\frac{2}{\epsilon_1}
	\EE|\bm{U}_n^{boots}-\bm{U}_n^c|\\
	&\le 4c_n^{1/4}\epsilon_1+\frac{C}{\epsilon_1}
	\left|(\widetilde{\bm{\Xi}}^c)^{1/2}-(\bm{\Xi}^c)^{1/2}\right|_F\\
	&\le 4c_n^{1/4}\epsilon_1+\frac{C}{\epsilon_1}
	\left|\widetilde{\bm{\Xi}}^c-\bm{\Xi}^c\right|_F\\
	&\le 4c_n^{1/4}\epsilon_1+\frac{Cc_n}{\epsilon_1}n^{-1/3}h_n,
	\end{align*}
	where the third inequality is due to the fact $\EE|\bm{U}_n^{boots}-\bm{U}_n^c|\le \sqrt{{\rm Tr} [(\widetilde{\bm{\Xi}}^c)^{1/2}-(\bm{\Xi}^c)^{1/2})]^2}=
	|(\widetilde{\bm{\Xi}}^c)^{1/2}-(\bm{\Xi}^c)^{1/2}|_F$. The fourth inequality uses the inequality $|\bm{R}_1^{1/2}-\bm{R}_2^{1/2}|_F\le C|\bm{R}_1-\bm{R}_2|_F$ for any positive definite matrices $\bm{R}_1$ and $\bm{R}_2$ (see Lemma 2.2 in \cite{Schmitt92} for more details).
	
	By choosing $\epsilon_1=\bigO(c_n^{3/8}n^{-1/6}\sqrt{h_n})$, we are able to derive that
	\begin{equation}\label{distance}
	P_1\le Cc_n^{5/8}n^{-1/6}h_n^{1/2}.
	\end{equation}
	In summary, we combine all three error bounds and obtain the approximation rate in \eqref{eq_compare}.
	
	Now, we will construct JSCB in \eqref{cover2} by an under-smoothed estimator $\widetilde{\beta}_j(t)$, which means the uniform convergence rate for the standard deviation term dominates those for bias terms. 
	%Let $r_k$ and $\rho_k$ be the $k$th smallest eigenvalue of $\bm{R}_j(\lambda)$ and $\bm{\Sigma}_c^{-1}(\lambda)$, respectively. Here we only consider the simple case that $\bm{R}(\lambda)$ is a diagonal matrix and the proof for non-diagonal case is straightforward. By Assumption 4, notice that for some positive constant $C$, $$\rho_k\le \min(C,\lambda^{-1}k^{-2(\gamma+d_2+1)}).$$
	
	Next, we can calculate the $\mathcal{L}^\infty$ rate of the bias term,
	\begin{align}\label{bias}
	&\left|\EE[\widetilde{\beta}_j(t)]-\beta_{j}(t)\right|_\infty \notag \\
	\le &Cc_n^{-d_1}+\sup_{t\in[0,1]}\left|\EE\left\{\bm{E}_j^\top\bm{C}_f(t)
	\widetilde{\bm{\Sigma}}_c^{-1}(\lambda)	\frac{\bm{X}_c^\top\widetilde{\bm{\epsilon}}}{n}\right\}\right|+\sup_{t\in[0,1]}
	\left|\bm{E}_j^\top\bm{C}_f(t)
	\widetilde{\bm{\Sigma}}_c^{-1}(\lambda)
	\bm{R}(\lambda)\bm{\theta}_c\right| \notag\\
	\le& C\left(c_n^{-d_1}+c_n^{\psi-d_1+1}/\sqrt{n}\right)+\sup_{t\in[0,1]}\left|\bm{E}_j^\top\bm{C}_f(t)
	\bm{\Sigma}_c^{-1}(\lambda)\bm{R}(\lambda)\bm{\theta}_c\right| 
	\end{align}
	where the second inequality uses Cauchy-Schwarz inequality. 
	Now, we turn to deal with the last term in \eqref{bias}. Notice that 
	$\widetilde{\bm{\Sigma}}_c(\lambda)=
    \frac{\bm{X}_c^\top\bm{X}_c}{n}+\bm{R}(\lambda)$, we decompose it as $\widetilde{\bm{\Sigma}}_c(\lambda)=\bm{D}+\bm{P}$ where $\bm{D}$ is the diagonal matrix with elements $\{c_0+\lambda k^{2(\gamma+d_2+1)}\}_{k=1}^{c_n}$ with $0<c_0<\lambda_{\min}( \bm{X}_c^\top\bm{X}_c/n)$ and $\bm{P}:=\bm{X}_c^\top\bm{X}_c/n-c_0\bm{I}_{c_n}$ can be viewed as the perturbation matrix with the smallest eigenvalue bounded away from zero. We will employ the special case of the Woodbury matrix identity $(\bm{D}+\bm{P})^{-1}=
	\bm{D}^{-1}-(\bm{I}_{c_n}+\bm{P}^{-1}\bm{D})^{-1}\bm{D}^{-1}$. In particular, since $\bm{P}^{-1}\bm{D}$ is semi positive definite, we find that $|(\bm{I}_{c_n}+\bm{P}^{-1}\bm{D})^{-1}|\le 1$. Consequently, we can obtain
	\begin{align*}
	&\sup_{t\in[0,1]}\left|\bm{E}_j^\top\bm{C}_f(t)
	(\bm{D}+\bm{P})^{-1}\bm{R}(\lambda)\bm{\theta}_c\right|\\
	\le &\sup_{t\in[0,1]}\left|\bm{E}_j^\top\bm{C}_f(t)\bm{D}^{-1}\bm{R}(\lambda)
	\bm{\theta}_c\right|+\sup_{t\in[0,1]}\left|\bm{E}_j^\top\bm{C}_f(t)(\bm{I}_{c_n}+\bm{P}^{-1}\bm{D})^{-1}\bm{D}^{-1}\bm{R}(\lambda)\bm{\theta}_c\right|\\
	\le &\sum_{k=1}^{\lfloor\lambda^{-\frac{1}{2(\gamma+d_2+1)}}\rfloor}\lambda
	k^{\psi-(d_1+1)+2(\gamma+d_2+1)}+
	\sum_{k=\lfloor\lambda^{-\frac{1}{2(\gamma+d_2+1)}}\rfloor}^{c_n}
	k^{\psi-(d_1+1)}\\
	\le &C\lambda^{\frac{d_1-\psi}{2(\gamma+d_2+1)}}.
	\end{align*}
    As a result, the $\mathcal{L}^\infty$ rate of the bias term turns to be 
	$$\left|\EE[\widetilde{\beta}_j(t)]-\beta_{j}(t)\right|_\infty\le 
	C\left(c_n^{-d_1}+c_n^{\psi-d_1+1}/\sqrt{n}+\lambda^{\frac{d_1-\psi}{2(\gamma+d_2+1)}}\right).$$
	On the other hand, note that $$\sup_{t\in[0,1]}|\widetilde{\beta}_j(t)-\EE\widetilde{\beta}_j(t)|\ge
	\sup_{t\in[0,1]}\left|\bm{E}_j^\top\bm{C}_f(t)
	\widetilde{\bm{\Sigma}}_c^{-1}(\lambda)\bm{Z}_n^c/\sqrt{n}\right|.$$ With the Gaussian approximation result constructed in \cref{gau}, it suffices to derive the lower bound of $\sup_{t\in[0,1]}
	\left|\bm{E}_j^\top\bm{C}_f(t)
	\widetilde{\bm{\Sigma}}_c^{-1}(\lambda)\bm{U}_n^c/\sqrt{n}\right|$. Denote the discrete time points in $[0,1]$ as $\{t_{i,n}=i/r_n\}_{i=0}^{r_n}$ where $r_n=\bigO(n^\nu), \nu$ is a positive integer that diverges to infinity. By the Orlicz norm with $\Psi(x)=e^{x^2}-1$, we can obtain for any $j=1,...,p$,
	\begin{align*}
	& \left\Vert\sup_{t\in [0,1]}\left|\bm{E}_j^\top\bm{C}_f(t)
	\widetilde{\bm{\Sigma}}_c^{-1}(\lambda)\bm{U}_n^c/\sqrt{n}\right|\right\Vert_{\Psi}\\
	\ge &\left\Vert\max_{0\le i\le r_n}\left|\bm{E}_j^\top\bm{C}_f(t_{i,n})
	\widetilde{\bm{\Sigma}}_c^{-1}(\lambda)\bm{U}_n^c/\sqrt{n}\right|\right\Vert_{\Psi}\\
	\ge &C\sqrt{\frac{\log n}{n}} \max_{0\le i\le r_n}\left\Vert
	\bm{E}_j^\top\bm{C}_f(t_{i,n})
    \widetilde{\bm{\Sigma}}_c^{-1}(\lambda)\bm{U}_n^c
	\right\Vert_\Psi \\
    \ge &C\sqrt{\frac{\log n}{n}} \max_{0\le i\le r_n}\left\{
	{\rm Tr}\left(\bm{E}_j^\top\bm{C}_f(t_{i,n})
    \widetilde{\bm{\Sigma}}_c^{-1}(\lambda)
    \bm{\Xi}^c\widetilde{\bm{\Sigma}}_c^{-1}(\lambda)
	\bm{C}_f^\top(t_{i,n})\bm{E}_j\right)\right\}^{1/2}\\
		\ge &C\sqrt{\frac{\log n}{n}} \left\{{\rm Tr}
		\left(\widetilde{\bm{\Sigma}}_c^{-2}(\lambda)\bm{D}_f\right)\right\}^{1/2}\\
		\ge &C\sqrt{\frac{\log n}{n}}\sqrt{\sum_{k=1}^{c_n}\rho_k(
			\widetilde{\bm{\Sigma}}_c^{-2}(\lambda))\rho_{c_n-k+1}(\bm{D}_f)}\\
		\ge &C\sqrt{\frac{\log n}{n}}
		\sqrt{\sum_{k=1}^{\lfloor\lambda^{-\frac{1}{2(\gamma+d_2+1)}}\rfloor}k^{2(d_2+1)}+\sum_{k=
					\lfloor\lambda^{-\frac{1}{2(\gamma+d_2+1)}}\rfloor}^{c_n}
				\lambda^{-2}
				k^{-4(\gamma+d_2+1)+2(d_2+1)}}\\
		\ge &C\lambda^{-\frac{2d_2+3}{4(\gamma+d_2+1)}}\sqrt{\frac{\log n}{n}},
	\end{align*}
	where $\lambda_k(\cdot)$ denotes the $k$th largest eigenvalue of the matrix and $\bm{D}_f$ is a block diagonal matrix with the $j$th diagonal block ${\rm diag}(1/f_{j1}^2,...,1/f_{jc_{j,n}}^2)$ for $j=1,\ldots,p$. Now we comment on the above deductions. The first inequality uses chaining technique and the second inequality follows by the fact $\Vert \max_{1\le i\le r_n} X_i\Vert_\Psi\ge C_\Psi \Psi^{-1}(r_n)\max_i\Vert X_i\Vert_\Psi$ for Gaussian random variables $\{X_i\}_{i=1}^{r_n}$. Due to the assumption $\lambda_{\min}(\bm{\Xi}^c)\ge b>0$ and the fact ${\rm Tr}(\bm{AB})={\rm Tr}(\bm{BA})$ for any matrices $\bm{A},\bm{B}$, the third inequality holds. The fourth inequality follows by $\sup_{t\in[0,1]}\alpha_k^2(t)\ge \sum_{i=1}^{r_n} \alpha_k^2(t_{i,n})/r_n$, the statement ${\rm Tr}(\bm{AB})\ge \sum_{k=1}^{c_n}\lambda_k(\bm{A})\lambda_{c_n-k+1}(\bm{B})$ is used for the fifth inequality. Finally, the sixth inequality follows by the fact $\rho_k(\widetilde{\bm{\Sigma}}_c(\lambda))
	\le \max\{C,\lambda k^{2(\gamma+d_2+1)}\}$ and by elementary calculations, we obtain the last inequality.

In summary, we have $\sup_{t\in[0,1]}\left|\bm{E}_j^\top\bm{C}_f(t)
\widetilde{\bm{\Sigma}}_c^{-1}(\lambda)\bm{U}_n^c/\sqrt{n}\right|\ge C\lambda^{-\frac{2d_2+3}{4(\gamma+d_2+1)}}\sqrt{\frac{\log n}{n}}$. Letting the above convergence rate of the variance larger than that of each bias term, we need to satisfy the Condition \eqref{c4}~$c_n\gg\left(n/\log(n)\right)^{\frac{1}{2(d_1+d_2-\psi)+3}}$ and 
$\lambda\ll \left(\log(n)/n\right)^{\frac{2(\gamma+d_2+1)}{2(d_1+d_2-\psi)+3}}$. Combining all these derivations, we complete the proof. 
	\end{proof}
	\medskip
	
	\noindent
	$\bullet$ {\bf Consistency properties of estimators}.
	\medskip
	
	\noindent
	Next, we will show additional results on the consistency of the estimated quantities. Here, consider
	\begin{align*}
		\widehat{\epsilon}_i&=Y_i-\sum_{j=1}^p\sum_{k=1}^{c_{j,n}}
		\widetilde{\theta}_{jk}\widehat{x}_{ij,k},\\
		\widehat{f}_{jk}&=\frac{1}{n}\sum_{i=1}^n \left(\widetilde{x}_{ij,k}-\frac{1}{n}\sum_{i=1}^n \widetilde{x}_{ij,k}\right)^2,
	\end{align*}
	where $\widehat{x}_{ij,k}=\widetilde{x}_{ij,k}/\widehat{f}_{jk}$.
	
	\begin{lemma}\label{consis}
    
		Under Assumptions 1--6 and 8 in the paper, 
		we have
		\begin{align*}
			&\widehat{f}_{jk}^2=f_{jk}^2\left(1+\bigO_\Pr\left(\sqrt{\frac{\log(n)}{n}}\right)\right)
			~\text{uniformly for~} k=1,\cdots,c_{j,n},~j=1,\cdots,p,\\
			&\left\Vert \max_{1\le i\le n}\left|\widehat{\epsilon}_i-\epsilon_i\right|\right\Vert_{q/2}
			=\bigO\left(c_nn^{-1/2+2/q}\right) ~\text{for~} q>4.
		\end{align*}
	\end{lemma}
	
	\begin{proof}
		Without loss of generality, we assume $\EE(x_{ij,k})=0$ and observe that $f_{jk}^2=\EE(\widetilde{x}_{ij,k}^2)$. Further note that
		$$\widehat{f}_{jk}^2=\frac{1}{n}\sum_{i=1}^n \left(\widetilde{x}_{ij,k}-\frac{1}{n}\sum_{i=1}^n\widetilde{x}_{ij,k}\right)^2
		=\frac{f_{jk}^2}{n}\sum_{i=1}^n \left(x_{ij,k}-\frac{1}{n}\sum_{i=1}^nx_{ij,k}\right)^2.$$
		By elementary calculations, we will obtain
		\begin{equation*}
			\widehat{f}_{jk}^2=f_{jk}^2\left(1+\frac{1}{n}\sum_{i=1}^n (x_{ij,k}^2-1)-
			\left(\frac{1}{n}\sum_{i=1}^n x_{ij,k}\right)^2\right).
		\end{equation*}
		Let $y_{ij,k}=x_{ij,k}^2-1$, denote by $\delta_y(l,\cdot)$ the physical dependence measure of $y_{ij,k}$, then we have
		$$\delta_y(l,q/2)=\Vert y_{ij,k}-y_{ij,k}^\ast \Vert_{q/2} \le \Vert x_{ij,k}+x_{ij,k}^\ast \Vert_q
		\Vert x_{ij,k}-x_{ij,k}^\ast \Vert_q \le C(l+1)^{-\tau},$$ where $y_{ij,k}^\ast$ is the i.i.d. copy of $y_{ij,k}$. Using the Gaussian approximation result for the above partial sum process, we obtain that
		\begin{align*}
			&\max_{1\le j \le p}\max_{1\le k\le c_{j,n}}\left|\frac{1}{n}\sum_{i=1}^n (x_{ij,k}^2-1)\right|=\bigO_\Pr\left(\sqrt{\frac{\log(n)}{n}}\right),\\
			&\max_{1\le j \le p}\max_{1\le k\le c_{j,n}}\left|\left(\frac{1}{n}\sum_{i=1}^n x_{ij,k}\right)^2\right|=\bigO_\Pr\left(\frac{\log(n)}{n}\right).
		\end{align*}
		Thus, $\widehat{f}_{jk}^2=f_{jk}^2\left(1+\bigO_\Pr\left(\sqrt{\frac{\log(n)}{n}}\right)\right)$ uniformly in $j$ and $k$.
		
		On the other hand, consider the vectorized representation of the residuals $\widehat{\bm{\epsilon}}=\bm{Y}-\widehat{\bm{X}}_c\widehat{\bm{\theta}}_c$ where $\widehat{\bm{X}}_c, \widehat{\bm{\theta}}_c$ are constructed in the same manner as $\bm{X}_c$ and $\bm{\theta}_c$ with $f_{jk}$ replaced by its estimate $\widehat{f}_{jk}$. Let $\bm{E}_i$ be an $n$-dimensional vector with $i$th element being 1 and others being 0, then we have for $q>4$,
		\begin{align*}
			&\left\Vert\widehat{\epsilon}_i-\epsilon_i\right\Vert_{q/2}\\
			\le& Cc_n^{-(d_1+d_2+1)}+ \left\Vert \bm{E}_i^\top(\bm{X}_c-
            \widehat{\bm{X}}_c)\bm{\theta}_c+\bm{E}_i^\top
			\widehat{\bm{X}}_c\widehat{\bm{\Sigma}}_c^{-1}(\lambda)
			\widehat{\bm{R}}(\lambda)\bm{\theta}_c\right.\\
			&{}\qquad \qquad \qquad \qquad \left.+\bm{E}_i^\top\widehat{\bm{X}}_c
			\widehat{\bm{\Sigma}}_c^{-1}(\lambda)\widehat{\bm{X}}_c^\top
			(\bm{X}_c-\widehat{\bm{X}}_c)\bm{\theta}_c/n-
			\bm{E}_j^\top\widehat{\bm{X}}_c
			\widehat{\bm{\Sigma}}_c^{-1}(\lambda)\widehat{\bm{X}}_c^\top
			\bm{\epsilon}/n\right\Vert_{q/2}\\
			=& \bigO\left(c_n^{-(d_1+d_2+1)}+
            \sqrt{c_n\log n/n}+\sqrt{c_n}
			\lambda^{\frac{d_1+d_2+1}{2(\gamma+d_2+1)}}+c_n/\sqrt{n}
			\right)=\bigO\left(\frac{c_n}{\sqrt{n}}\right),
		\end{align*}
		where the second inequality follows by H{\"o}lder's inequality. Using the $L_q$ maximal inequality, it yields that $$\left\Vert\max_{1\le i\le n}|\widehat{\epsilon}_i-\epsilon_i|\right\Vert_{q/2}=
  \bigO(c_nn^{-1/2+2/q}).$$
	\end{proof}
	
	\noindent
	\textbf{Proof of Theorem 1 of the paper}. The proof of $\Pr(\mathcal{B}_n^\epsilon)=1-o(1)$ can be found in the proof of \cref{prop_boots}. 
    
	Based on the uniform consistency of $\widehat{\epsilon}_i$ and $\widehat{f}_{jk}$ from \cref{consis}, one can calculate
	$$\left|\widehat{\bm{\Xi}}^c-\widetilde{\bm{\Xi}}^c
	\right|_F=\bigO\left(c_n^2n^{-1/2+2/q}\right).$$ 
    Note that 
    \begin{align*}
		&\widehat{\mathcal{K}}(\widehat{\bm{U}}_n^{boots},\bm{Z}_n^c)\\
		=&\sup \limits_{\bm{g}_n\in \mathcal{G},x\in\mathbb{R}}\Big|
		\Pr\Big(\sup_{t\in[0,1]}\big|\widehat{\bm{Q}}
        _n^{boots}(t,\lambda)\big|_{\bm{g}_n(t)}\le x\bigg|\bm{Z}_1^n\Big)-\Pr\Big(\sup_{t\in[0,1]}\big|
		\bm{Q}_n^z
        (t,\lambda)\big|_{\bm{g}_n(t)}\le x\Big)\Big|\\
        \le &\sup \limits_{\bm{g}_n\in \mathcal{G},x\in\mathbb{R}}\Big|
		\Pr\Big(\sup_{t\in[0,1]}\big|\widehat{\bm{Q}}
        _n^{boots}(t,\lambda)\big|_{\bm{g}_n(t)}\le x\bigg|\bm{Z}_1^n\Big)-\Pr\Big(\sup_{t\in[0,1]}\big|\bm{Q}
        _n^{boots}(t,\lambda)\big|_{\bm{g}_n(t)}\le x\bigg|\bm{Z}_1^n\Big)\Big|\\
        & + \sup \limits_{\bm{g}_n\in \mathcal{G},x\in\mathbb{R}}\Big|
		\Pr\Big(\sup_{t\in[0,1]}\big|\bm{Q}
        _n^{boots}(t,\lambda)\big|_{\bm{g}_n(t)}\le x\bigg|\bm{Z}_1^n\Big)
        -\Pr\Big(\sup_{t\in[0,1]}\big|
		\widetilde{\bm{Q}}_n^z
        (t,\lambda)\big|_{\bm{g}_n(t)}\le x\Big)\Big|\\
        &+\sup \limits_{\bm{g}_n\in \mathcal{G},x\in\mathbb{R}}\Big|
        \Pr\Big(\sup_{t\in[0,1]}\big|
		\widetilde{\bm{Q}}_n^z
        (t,\lambda)\big|_{\bm{g}_n(t)}\le x\Big)\Big|
        -\Pr\Big(\sup_{t\in[0,1]}\big|
		\bm{Q}_n^z
        (t,\lambda)\big|_{\bm{g}_n(t)}\le x\Big)\Big|\\
        =:& P_3+ \mathcal{K}'(\bm{U}_n^{boots},
        \bm{Z}_n^c)
        + P_4.
        \end{align*}

        Next, we start with the error bound of $P_3$. By elementary calculations, we have
        {\footnotesize
        \begin{align*}
            P_3  \le& \sup \limits_{\bm{g}_n\in \mathcal{G},x\in\mathbb{R}}\Big|
		\Pr\Big(\sup_{t\in[0,1]}\big|\bm{C}_f(t)
        \widehat{\bm{\Sigma}}_c^{-1}(\lambda)\widehat{\bm{U}}_n^{boots}\big|
        _{\bm{g}_n(t)}\le x + \delta_1 \bigg|\bm{Z}_1^n\Big)
        -\Pr\Big(\sup_{t\in[0,1]}\big|\bm{C}_f(t)
        \widetilde{\bm{\Sigma}}_c^{-1}(\lambda)\bm{U}_n^{boots}\big|
		_{\bm{g}_n(t)}\le x\bigg|\bm{Z}_1^n\Big)\Big|\\
        & {} + \sup \limits_{\bm{g}_n\in \mathcal{G},x\in\mathbb{R}}\Big|
		\Pr\Big(\sup_{t\in[0,1]}\big|[\widehat{\bm{C}}_f(t)-
        \bm{C}_f(t)]\widehat{\bm{\Sigma}}_c^{-1}(\lambda)\widehat{\bm{U}}_n^{boots}\big|
        _{\bm{g}_n(t)}\ge \delta_1 \bigg|\bm{Z}_1^n\Big)\\
        \le & \sup \limits_{\bm{g}_n\in \mathcal{G},x\in\mathbb{R}}\Big|
		\Pr\Big(\sup_{t\in[0,1]}\big|\bm{C}_f(t)
        \widehat{\bm{\Sigma}}_c^{-1}(\lambda)\widehat{\bm{U}}_n^{boots}\big|
        _{\bm{g}_n(t)}\le x + \delta_1 \bigg|\bm{Z}_1^n\Big)
        -\Pr\Big(\sup_{t\in[0,1]}\big|\bm{C}_f(t)
        \widetilde{\bm{\Sigma}}_c^{-1}(\lambda)\bm{U}_n^{boots}\big|
		_{\bm{g}_n(t)}\le x + \delta_1 \bigg|\bm{Z}_1^n\Big)\Big|\\
        & + \sup \limits_{\bm{g}_n\in \mathcal{G},x\in\mathbb{R}}\Big|
		\Pr\Big(\sup_{t\in[0,1]}\big|\bm{C}_f(t)
        \widetilde{\bm{\Sigma}}_c^{-1}(\lambda)\bm{U}_n^{boots}\big|
		_{\bm{g}_n(t)}\le x + \delta_1 \bigg|\bm{Z}_1^n\Big)
        -\Pr\Big(\sup_{t\in[0,1]}\big|\bm{C}_f(t)
        \widetilde{\bm{\Sigma}}_c^{-1}(\lambda)\bm{U}_n^{boots}\big|
		_{\bm{g}_n(t)}\le x \bigg|\bm{Z}_1^n\Big)\Big|\\
        & + \sup \limits_{\bm{g}_n\in \mathcal{G},x\in\mathbb{R}}\Big|
		\Pr\Big(\sup_{t\in[0,1]}\big|[\widehat{\bm{C}}_f(t)-
        \bm{C}_f(t)]\widehat{\bm{\Sigma}}_c^{-1}(\lambda)\widehat{\bm{U}}_n^{boots}\big|
        _{\bm{g}_n(t)}\ge \delta_1 \bigg|\bm{Z}_1^n\Big)\\
        =:& I+ II + III,
        \end{align*}}
        where $\delta_1=\delta_{1,n}$ is a positive sequence that converges to zero as $n$ diverges to infinity.
        
        To derive the error bound of $I$, we follow the comparison results discussed in the proof of \cref{prop_boots}. Specifically, denote $$\widehat{\bm{S}}_n(\lambda) = \widehat{\bm{\Sigma}}_c^{-1}(\lambda)\widehat{\bm{U}}_n^{boots}, \quad \bm{S}_n(\lambda) = \widetilde{\bm{\Sigma}}_c^{-1}(\lambda)\bm{U}_n^{boots}, \quad \bm{S}_n^\ast(\lambda) = \widehat{\bm{\Sigma}}_c^{-1}(\lambda)\bm{U}_n^{boots}.$$
        Then by \cref{lemma3} again, we have
        \begin{align*}
            I \le & 4c_n^{1/4}\epsilon_1 + \frac{C}{\epsilon_1} \EE\left[ \left|\widehat{\bm{S}}_n(\lambda) - \bm{S}_n(\lambda)\right| \mid \bm{Z}_1^n\right]\\
            \le & 4c_n^{1/4}\epsilon_1 + \frac{C}{\epsilon_1} \left(\EE\left[ \left|\widehat{\bm{S}}_n(\lambda) - \bm{S}_n^\ast(\lambda)\right| \mid \bm{Z}_1^n\right]
            +\EE\left[ \left|\bm{S}_n^\ast(\lambda) - \bm{S}_n(\lambda)\right| \mid \bm{Z}_1^n\right]
            \right).
            \end{align*}
            By optimizing $\epsilon_1$, we obtain
            $$I\le Cc_n^{1/8}\left(\EE\left[ \left|\widehat{\bm{S}}_n(\lambda) - \bm{S}_n^\ast(\lambda)\right| \mid \bm{Z}_1^n\right]
            +\EE\left[ \left|\bm{S}_n^\ast(\lambda) - \bm{S}_n(\lambda)\right| \mid \bm{Z}_1^n\right]
            \right)^{1/2}.$$

            By \cref{consis} and Jensen's inequality, we have
            \begin{align*}
                &\EE\left[ \left|\widehat{\bm{S}}_n(\lambda) - \bm{S}_n^\ast(\lambda)\right| \mid \bm{Z}_1^n\right] \\
                \le & \sqrt{\EE\left[ \left(\widehat{\bm{S}}_n(\lambda) - \bm{S}_n^\ast(\lambda)\right)^\top
                \left(\widehat{\bm{S}}_n(\lambda) - \bm{S}_n^\ast(\lambda)\right) \mid \bm{Z}_1^n\right]}\\
                = &\sqrt{\EE\left[ {\rm Tr}\left\{\left(\widehat{\bm{U}}_n^{boots} - \bm{U}_n^{boots}\right)^\top\widehat{\bm{\Sigma}}_c^{-2}(\lambda)
                \left(\widehat{\bm{U}}_n^{boots} - \bm{U}_n^{boots}\right)\right\}\mid \bm{Z}_1^n\right]}\\
                \le & \left|\widehat{\bm{\Sigma}}_c^{-1}(\lambda)\right| \sqrt{\frac{1}{n-m+1}\sum_{i=1}^{n-m+1}\left|\frac{1}{\sqrt{m}}\sum_{j=i}^{i+m-1}(\widehat{\bm{x}}_{cj}\widehat{\epsilon}_j-\bm{x}_{cj}\epsilon_j)\right|_F^2}\\
                \le & C\left(\sqrt{\frac{c_n\log n}{n}} + c_n^{3/2} n^{-1/2+2/q}\right)=\bigO\left(c_n^{3/2} n^{-1/2+2/q}\right).
            \end{align*}

            Similarly,
            \begin{align*}
                & \EE\left[ \left|\bm{S}_n^\ast(\lambda) - \bm{S}_n(\lambda)\right| \mid \bm{Z}_1^n\right] \\
                \le & \sqrt{\EE\left[ {\rm Tr}\left\{(\bm{U}_n^{boots})^\top
                \left[\widetilde{\bm{\Sigma}}_c^{-1}(\lambda)-\widehat{\bm{\Sigma}}_c^{-1}(\lambda)\right]^2
                \bm{U}_n^{boots}\right\}\mid \bm{Z}_1^n\right]}\\
                 \le & \left|\widetilde{\bm{\Sigma}}_c^{-1}(\lambda)-\widehat{\bm{\Sigma}}_c^{-1}(\lambda)\right|_F \left| \bm{U}_n^{boots}
                (\bm{U}_n^{boots})^\top\right|\\
                \le & Cc_n^2n^{-1/2+2/q}.
            \end{align*}
            Therefore, we conclude 
            $$I\le Cc_n^{9/8}n^{-1/4+1/q}.$$
    For $II$, we consider the discrete time points $\{t_{i,n}\}_{i=0}^{r_n}$ again and employ the anti-concentration inequality \citep[Corollary 1]{chernozhukov2015comparison} to control the error bound. Note that conditional on $\bm{Z}_1^n$,
    \begin{align*}
    &\left\Vert \max_{0\le i\le r_n-1}\sup_{s\in [t_{i,n},t_{i+1,n}]} \left|[\bm{C}_f(s)-\bm{C}_f(t_{i,n})]\widetilde{\bm{\Sigma}}_c^{-1}(\lambda)
    \bm{U}_n^{boots}\right|_{\bm{g}(s)}\right\Vert_\Psi \\
    \le & \sqrt{\log(r_n)}\left\Vert \sup_{s\in [t_{i,n},t_{i+1,n}]}\left|[\bm{C}_f(s)-\bm{C}_f(t_{i,n})]\widetilde{\bm{\Sigma}}_c^{-1}(\lambda)
    \bm{U}_n^{boots}\right|_{\bm{g}(s)}\right\Vert_\Psi\\
    \le & C\sqrt{\log n}
    \int_{t_{i,n}}^{t_{i+1,n}}
    \left\Vert\bm{E}_j^\top
	\bm{C}_f'(s)\widetilde{\bm{\Sigma}}_c^{-1}(\lambda)\bm{U}_n^{boots}
    \right\Vert_\Psi \dee s\\
    \le & C\sqrt{\log n} \sup_{s\in[0,1]}\left|\bm{E}_j^\top\bm{C}_f'(s)\right| \left|\widetilde{\bm{\Sigma}}_c^{-1}(\lambda)\right| \Vert \bm{U}_n^{boots}\Vert_2\big/ r_n\\
    \le & Cc_n^{\phi+d_2+2}\sqrt{\log n}\big/r_n.
    \end{align*}

     Due to the relation $\Pr(A)\le \Pr(A\bigcap B)+\Pr(B^{c})$ and given $\delta_2=\delta_{2,n}$ a positive sequence that converges to zero as $n$ diverges to infinity, we have
     \begin{align*}
         II  \le & \sup \limits_{\bm{g}_n\in \mathcal{G},x\in\mathbb{R}}\Big|
		\Pr\Big(\max_{0\le i\le r_n}\big|\bm{C}_f(t_{i,n})
        \widetilde{\bm{\Sigma}}_c^{-1}(\lambda)\bm{U}_n^{boots}\big|
		_{\bm{g}_n(t_{i,n})}\le x + \delta_1 + \delta_2 \mid \bm{Z}_1^n\Big)\\
        & -\Pr\Big(\max_{0\le i\le r_n}\big|\bm{C}_f(t_{i,n})
        \widetilde{\bm{\Sigma}}_c^{-1}(\lambda)\bm{U}_n^{boots}\big|
		_{\bm{g}_n(t_{i,n})}\le x-\delta_2 \mid \bm{Z}_1^n\Big)\Big|\\
        & + 2\sup \limits_{\bm{g}_n\in \mathcal{G},x\in\mathbb{R}}\Big|
		\Pr\Big(\max_{0\le i\le r_n-1}\sup_{s\in[t_{i,n},t_{i+1,n}]}
        \big|[\bm{C}_f(s)-\bm{C}_f(t_{i,n})]
        \widetilde{\bm{\Sigma}}_c^{-1}(\lambda)\bm{U}_n^{boots}\big|
		_{\bm{g}_n(s)}\ge \delta_2 \mid \bm{Z}_1^n\Big)\Big|\\
        \le & C\left\{(\delta_1+\delta_2)\sqrt{\log(r_n)} +
        \frac{c_n^{\phi+d_2+2}\sqrt{\log n}
        \big/r_n}{\delta_2}\right\}.
     \end{align*}
     By optimizing $\delta_2$, we obtain that
     $$II\le C\left(\delta_1\sqrt{\log n} + 
     \sqrt{c_n^{\phi+d_2+2}\log n/r_n}\right).$$

     Next, we turn to $III$. Denote $\bm{C}_f^\ast(t) = \widehat{\bm{C}}_f(t) - \bm{C}_f(t)$. Notice that conditional on $\bm{Z}_1^n$,
     {\footnotesize
     \begin{align*}
         & \left\Vert \sup_{t\in[0,1]}\big|
        \bm{C}_f^\ast(t)
        \widehat{\bm{\Sigma}}_c^{-1}(\lambda)\widehat{\bm{U}}_n^{boots}\big|
        _{\bm{g}_n(t)}\right\Vert_\Psi\\
        \le & \max\left\{ \left\Vert \max_{0\le i\le r_n}\big|\bm{C}_f^\ast(t_{i,n})
        \widehat{\bm{\Sigma}}_c^{-1}(\lambda)\widehat{\bm{U}}_n^{boots}\big|
        _{\bm{g}_n(t_{i,n})}\right\Vert_\Psi, \left\Vert \max_{0\le i\le r_n-1}\sup_{s\in [t_{i,n},t_{i+1,n}]}
        \left|\left[\bm{C}_f^\ast(s)-\bm{C}_f^\ast(t_{i,n})\right]
        \widehat{\bm{\Sigma}}_c^{-1}(\lambda)\widehat{\bm{U}}_n^{boots}\right|_{\bm{g}_n(s)}\right\Vert_\Psi\right\} \\
        \le & C\sqrt{c_n \log^2(n)/n}.
     \end{align*}}
     Consequently, we have
     $III\le C\frac{\sqrt{c_n\log^2(n)/n}}{\delta_1}$. By optimizing $\delta_1$ in terms $II$ and $III$, we conclude that 
     $$P_3 \le C\left(c_n^{9/8}n^{-1/4+1/q} + c_n^{-1/4}n^{-1/2}\log^{3/4}(n) + \sqrt{c_n^{\phi+d_2+2}\log (n)/r_n}\right)=\bigO\left(
     c_n^{9/8}n^{-1/4+1/q}\right).$$
        
	Finally, we will derive the error bound of $P_4$. Denote $\bm{V}_n^c=\bm{\Sigma}_c^{-1}(\lambda)\bm{Z}_n^c$ and $\widetilde{\bm{V}}_n^c=
    \widetilde{\bm{\Sigma}}_c^{-1}(\lambda)\bm{Z}_n^c$. Following the proof of \cref{thm2}, we can obtain
    \begin{align*}
    P_4 \le & \Pr(|\widetilde{\bm{V}}_n^c
    -\bm{V}_n^c| > \epsilon) + 
    \sup_{B\in\mathcal{B}_n}\left\{
    \Pr(\bm{V}_n^c \in B^\epsilon\backslash B), \Pr(
    \bm{V}_n^c \in B\backslash B^{-\epsilon})\right\}\\
    \le & \frac{Cc_n^{3/2}\log(n)}{\epsilon \sqrt{n}} + 4c_n^{1/4}\epsilon + 2\mathcal{K}(\bm{Z}_n^c,\bm{U}_n^c)\\
    \le & C\left(c_n^{\frac{7}{8}}n^{-\frac{1}{4}}\log^{1/2}(n)+
    c_n^{\frac{7}{4}}
	n^{-\frac{1}{2}+\frac{9}{2q}+
 \frac{2}{\tau-1}}\right).
    \end{align*}
    
	In summary, we conclude 
	\begin{align*}
	\widehat{\mathcal{K}}(\widehat{\bm{U}}_n^{boots},\bm{Z}_n^c)\le& C\left(c_n^{\frac{7}{4}}n^{-\frac{1}{2}+\frac{9}{2q}+\frac{2}{\tau-1}}+
	c_n^{\frac{3}{8}}\lambda^{\frac{d_2+\psi+2}{2(2\gamma+d_2-\psi)}}+
	c_n^{\frac{5}{8}}n^{-\frac{1}{6}}h_n^{1/2} + 
    c_n^{\frac{9}{8}}n^{-\frac{1}{4}+\frac{1}{q}}\right).
	\end{align*}
	\qed
	
	\subsection{Proof of Theoretical Results in Section 5.3 of the Article}
	To prove Proposition 2 of the paper, here we consider the situation where FPCs are employed as basis functions, denoted by $\{\widetilde{\alpha}_k(t)\}_{k=1}^\infty$. Now we let $\underline{x}_{ij,k}=\langle X_{ij}(t),\widetilde{\alpha}_k(t)\rangle$, ${x}_{ij,k}^\ast=\underline{x}_{ij,k}/\underline{f}_{jk}$ where $\underline{f}_{jk}={\rm Std}(\underline{x}_{ij,k})$ and
	$\theta_{jk}^\ast=\beta_{jk}\underline{f}_{jk}$.
	\bigskip
	
	\noindent
	\textbf{Proof of Proposition 2}.
	Armed with FPCs, we denote the least squares estimator as $\bm{\theta}_c^\ast=
	[\widetilde{\bm{X}}_c^\top\widetilde{\bm{X}}_c/n+\widetilde{\bm{R}}
	(\lambda)]^{-1}\widetilde{\bm{X}}_c^\top\bm{Y}/n$ where $\widetilde{\bm{X}}_c$ and $\widetilde{\bm{R}}(\lambda)$ have similar definitions to $\bm{X}_c, \bm{R}(\lambda)$ with empirical FPCs as bases and $\underline{f}_{jk}$ to be estimated. Then we obtain
	$$\bm{\beta}^\ast(t)=\widetilde{\bm{C}}_f(t)\bm{\theta}_c^\ast,$$
	where $\widetilde{\bm{C}}_f(t)$ also has similar definition to $\bm{C}_f(t)$ with empirical FPCs and $\underline{f}_{jk}$ replaced by its estimate. In consequence, we have
	\begin{align}
	&\left|\EE[\bm{\beta}^\ast(t)]-\bm{\beta}(t)\right|\notag \\
	\le&
	\left|[\widetilde{\bm{C}}_f(t)-\bm{C}_f(t)]
	\bm{\theta}_c\right|+\left|\widetilde{\bm{C}}_f(t)
	\left(\frac{\widetilde{\bm{X}}_c^\top\widetilde{\bm{X}}_c}{n}+
	\widetilde{\bm{R}}(\lambda)\right)^{-1}\widetilde{\bm{R}}(\lambda)
	\bm{\theta}_c^\ast\right| \label{FPC1} \\
	&~+\left|\EE\widetilde{\bm{C}}_f(t)
	\left(\frac{\widetilde{\bm{X}}_c^\top\widetilde{\bm{X}}_c}{n}+
	\widetilde{\bm{R}}(\lambda)\right)^{-1}\frac{\widetilde{\bm{X}}_c^\top
		\widetilde{\bm{\epsilon}}}{n}\right|+\bigO(c_n^{-d_1}). \label{FPC2}
	\end{align}
	The first term of \cref{FPC1} describes the bias from empirical estimation for eigenfunctions, the second term captures the standard deviation of the estimation and the last term denotes the truncation error in the basis expansion of $\bm{\beta}(t)$.
	
	First, for any $t,s\in[0,1]$, let $\Gamma_{j}(t,s)={\rm cov}(X_{ij}(t),X_{ij}(s))$ be the covariance of $X_{ij}(\cdot)$ and $\widetilde{\Gamma}_{j}(t,s)$ be its sample covariance. These two quantities can be written in the eigen-decomposition (also known as the Karhunen-Lo{\`e}ve expansion)
	\begin{align*}
	\Gamma_{j}(t,s)&=\EE[X_{ij}(t)X_{ij}(s)]=\sum_{k=1}^\infty f_{jk}^2\alpha_k(t)\alpha_k(s),\\
	\widehat{\Gamma}_{j}(t,s)&=\frac{1}{n}\sum_{i=1}^nX_{ij}(t)X_{ij}(s)
	=\sum_{k=1}^\infty\widetilde{f}_{jk}^2
	\widetilde{\alpha}_k(t)\widetilde{\alpha}_k(s),
	\end{align*}
	where the sequences $f_{j1}^2>f_{j2}^2>\cdots>f_{jc_{j,n}}^2>0$ and $\widetilde{f}_{j1}^2\ge\widetilde{f}_{j2}^2\ge
 \cdots$ are the population and sample eigenvalues, $(\alpha_k(t),k\ge 1)$ and $(\widetilde{\alpha}_k(t),k\ge 1)$ are the corresponding eigenfunctions. Denote $\widetilde{\delta}_k=\min_{1\le j\le p}(f_{jk}^2-f_{j,k+1}^2)$, with Assumptions 1 and 8 in the paper, we have $\widetilde{\delta}_k \ge Ck^{-2(d_2+1)}$ for $k=1,2,\cdots$. Next, we need to prove the following statements:
	
	\begin{itemize}
		\item $\max_{1\le j\le p}\sup_{t,s\in[0,1]}\left|\widetilde{\Gamma}_{j}(t,s)-
		\Gamma_{j}(t,s)\right|=\bigO_\Pr(1/\sqrt{n})$,
		\item $\left|\widetilde{\alpha}_k(t)
		-\alpha_k(t)\right|_{\mathcal{L}^2}=\bigO\left(k^{2(d_2+1)}/\sqrt{n}\right)$.
	\end{itemize}
	Note that
	\begin{align*}
	\widetilde{\Gamma}_{j}(t,s)-\Gamma_{j}(t,s)=&\frac{1}{n}\sum_{i=1}^n[X_{ij}(t)X_{ij}(s)-\EE X_{ij}(t)X_{ij}(s)]\\
	=&\frac{1}{n}\sum_{i=1}^n\left(\sum_{k=1}^\infty\sum_{h=1}^\infty
	x_{ij,k}x_{ij,h}f_{jk}f_{jh}\alpha_k(t)\alpha_h(s)-\sum_{k=1}^\infty
	f_{jk}^2\alpha_k(t)\alpha_k(s)\right)\\
	:=&\frac{1}{n}\sum_{i=1}^n\left(\sum_{\substack{k=1\\ k=h}}^\infty b_{ij,k}f_{jk}^2\alpha_k(t)\alpha_k(s)+\sum_{\substack{k,h=1\\ k\neq h}}^\infty x_{ij,k}x_{ij,h}f_{jk}f_{jh}\alpha_k(t)\alpha_h(s)\right),
	\end{align*}
	where $b_{ij,k}=x_{ij,k}^2-1$. Further denote $\tilde{b}_{ij}^{kh}=x_{ij,k}x_{ij,h}$,
	then we can deduce the corresponding dependence measures as 
	\begin{align*}
	\delta_b(l,q/2)&=\Vert b_{ij,k}-b_{ij,k}^\ast\Vert_{q/2}\le \Vert x_{ij,k}+x_{ij,k}^\ast\Vert_q \Vert x_{ij,k}-x_{ij,k}^\ast\Vert_q \le C(l+1)^{-\tau},\\
	\delta_{\tilde{b}}(l,q/2)&=\Vert \tilde{b}_{ij}^{kh}-(\tilde{b}_{ij}^{kh})^{\ast}\Vert_{q/2}\le \Vert x_{ij,k}-x_{ij,k}^\ast\Vert_q \Vert x_{ij,h}\Vert_q +\Vert x_{ij,k}^\ast \Vert_q \Vert x_{ij,h}-x_{ij,h}^\ast\Vert_q \le C(l+1)^{-\tau},
	\end{align*} 
	where $x_{ij,k}^\ast, x_{ij,h}^\ast$ are i.i.d. copies of $x_{ij,k}, x_{ij,h}$.
	Furthermore let $B_{jk}=\sum_{i=1}^n b_{ij,k}/\sqrt{n}$ and $\tilde{B}_{j,kh}=\sum_{i=1}^n \tilde{b}_{ij}^{kh}/\sqrt{n}$, define the projection operator $$\mathcal{P}_j(\cdot)=\EE(\cdot|\mathcal{F}_j)-\EE(\cdot|\mathcal{F}_{j-1}),$$ with the result of Theorem 3 in \cite{Wu05}, we can obtain that for any $j,k,h$,
	\begin{align*}
	\Vert B_{jk}\Vert_q&\le C\sum_{l=0}^\infty\Vert\mathcal{P}_0(b_{lj,k})
	\Vert_q\le C\sum_{l=0}^\infty\Vert b_{lj,k}-b_{lj,k}^\ast\Vert_q\le C,\\
	\Vert \tilde{B}_{j,kh}\Vert_q&\le C\sum_{l=0}^\infty\Vert\mathcal{P}_0(\tilde{b}_{lj}^{kh})
	\Vert_q\le C.
	\end{align*}
	Consequently, we can deduce
	\begin{align*}
	&\left\Vert \max_{1\le j\le p}\sup_{t,s\in[0,1]}\left|\widetilde{\Gamma}_j(t,s)-\Gamma_j(t,s)\right|\right\Vert_q\\
	\le &\sum_{k=1}^\infty \sup_{t,s\in[0,1]}|\alpha_k(t)\alpha_k(s)|f_{jk}^2\Vert B_{jk}\Vert_q/\sqrt{n}+\sum_{\substack{k,h=1 \\ k\neq h}}^\infty\sup_{t\in[0,1]}|\alpha_k(t)|\sup_{s\in[0,1]}|\alpha_h(s)|f_{jk}f_{jh}\Vert \tilde{B}_{j,kh}\Vert_q/\sqrt{n}\\
	\le &C/\sqrt{n},
	\end{align*}
	where the last inequality follows by the Assumption $\sup_{t\in[0,1]}|\alpha_k(t)|\le Ck^\psi$ for any $k\ge 1$ and $f_{jk}\le Ck^{-(d_2+1)}$. Define $|g(t,s)|_{\mathcal{S}}:=\left(\int_0^1\int_0^1 g^2(t,s)\dee t\dee s\right)^{1/2}$ for some function $g(t,s)\in\mathcal{L}([0,1]^2)$, by \citet[Eq. (5.2)]{Hall07}, we conclude for any $k$,
	\begin{equation*}
	\left|\widetilde{\alpha}_k(t)-\alpha_k(t)\right|_{\mathcal{L}^2}\le  \widetilde{\delta}_k^{-1}\left|\widetilde{\Gamma}_{j}-\Gamma_{j}\right|_{\mathcal{S}}=
	\bigO\left(\frac{k^{2(d_2+1)}}{\sqrt{n}}\right).
	\end{equation*}
	Further by the result of \cref{consis} in \cref{sec2}, we can deduce that
	$$\max_{1\le j\le p}\sup_{k\ge 1}\left|\widetilde{f}_{jk}^2-f_{jk}^2\right|=\bigO_\Pr(\sqrt{\log(n)/n}).$$
	Followed by the proof of Theorem 3.6 $(c)$ in \cite{li2010uniform} we have for any $t\in[0,1]$,
	\begin{align*}
	&\widetilde{f}_{jk}^2\widetilde{\alpha}_k(t)-f_{jk}^2\alpha_k(t)\\
	=&\int_0^1\widetilde{\Gamma}_{j}(t,s)\widetilde{\alpha}_k(s)\dee s-
	\int_0^1\Gamma_{j}(t,s)\alpha_k(s)\dee s\\ =&
	\int_0^1\left[\widetilde{\Gamma}_{j}(t,s)-
	\Gamma_{j}(t,s)\right]\alpha_k(s)\dee s+
	\int_0^1\widetilde{\Gamma}_{j}(t,s)
	[\widetilde{\alpha}_k(s)-\alpha_k(s)]\dee s
	\end{align*}
	By the Cauchy–Schwartz inequality, uniformly for all $t\in[0,1]$,
	$$\int_0^1\widetilde{\Gamma}_{j}(t,s)
	[\widetilde{\alpha}_k(s)-\alpha_k(s)]\dee s\le
	C\left|\widetilde{\alpha}_k(s)-\alpha_k(s)\right|_{\mathcal{L}^2}\le Ck^{2(d_2+1)}/\sqrt{n}.$$
	On the other hand, since $\max_{1\le j\le p}\sup_{t,s\in[0,1]}\left|\widetilde{\Gamma}_j(t,s)-
	\Gamma_j(t,s)\right|=\bigO_\Pr(1/\sqrt{n})$, then we have $$\int_0^1\left[\widetilde{\Gamma}_{j}(t,s)-\Gamma_{j}(t,s)\right]\alpha_k(s)\dee s \le \frac{C}{\sqrt{n}}.$$
	In summary, $|\widetilde{f}_{jk}^2\widetilde{\alpha}_k(t)-f_{jk}^2\alpha_k(t)|=\bigO_\Pr(k^{2(d_2+1)}/\sqrt{n})$. By the triangle inequality and the above results, for any $1\le j\le p$ and $t\in[0,1]$,
	\begin{align*}
	&f_{jk}^2|\widetilde{\alpha}_k(t)-\alpha_k(t)|\\
	\le &|\widetilde{f}_{jk}^2\widehat{\alpha}_k(t)-f_{jk}^2\alpha_k(t)|+
	\sup_{k\ge 1} |\widetilde{f}_{jk}^2-f_{jk}^2|
	\sup_{t\in[0,1]}|\widetilde{\alpha}_k(t)|\\
	\le& \frac{Ck^{2(d_2+1)}}{\sqrt{n}}.
	\end{align*}
	For any $i=1,..,n,~j=1,...,p$, since $|f_{jk}^2|\ge Ck^{-2(d_2+1)}$ by Assumption 3 of the main paper, then we have for $k\ge 1$,  \begin{equation}\label{rate}
	|\widetilde{\alpha}_k(t)-\alpha_k(t)|_\infty\le Ck^{4(d_2+1)}/\sqrt{n}.
	\end{equation}
	Next, we will identify the uniform convergence property for estimator ${\beta}_j^\ast(t)$. With the above $\mathcal{L}^\infty$ convergence rate in \eqref{rate}, the bias resulted from estimated eigenfunctions at the first term of \cref{FPC1} can be obtained as
	$$\sup_{t\in[0,1]}\left|\bm{E}_j^\top\left[\widetilde{\bm{C}}_f(t)-
	\bm{C}_f(t)\right]\bm{\theta}_c\right|\le
	\sum_{k=1}^{c_{j,n}}|\widetilde{\alpha}_k(t)-\alpha_k(t)|_\infty\beta_{jk}
	=\bigO\left(\frac{c_n^{4(d_2+1)-d_1}}{\sqrt{n}}\right),$$ where $\bm{E}_j$ is a $p$-dimensional vector with $j$th element being 1 and others being 0. For the other biases (the second term in \cref{FPC1} and the first term in \cref{FPC2}) together with the standard deviation terms, one can derive the same orders as those in the proof of \cref{thm2}. Hence with Assumptions in \cref{thm2} and the extra condition $$\lambda^{-\frac{2d_2+3} {4(\gamma+d_2+1)}}>c_n^{4(d_2+1)-d_1}/\sqrt{n}$$ holds to guarantee the standard deviation of the estimation dominates, then we finish the statement of Proposition 2. \qed
	
	\subsection{Additional Lemmas}\label{sec_c5}
	In this subsection, we aim to derive the approximation of  $\widetilde{\bm{\Sigma}}_c^{-1}(\lambda)$ to $\bm{\Sigma}_c^{-1}(\lambda)$. Recall $\bm{\Sigma}_c(\lambda)=\EE\bm{X}_c^\top\bm{X}_c/n+\bm{R}(\lambda)$, then we have the following lemma.	
	\begin{lemma}\label{lemma0}
		Under Assumption 2 in the main paper, we have $$|\bm{\Sigma}_c(\lambda)-\widetilde{\bm{\Sigma}}_c(\lambda)|
		=\bigO_\Pr\left(\frac{c_n\log(n)}{\sqrt{n}}\right).$$
	\end{lemma}
	
	\begin{proof}
		Rewrite $\widetilde{\bm{\Sigma}}_c(\lambda)=\frac{1}{n}\sum_{i=1}^n\bm{x}_{ci}
		\bm{x}_{ci}^\top+\bm{R}(\lambda)$, where $\bm{x}_{ci}\in \mathbb{R}^c$ is the $i$th column of $\bm{X}_c^\top$. This proof mainly uses a Bernstein-type inequality for sums of random matrices to establish the convergence rate. For completeness, we present this inequality in the following lemma (\cite{tropp2012user}).
		
		\begin{lemma}\label{bernstein}
			Let $\{\bm{\Xi}_i\}_{i=1}^n$ be a finite sequence of independent random matrices with dimensions $d_1\times d_2$. Assume $\EE(\bm{\Xi}_i)=\bm{0}$ for each $i$,  $\max_{1\le i\le n}|\bm{\Xi}_i|\le R_n$ and define
			$$\sigma_n^2=\max\left\{\left|\sum_{i=1}^n\EE\left(\bm{\Xi}_i
			\bm{\Xi}_i^\top\right)\right|,\left|\sum_{i=1}^n\EE\left(
			\bm{\Xi}_i^\top\bm{\Xi}_i\right)\right|\right\}.$$ Then for all $t>0$,
			$$\Pr\left(\left|\sum_{i=1}^n\bm{\Xi}_i\right|\ge t\right)\le (d_1+d_2)\exp\left(\frac{-t^2/2}{\sigma_n^2+R_nt/3}\right).$$
		\end{lemma}
		To employ the above lemma, first we introduce the $m$-dependent approximation sequence to deal with the issue of independence. To be more specific, we denote
		$$\bm{x}_{ci}^m=\EE(\bm{x}_{ci}|\eta_{i-m},...,\eta_i),~i=1,\cdots,n.$$
		It is easy to find that $\bm{x}_{ci}^m$ and $\bm{x}_{cj}^m$ will be independent if $|i-j|>m$. Denote $\widetilde{\bm{\Sigma}}_c^m(\lambda):=\frac{1}{n}\sum_{i=1}^n\bm{x}_{ci}^m
		(\bm{x}_{ci}^m)^\top+\bm{R}(\lambda)$, $\mathbf{X}_i={\rm vec}(\bm{x}_{ci}\bm{x}_{ci}^\top)$ and $\mathbf{X}_i^m={\rm vec}(\bm{x}_{ci}^m(\bm{x}_{ci}^m)^\top)$. By Assumption 2 of the paper and the discussion of \cite[Remark 2.3]{Zhang18}, we can derive that 
		$\Omega_{m,q}=\bigO(c_n^{1/q}m^{-\tau+1})$. Consequently, we have 
		\begin{align}\label{proof1}
		\Pr\left(\left|\widetilde{\bm{\Sigma}}_c(\lambda)-
		\widetilde{\bm{\Sigma}}_c^m(\lambda)\right|\ge t\right)&\le \Pr\left(
		c_n\left|\widetilde{\bm{\Sigma}}_c(\lambda)-
		\widetilde{\bm{\Sigma}}_c^m(\lambda)\right|_{\max}\ge t\right) \notag \\
		&\le \Pr\left(\left|\frac{1}{\sqrt{n}}\sum_{i=1}^n\left(\mathbf{X}_i-
		\mathbf{X}_i^m\right)\right|_\infty \ge 
		\frac{t\sqrt{n}}{c_n}\right)\notag \\
		&\le \frac{C\{\log(c_n^2)\}^{q/2}c_n(m+1)^{-q(\tau-1)}}{(t\sqrt{n}/c_n)^q}.
		\end{align}
		By choosing $t\sqrt{n}/c_n=c_n^{1/q}m^{-\tau+2}$ as well as $m$ sufficiently large, armed with \cref{proof1}, we have
		$\Pr\left(\left|\widetilde{\bm{\Sigma}}_c(\lambda)-
		\widetilde{\bm{\Sigma}}_c^m(\lambda)\right|\ge t\right)=o(1)$. Furthermore, by Jensen's inequality, we conclude that
		$$|\EE(\widetilde{\bm{\Sigma}}_c(\lambda)-
		\widetilde{\bm{\Sigma}}_c^m(\lambda))| \le \EE|\widetilde{\bm{\Sigma}}_c(\lambda)-
		\widetilde{\bm{\Sigma}}_c^m(\lambda)|
		\le \frac{Cc_n^{1+1/q}m^{-\tau+2}}{\sqrt{n}}=o(1).$$
		Hence, we only need to control the $m$-dependence approximation sequence. First, we calculate some relative quantities as follows,
		\begin{align*}
		R_m&=\frac{1}{n}\sup_i\left| \bm{x}_{ci}^m(\bm{x}_{ci}^m)^\top-
		\EE\bm{x}_{ci}^m(\bm{x}_{ci}^m)^\top\right|\\
		&\le \frac{1}{n}\sup_i\left|\bm{x}_{ci}^m(\bm{x}_{ci}^m)^\top\right|\le \frac{Cc_n}{n}.
		\end{align*}
		Furthermore, we define $k_0=\lfloor\frac{n}{m}\rfloor$ and the index set sequences for $i=1,...,m$ by
		\begin{align*}
		\mathcal{I}_i=\begin{cases}
		\{i+km: k=0,1,\cdots,k_0\}, &\text{if}~i+k_0m\le n,\\
		\{i+km: k=0,1,\cdots,k_0-1\}, &\text{otherwise}.
		\end{cases}
		\end{align*}
		Then we have $\sigma_m^2\le \frac{Cc_n^2}{mn}$. Consequently, we apply the above bounds to Bernstein-type inequality, i.e.,
		\begin{align*}
		&\Pr\left(\left|\widetilde{\bm{\Sigma}}_c^m(\lambda)-
		\EE\widetilde{\bm{\Sigma}}_c^m(\lambda)\right|\ge t\right)\\ \le& Cm\sup_i\Pr\left(\left|\frac{1}{n}\sum_{k\in\mathcal{I}_i}
		\bm{x}_{ci}^m(\bm{x}_{ci}^m)^\top-\frac{1}{n}\sum_{k\in\mathcal{I}_i}
		\EE\bm{x}_{ci}^m(\bm{x}_{ci}^m)^\top\right| \ge \frac{t}{m}\right)\\
		\le &Cmc\exp\left(\frac{-t^2/(2m^2)}{\sigma_m^2+R_mt/3m}\right).
		\end{align*}
		Now, by choosing $m=\bigO(\log(n))$, $t=\bigO(\frac{c_n\log(n)}{\sqrt{n}})$ and using triangle inequality, we conclude that
		\begin{align*}
		&\left|\widetilde{\bm{\Sigma}}_c(\lambda)-\bm{\Sigma}_c(\lambda)\right| \\
		\le & \left|\widetilde{\bm{\Sigma}}_c(\lambda)-
		\widetilde{\bm{\Sigma}}_c^m(\lambda)\right|+ \left|\widetilde{\bm{\Sigma}}_c^m(\lambda)-
		\EE\widetilde{\bm{\Sigma}}_c^m(\lambda)\right|+
		\left|\EE\widetilde{\bm{\Sigma}}_c^m(\lambda)-
		\bm{\Sigma}_c(\lambda)\right|\\
		= &\bigO_\Pr\left(\frac{c_n\log(n)}{\sqrt{n}}\right).
		\end{align*}
	\end{proof}
	
	Next Lemma provides an approximation to the random term $\widetilde{\bm{\Sigma}}_c^{-1}(\lambda)$.
	\begin{lemma}\label{lemma1}
		Under Assumptions 2 and 3 in the main article, we will obtain $$|\bm{\Sigma}_c^{-1}(\lambda)-\widetilde{\bm{\Sigma}}_c^{-1}(\lambda)|
		=\bigO_\Pr\left(\frac{c_n\log(n)}{\sqrt{n}}\right).$$
	\end{lemma}
	
	\noindent
	\textbf{Proof of \cref{lemma1}}.
	Armed with the result of \cref{lemma0}, we have
	\begin{align*}
	\left|\widetilde{\bm{\Sigma}}_c^{-1}(\lambda)-
	\bm{\Sigma}_c^{-1}(\lambda)\right|&=\left|
	\widetilde{\bm{\Sigma}}_c^{-1}(\widetilde{\bm{\Sigma}}_c(\lambda)-
	\bm{\Sigma}_c(\lambda))\bm{\Sigma}_c^{-1}(\lambda)\right|\\ &\le \left|\widetilde{\bm{\Sigma}}_c^{-1}(\lambda)\right| \left|\widetilde{\bm{\Sigma}}_c(\lambda)-\bm{\Sigma}_c(\lambda)\right| \left|\bm{\Sigma}_c^{-1}(\lambda)\right|\\ &\le
	C\left(\frac{c_n\log(n)}{\sqrt{n}}\right).
	\end{align*} \qed
	
	\begin{lemma}\label{prop3}
		Suppose Assumptions 2 and 4 in the paper hold with $q=4$ and $m\to \infty$ with $m/n\to 0$. Then we have
		\begin{equation*}\label{Lambda}
			\left\Vert\widetilde{\Xi}_{jk}^c-
			\Xi_{jk}^c\right\Vert_2=\bigO\left(\frac{1}{m}+\sqrt{\frac{m}{n}}\right),
		\end{equation*}
		where $\widetilde{\Xi}_{jk}^c$ and $\Xi_{jk}^c$ are $(j,k)$th elements of $\widetilde{\bm{\Xi}}^c$ and $\bm{\Xi}^c$, respectively.
	\end{lemma}
	
	To prove the above lemma, we will follow the proof strategy of \cite{Zhou13}[Theorem 3] and following lemmas are needed. Denote $\bm{S}_{i,m}=\sum_{j=i}^{i+m-1}\bm{z}_{cj}$ and its $k$th entry by $S_{i,k,m}$.  Recall $$\widetilde{\bm{\Xi}}^c=\frac{1}{(n-m+1)m}\sum_{i=1}^{n-m+1}\bm{S}_{i,m}
	\bm{S}_{i,m}^\top,$$ for $k\in\mathbb{Z}$, define the projection operator $$\mathcal{P}_k\cdot=\EE(\cdot|\mathcal{F}_k)-\EE(\cdot|\mathcal{F}_{k-1}).$$
	
	\begin{lemma}\label{lemma5}
		Under conditions of \cref{prop3}, for any $1\le j,k\le c_n$, we have $\Vert\widetilde{\Xi}_{jk}-
		\EE[\widetilde{\Xi}_{jk}]\Vert_2=\bigO(\sqrt{m/n})$.
	\end{lemma}
	\begin{proof}
		Since $S_{i,k,m}$ is $\mathcal{F}_{i+m-1}$ measurable, it can be written as $f(\mathcal{F}_{i+m-1})$. For $j,l\in \mathbb{Z}$, define
		$$S_{i,k,m}^l=f(\mathcal{F}_{i+m-1,\{i-l\}}),~~~\mathcal{F}_{j,\{i-l\}}=
		(\cdots,\eta_{i-l-1},\eta_{i-l}',\eta_{i-l+1},\cdots,\eta_j),$$ where $\mathcal{F}_{j,\{i-l\}}$ is obtained by replacing $\eta_{i-l}$ in $\mathcal{F}_j$ by an i.i.d. copy $\eta_{i-l}'$ and $\mathcal{F}_{j,\{i-l\}}=\mathcal{F}_j$ if $i-l>j$. By \cite{Zhou13}[Lemma 6(i)], Assumptions 1 and 4 with $q=4$, we have $\sup_{i,k}\Vert S_{i,k,m}\Vert_4=\sup_{i,k}\Vert S_{i,k,m}^l\Vert_4=\bigO(\sqrt{m})$. Since
		\begin{equation*}
			\left\Vert S_{i,k,m}-S_{i,k,m}^l\right\Vert_4\le \sum_{j=l}^{l+m-1}\delta_z(j,4),
		\end{equation*}
		we have
		\begin{align*}
			\left\Vert S_{i,j,m}S_{i,k,m}-S_{i,j,m}^lS_{i,k,m}^l\right\Vert_2&\le
			\left\Vert S_{i,j,m}\right\Vert_4\left\Vert S_{i,k,m}-S_{i,k,m}^l\right\Vert_4+
			\left\Vert S_{i,j,m}-S_{i,k,m}^l\right\Vert_4\left\Vert S_{i,k,m}^l\right\Vert_4\\
			&=\bigO(\sqrt{m})\sum_{i=l}^{l+m-1}\delta_z(i,4).
		\end{align*}
		By \cite{Wu05}[Theorem 1], $\Vert\mathcal{P}_{i-l}(S_{i,j,m}S_{i,k,m})\Vert_2 \le \Vert S_{i,j,m}S_{i,k,m}-S_{i,j,m}^l S_{i,k,m}^l\Vert_2$. Further denote $$\Psi_{jk}^l=\frac{1}{(n-m+1)m}\sum_{i=1}^{n-m+1}\mathcal{P}_{i-l}
		(S_{i,j,m}S_{i,k,m}).$$ Note that $\mathcal{P}_{i-l}(S_{i,j,m} S_{i,k,m})$ for $1\le i\le n-m+1$ are martingale differences, then we have $\Vert\Psi_{jk}^l\Vert_2^2=\bigO\left(\frac{1}{m(n-m+1)}\right)
		\left\{\sum_{i=l}^{l+m-1}\delta(i,4)\right\}^2$. Since $\widetilde{\Xi}_{jk}-\EE[\widetilde{\Xi}_{jk}]=\sum_{i=0}^\infty
		\Psi_{jk}^i$ and $\sum_{i=0}^\infty\delta_z(i,4)<\infty$, we have  $\Vert\widetilde{\Xi}_{jk}-\EE[\widetilde{\Xi}_{jk}]\Vert_2
		=\bigO(\sqrt{m/n})$.
	\end{proof}
	
	\begin{lemma}\label{lemma6}
		Under conditions of \cref{prop3}, for any $1\le j,k\le c_n$, we have
		$$|\EE[\widetilde{\Xi}_{jk}]-\Xi_{jk}|=\bigO(1/m).$$
	\end{lemma}
	
	\begin{proof}
		Notice that $z_{ci,j}=H_j(\mathcal{F}_i)$ is a stationary time series. Let $\Gamma_i(l)$ be its $l$th autocovariance, then we obtain that for any $1\le i\le n-m+1$, $|\Gamma_i(l)|\le C(l+1)^{-\tau}$ by Assumption 2 and 4 of the paper. Therefore,
		\begin{align*}
			&|\EE[\widetilde{\Xi}_{jk}]-\Xi_{jk}|\\
			=&\left|\frac{1}{(n-m+1)m}\sum_{i=1}^{n-m+1}\left\{\EE(S_{i,j,m}
			S_{i,k,m})-\frac{m}{n}\EE\left(\sum_{i=1}^nz_{ci,j}\right)
			\left(\sum_{i=1}^nz_{ci,k}\right)\right\}\right|\\
			\le &\frac{2}{m}\max_{1\le i\le n-m+1}\left\{\sum_{l=0}^{m-1}j|\Gamma_i(l)|
			+m\sum_{l\ge m}|\Gamma_i(l)| \right\}=\bigO(1/m).
		\end{align*}
	\end{proof}
	
	\noindent
	\textbf{Proof of \cref{prop3}}.
	Combining the results of \cref{lemma5} and \cref{lemma6}, we complete the proof. \qed
 
% \small
 %\bibliography{scalar} 

\end{document}